***Roadmap: 2D Materials for Quantum Technologies***

Qimin Yan[1*], Tongcang Li[2,3,4,5], Xingyu Gao[2], Sumukh Vaidya[2], Saakshi Dikshit[3], Yue Luo[6], Stefan Strauf[7], Reda Moukaouine[8,9], Anton Pershin[9,10], Adam Gali[9,10,11], Zhenyao Fang[1], Harvey Stanfield[12], Ivan J. Vera-Marun[12], Michael Newburger[13], Simranjeet Singh[14], Tiancong Zhu[2,4,5], Mauro Brotons-Gisbert[15], Klaus D. Jöns[16], Brian D. Gerardot[15], Brian S. Y. Kim[17,18], John R. Schaibley[18], Kyle L. Seyler[19], Jesse Balgley[20], James Hone[20], Kin Chung Fong[1,21,22], Lin Wang[23], Guido Burkard[24], Yihang Zeng[2], Tobias Heindel[25], Serkan Ateş[26], Tobias Vogl[27] and Igor Aharonovich[28, 29].

[1] Department of Physics, Northeastern University, Boston, USA
[2] Department of Physics and Astronomy, Purdue University, West Lafayette, IN, 47907, USA
[3] Elmore Family School of Electrical and Computer Engineering, Purdue University, West Lafayette, IN, 47907, USA
[4] Purdue Quantum Science and Engineering Institute, Purdue University, West Lafayette, IN, 47907, USA
[5] Birck Nanotechnology Center, Purdue University, West Lafayette, IN, 47907, USA
[6] School of Electronic Science and Engineering, Southeast University, Nanjing, 211189, China
[7] Department of Physics, Stevens Institute of Technology, Hoboken, New Jersey 07030, United States
[8] György Hevesy Doctoral School, ELTE Eötvös Loránd University, Institute of Chemistry, Budapest, Hungary.
[9] HUN-REN Wigner Research Center for Physics, Institute for Solid State Physics and Optics, Budapest, Hungary.
[10] Department of Atomic Physics, Budapest University of Technology and Economics, Budapest, Hungary.
[11] MTA-WFK Lendület "Momentum" Semiconductor Nanostructures Research Group, Budapest, Hungary.
[12] Department of Physics and Astronomy, The University of Manchester, Manchester, United Kingdom
[13] Materials and Manufacturing Directorate, Air Force Research Laboratory, Wright-Patterson AFB, OH, 45433, USA
[14] Department of Physics, Carnegie Mellon University, Pittsburgh, PA, 15213, USA
[15] Institute of Photonics and Quantum Sciences, SUPA, Heriot-Watt University, EH14 4AS, United Kingdom
[16] Institute for Photonic Quantum Systems (PhoQS), Center for Optoelectronics and Photonics Paderborn (CeOPP) and Department of Physics, Paderborn University, 33098 Paderborn, Germany
[17] Department of Materials Science & Engineering, University of Arizona, Tucson, Arizona 85721, USA
[18] Department of Physics, University of Arizona, Tucson, Arizona 85721, USA
[19] College of Optical Sciences, University of Arizona, Tucson, Arizona 85719, USA
[20] Department of Mechanical Engineering, Columbia University, New York, NY 10027, USA
[21] Department of Electrical and Computer Engineering, Northeastern University, Boston, MA 02115, USA
[22] Quantum Materials and Sensing Institute, Burlington, MA 01803, USA
[23] Institute for Advanced Simulation (IAS-4), Forschungszentrum Jülich, Germany
[24] Department of Physics, University of Konstanz, D-78457 Konstanz, Germany
[25] Department for Quantum Technology, University of Münster, Heisenbergstraße 11, 48149 Münster, Germany
[26] Faculty of Engineering and Natural Sciences, Sabanci University, 34956, Istanbul, Turkey
[27] TUM School of Computation, Information and Technology, Technical University of Munich, 80333 Munich, Germany

[28] School of Mathematical and Physical Sciences, University of Technology Sydney, Ultimo, New South Wales 2007, Australia
[29] ARC Centre of Excellence for Transformative Meta-Optical Systems, University of Technology Sydney, Ultimo, New South Wales 2007, Australia

*E-mails: q.yan@northeastern.edu

**Abstract**

Two-dimensional (2D) materials have emerged as a versatile and powerful platform for quantum technologies, offering atomic-scale control, strong quantum confinement, and seamless integration into heterogeneous device architectures. Their reduced dimensionality enables unique quantum phenomena, including optically addressable spin defects, tunable single-photon emitters, low-dimensional magnetism, gate-controlled superconductivity, and correlated states in moiré superlattices. This Roadmap provides a comprehensive overview of recent progress and future directions in exploiting 2D materials for quantum sensing, computation, communication, and simulation. We survey advances spanning spin defects and quantum sensing, quantum emitters and nonlinear photonics, computational theory and data-driven discovery of quantum defects, spintronic and magnonic devices, cavity-engineered quantum materials, superconducting and hybrid quantum circuits, quantum dots, Moiré quantum simulators, and quantum communication platforms. Across these themes, we identify common challenges in defect control, coherence preservation, interfacial engineering, and scalable integration, alongside emerging opportunities driven by machine-learning-assisted design and integrated experiment–theory feedback loops. By connecting microscopic quantum states to mesoscopic excitations and macroscopic device architectures, this Roadmap outlines a materials-centric framework for integrating coherent quantum functionalities and positions 2D materials as foundational building blocks for next-generation quantum technologies.

# Table of Contents

# 1. Introduction

**Qimin Yan[1*] and Tongcang Li[2,3,4,5]**

[1] Department of Physics, Northeastern University, Boston, USA
[2] Department of Physics and Astronomy, Purdue University, West Lafayette, IN, 47907, USA
[3] Elmore Family School of Electrical and Computer Engineering, Purdue University, West Lafayette, IN, 47907, USA
[4] Purdue Quantum Science and Engineering Institute, Purdue University, West Lafayette, IN, 47907, USA
[5] Birck Nanotechnology Center, Purdue University, West Lafayette, IN, 47907, USA

*E-mail: q.yan@northeastern.edu

Comprising atomically thin crystals with exceptional tunability, two-dimensional (2D) materials exhibit diverse quantum phenomena arising from reduced dimensionality, strong electronic correlations, and symmetry-governed topological effects. Their structural flexibility and seamless heterointegration make them an ideal platform for realizing and manipulating quantum states of matter at the nanoscale. In the past decade, research on 2D materials has evolved from the discovery of graphene to an expanding family of semiconducting, magnetic, superconducting, and topological systems. Transition-metal dichalcogenides (TMDs), hexagonal boron nitride, black phosphorus, and various 2D oxides and nitrides have revealed rich phase diagrams, including charge-density-wave, Mott insulating, and unconventional superconducting states. The ability to stack different 2D layers with controlled twist angles—forming moiré superlattices—has further enabled flat-band physics, correlated electron behavior, and artificial quantum confinement with unprecedented precision.

Quantum technologies aim to control and exploit the fundamental principles of quantum mechanics for computation, communication, sensing, and simulation. Realizing these goals requires materials that can host coherent quantum states, support precise manipulation, and enable scalable device integration. 2D materials offer distinct advantages across multiple modalities for quantum technologies. In quantum computing, they enable spin–valley qubits, superconducting Josephson junctions, and hybrid van der Waals architectures. In quantum communication, their optically addressable defect states and single-photon emitters promise scalable quantum light sources. In quantum sensing, spin defects in 2D materials provide highly sensitive probes of magnetic, electric, and thermal fields with nanoscale resolution. Moreover, 2D materials serve as versatile interfaces in heterostructures coupling quantum systems such as color centers, superconducting circuits, and photonic cavities. Their reduced dimensionality, tunable electronic structure, and clean, reconfigurable interfaces make them ideal for engineering quantum states and couplings from the atomic to the macroscopic scale. Over the past decade, 2D materials have moved from model systems for condensed-matter physics to central components in the design of quantum technologies.

At the atomic level, 2D materials contain localized excitonic and electronic states that can act as qubits, photon sources, or quantum sensors. Point defects in wide-band-gap 2D materials such as hexagonal boron nitride (hBN) exhibit stable optical transitions and spin coherence analogous to nitrogen-vacancy centers in diamond, but with surface accessibility and integration flexibility. Controlled defect creation—by ion implantation, electron irradiation, or laser processing—has achieved near-deterministic positioning and charge control. In TMDs, strain-localized excitons act as bright, tunable quantum emitters, with energies adjusted by local strain and electrostatic environment.

Spin defects in 2D materials also serve as sensitive probes of local fields. Their planar geometry places active spins within angstroms of a surface, enabling nanoscale imaging of magnetic textures, strain, and temperature distributions. Optically detected magnetic resonance of boron-vacancy centers in hBN has already visualized 2D magnetization and spin-wave dynamics in nearby magnets. Although coherence times are limited by nuclear-spin noise and disorder, isotopic purification and dynamical-decoupling schemes continue to improve sensitivity.

Advances in electronic-structure computational theory are refining our understanding of how reduced screening, strong excitonic effects, substrate interactions, and spin–phonon processes determine defect energetics, optical transitions, and coherence properties. At the same time, high-throughput calculations and machine-learning workflows are opening the door to broad surveys of hosts, defect types, and charge states, enabling researchers to navigate the large configuration space and identify realistic candidates with favorable formation energetics, electronic structure, and spin signatures. When taken together, these perspectives point toward a more integrated framework in which accurate modelling of individual defect centers and large-scale screening efforts reinforce one another, which is essential for accelerating the development of quantum defects as reliable building blocks for emerging 2D quantum technologies.

At a collective level, intrinsic magnetic 2D crystals such as $CrI_3$ and $Fe_5GeTe_2$ support magnon excitations that can couple to microwave photons and spin qubits. Their magnetic anisotropy and damping are tunable by gating or strain, creating reconfigurable magnonic elements. Graphene-based spin valves, meanwhile, demonstrate long spin diffusion lengths and efficient spin injection, forming a bridge between conventional spintronics and quantum transport.

Strong excitonic effects and nonlinear optical responses make 2D materials key components in quantum photonics. Broken inversion symmetry and valley-dependent selection rules yield highly efficient single-photon generation and second-harmonic emission. Coupling 2D materials to photonic cavities or plasmonic resonators enhances emission and enables strong exciton–photon coupling. Twisted and moiré heterostructures create periodic confinement for excitons, producing arrays of near-identical quantum emitters. Beyond emission control, cavity coupling can modify ground-state properties through vacuum-field interactions, potentially inducing polaritonic phases or light-driven superconductivity. These developments transform 2D materials from passive emitters into active platforms for quantum light generation and control.

Atomically thin superconductors and semiconductors now play integral roles in superconducting and hybrid quantum circuits. Materials such as $NbSe_2$, $TaS_2$, and $MoTe_2$ form clean Josephson junctions and gate-tunable weak links, while hBN and other 2D insulators provide low-loss dielectrics for transmon and fluxonium qubits. Their crystalline uniformity and sharp interfaces reduce decoherence relative to amorphous oxides. Graphene's small heat capacity and weak electron–phonon coupling enable broadband single-photon detection and microwave-to-optical transduction, linking disparate quantum subsystems. Together, these results show how 2D materials contribute directly to scalable, multifunctional quantum hardware.

Moiré superlattices—formed by twisting or lattice-mismatching adjacent 2D layers—provide a powerful platform for simulating correlated and topological quantum phases. Flat electronic bands in twisted bilayer graphene and TMD heterostructures enhance electron–electron interactions, producing Mott insulators, superconductors, and Chern states that can be tuned by electrostatic gating or pressure. These systems combine atomic-scale structural control with macroscopic coherence, offering programmable realizations of strongly interacting Hamiltonians. They represent a natural bridge between material design and quantum simulation.

Recent progress in quantum communication using 2D materials demonstrates the growing breadth of quantum functionalities achievable in atomically thin systems. Single-photon emitters in hBN and strain-localized excitons in $WSe_2$ have enabled the first demonstrations of quantum key distribution (QKD) using 2D materials, achieving competitive key rates and low error ratios. Advances in deterministic emitter creation, spectral control, and nanophotonic coupling highlight the potential of 2D materials to support integrated, scalable quantum communication architectures.

2D materials thus enable a hierarchical view of quantum technology. At the microscopic level, defects and localized excitons provide addressable qubits and sensors. At the mesoscopic level, magnons, phonons, and excitons couple quantum subsystems. At the macroscopic level, moiré lattices and superconducting circuits realize programmable Hamiltonians and correlated quantum phases. Integrating these layers within a single 2D platform point toward multifunctional quantum devices capable of sensing, computation, and communication.

This *Roadmap on 2D Materials for Quantum Technologies* surveys these developments and outlines the opportunities and challenges. Developments across 12 topical areas are featured, including quantum sensing and spin dynamics, computational modeling and data-driven discovery of quantum defects, nonlinear and cavity quantum photonics, spintronic and magnonic transport, superconducting and hybrid quantum circuits, correlated and moiré-engineered quantum matter, and quantum communication using light sources in 2D materials.

Across all these topics, the unifying objective is the integration of coherent quantum functions into scalable devices. Looking ahead, integrating 2D materials into functional quantum devices requires a synergistic approach combining synthesis, characterization, theory, and machine learning–assisted design. Key challenges include atomic-level control of defects, interfacial coherence, and environmental stability, as well as scalable fabrication compatible with quantum architectures. The field is moving rapidly toward establishing a unified material–device framework, where quantum functionalities are predicted and realized through closed-loop design principles. As experiment, theory, computation, and synthesis continue to converge, the ability to assemble and couple quantum systems with atomic precision will redefine the relationship between material and device. As a result, 2D materials stand poised to become foundational building blocks for the next generation of quantum technologies.

We expect this Roadmap to be useful to both new and established researchers working with 2D materials in quantum science and technology. Our aim is to offer a clear overview of the current status of the field and to outline the opportunities and challenges that lie ahead. We also hope it will serve as a helpful reference for funding agencies and governmental organizations seeking to understand the scientific and technological directions emerging from this rapidly developing area.

## 2. Quantum sensing with spin defects in 2D materials

**Xingyu Gao[1], Sumukh Vaidya[1], Saakshi Dikshit[2], and Tongcang Li[1,2,3,4,*]**

[1] Department of Physics and Astronomy, Purdue University, West Lafayette, IN, 47907, USA
[2] Elmore Family School of Electrical and Computer Engineering, Purdue University, West Lafayette, IN, 47907, USA
[3] Purdue Quantum Science and Engineering Institute, Purdue University, West Lafayette, IN, 47907, USA
[4] Birck Nanotechnology Center, Purdue University, West Lafayette, IN, 47907, USA

*E-mail: tcli@purdue.edu

**Status**

Quantum sensing has emerged as a powerful technique that leverages the quantum properties of matter to detect physical, chemical, and biological signals with exceptional sensitivity. A particularly impactful class of quantum sensors is based on spin defects in solids, where the spin degree of freedom serves as a highly sensitive probe of external perturbations. The nitrogen-vacancy (NV) center in diamond has been a leading example, enabling nanoscale imaging with single-spin sensitivity. However, its performance is limited by the geometry and thickness of the bulk diamond host. In contrast, two-dimensional (2D) materials offer atomically thin platforms capable of hosting spin defects at the surface, allowing quantum sensors to be positioned just angstroms from the target. This proximity can dramatically boost sensitivity and spatial resolution for nanoscale sensing.

As shown in Figure 1, hexagonal boron nitride (hBN), an insulating 2D material with a wide bandgap of about 6 eV, has emerged as a new platform for quantum sensing. In 2020, Gottscholl *et al.* demonstrated optically detected magnetic resonance (ODMR) of spin defects in hBN at room temperature [1]. This defect, identified as the negatively charged boron vacancy ($V_B^-$), has a spin-1 ground state with a zero-field splitting of 3.47 GHz. This marks the first 2D analog to the diamond NV center. Since then, a different class of spin-active defects in hBN was also discovered [2-5], exhibiting negligible or no zero-field splitting and later identified as spin-1/2 centers associated with carbon impurities.

Today, **quantum sensing with hBN spin defects** is a rapidly expanding field. Optically addressable spin defects in hBN have been integrated into proof-of-concept sensors for detecting nanoscale magnetism, temperature, strain, and nuclear spins [6-9]. Multilayer hBN flakes hosting $V_B^-$ ensembles have been used to image ferromagnetic domains in 2D magnets [6,7], and a few-layer hBN sensor has detected spin wave excitations in a magnetic insulator (Fig. 1(a), 1(b)) [10]. The 2D sensor enables conformal contact with samples and flexible integration, surpassing bulk crystal limitations. Recent work also demonstrates the coexistence of spin-1 and spin-1/2 defects in hBN, enabling vector magnetometry and probing magnetic anisotropy using orientation-independent spin-1/2 centers (Fig. 1(c), 1(d)) [11]. Such multi-species spin architectures open new pathways toward advanced sensing protocols that are not achievable with single-type spin defects.

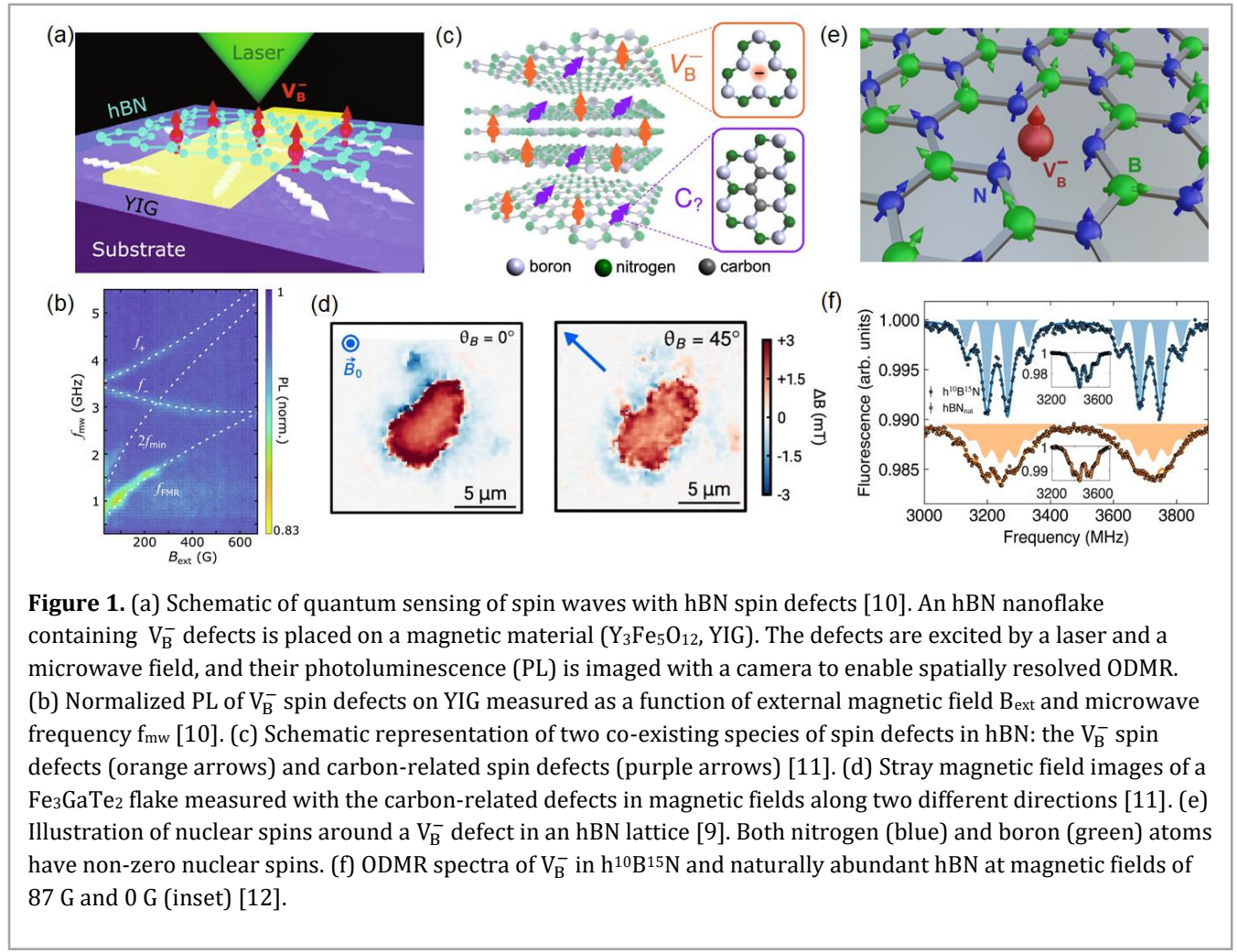

**Figure 1.** (a) Schematic of quantum sensing of spin waves with hBN spin defects [10]. An hBN nanoflake containing $V_B^-$ defects is placed on a magnetic material (Y$_3$Fe$_5$O$_{12}$, YIG). The defects are excited by a laser and a microwave field, and their photoluminescence (PL) is imaged with a camera to enable spatially resolved ODMR. (b) Normalized PL of $V_B^-$ spin defects on YIG measured as a function of external magnetic field B$_{ext}$ and microwave frequency f$_{mw}$ [10]. (c) Schematic representation of two co-existing species of spin defects in hBN: the $V_B^-$ spin defects (orange arrows) and carbon-related spin defects (purple arrows) [11]. (d) Stray magnetic field images of a Fe$_3$GaTe$_2$ flake measured with the carbon-related defects in magnetic fields along two different directions [11]. (e) Illustration of nuclear spins around a $V_B^-$ defect in an hBN lattice [9]. Both nitrogen (blue) and boron (green) atoms have non-zero nuclear spins. (f) ODMR spectra of $V_B^-$ in h$^{10}$B$^{15}$N and naturally abundant hBN at magnetic fields of 87 G and 0 G (inset) [12].

## Current and future challenges

Despite rapid progress, several challenges remain in optimizing spin defects in 2D materials for quantum sensing applications. A major hurdle is the short coherence times of spin defects in hBN, which are critical for sensing performance. At room temperature, the coherence times of hBN spin defects are typically range from tens to hundreds of nanoseconds, which are significantly shorter than those of diamond NV centers. This limitation arises from interactions with environmental nuclear spins, as both boron ($^{10}$B, $^{11}$B) and nitrogen ($^{14}$N, $^{15}$N) nuclei in hBN possess non-zero nuclear spins, forming a dense spin bath that causes decoherence (Fig. 1(e)) [9]. Isotopic engineering, such as utilizing enriched $^{10}$B and $^{15}$N hBN, has shown modest improvements in coherence times (Fig. 1(f)) [12,13], and dynamical decoupling techniques have been employed to further increase the coherence time [14, 15]. However, fully mitigating nuclear spin-induced decoherence remains unfeasible for hBN-based quantum sensors.

Another major challenge is the scalable and reproducible creation of spin defects at precise locations, particularly at the single-defect level. Existing methods such as neutron/electron irradiation and ion implantation typically yield stochastic defect distributions. $V_B^-$ defects also suffer from low quantum efficiency. Carbon-related defects exhibit substantially higher brightness and enable single-photon emission. However, their more complex chemical structure requires carbon ion implantation followed by high-temperature annealing. As a result, deterministic creation of single carbon-related defects with precise spatial control remains challenging.

Exploring alternative 2D host materials beyond hBN, especially those composed of nuclear spin-free isotopes, is vital for advancing quantum sensing. Transition metal dichalcogenides (TMDs), for example, have been theoretically predicted to host optically addressable spin defects with promising characteristics, such as spin-active ground states and strong zero-phonon coupling to optical transitions in the telecom band [16, 17]. However, experimental realization of such defects remains exclusive. One major hurdle is the reliable prediction of spin-active defects. First-principle computational methods often face limitations in modelling the complex interactions in 2D materials, leading to uncertainties in predicted defect properties. Additionally, deterministic creation of target defects remains a significant technical hurdle.

**Advances in science and technology to meet challenges**

Significant efforts are being made to address the current challenges related to coherence and functionality of spin defects in 2D materials. To enhance electron spin coherence, isotope engineering and dynamical decoupling techniques have been combined in hBN, further improving spin coherence and thus enabling enhanced sensing sensitivity [12]. Isotopic purification results in well-resolved hyperfine structures, allowing for significantly higher-fidelity control of individual electron-spin transitions. Consequently, dynamical decoupling sequences can be optimized further through strategies designed to suppress pulse errors. Beyond coherence improvement, nuclear spins associated with defects themselves have emerged as valuable quantum resources, often possessing much longer coherence times than electron spins. For example, the $V_B^-$ defects exhibit relatively strong hyperfine interactions with nearby nitrogen nuclear spins, facilitating advanced quantum control and memory capabilities [9, 12]. Furthermore, carbon-related defects introduced via controlled $^{13}C$ implantation demonstrate exceptionally strong electron-nuclear coupling (up to 300 MHz), enabling coherent control of single nuclear spins with coherence times exceeding one hundred microseconds at room temperature [18].

Progress in novel geometries also significantly expands the application potential of spin defects. Spin defects incorporated into boron nitride nanotubes (BNNT) have recently been employed for scanning probe magnetometry [19]. Due to their intrinsic quasi-1D geometry and orientation-independent spin response, BNNT-based sensors offer omnidirectional magnetic field detection, which is highly advantageous for scanning probe techniques.

Finally, exploration of new host materials beyond hBN is crucial for next-generation quantum sensors. The spin-pair model, initially proposed to explain spin dynamics of carbon-related defects in hBN, has offered new theoretical insights for exploring and identifying similar spin-pair defects in other layered materials. Recently, spin-1/2 defects have been experimentally observed in layered $GeS_2$ [20, 21]. Given that $GeS_2$ has much less nuclear spins than hBN, further improvements in coherence with better samples appear promising. Such new materials present exciting opportunities for quantum sensing, offering significant improvements over current state-of-the-art systems.

**Concluding remarks**

Quantum sensing with spin defects in 2D materials has rapidly progressed from a theoretical concept to a vibrant and maturing field. Milestones such as the discovery and control of spin defects in hBN and the demonstration of nanoscale magnetic imaging have confirmed both the feasibility and unique advantages of 2D quantum sensors. While coherence enhancement remains a central challenge, the overall trajectory is highly promising. Continued advancements are anticipated across multiple fronts, including materials development (e.g., higher purity, tailored defect engineering, exploration of new host materials), quantum control (e.g., longer T2 times, nuclear spin utilization), and device integration (e.g., on-chip photonics, novel sensor architectures).

**Acknowledgements**

T.L. acknowledges the support by the Gordon and Betty Moore Foundation, grant DOI 10.37807/gbmf12259.

**References**

[1] Gottscholl A, Kianinia M, Soltamov V, Orlinskii S, Mamin G, Bradac C, Kasper C, Krambrock K, Sperlich A, Toth M, Aharonovich I, Dyakonov V 2020 Initialization and read-out of intrinsic spin defects in a van der Waals crystal at room temperature *Nature materials* **19** 540

[2] Mendelson N, Chugh D, Reimers JR, Cheng TS, Gottscholl A, Long H, Mellor CJ, Zettl A, Dyakonov V, Beton PH, Novikov SV, Jagadish C, Tan HH, Ford MJ, Toth M, Bradac C, Aharonovich I 2021 Identifying carbon as the source of visible single-photon emission from hexagonal boron nitride *Nature materials* **20** 321

[3] Chejanovsky N, Mukherjee A, Geng J, Chen YC, Kim Y, Denisenko A, Finkler A, Taniguchi T, Watanabe K, Dasari DB, Auburger P, Gali A, Smet JH, Wrachtrup J 2021 Single-spin resonance in a van der Waals embedded paramagnetic defect *Nature materials* **20** 1079

[4] Stern HL, Gu Q, Jarman J, Eizagirre Barker S, Mendelson N, Chugh D, Schott S, Tan HH, Sirringhaus H, Aharonovich I, Atatüre M 2022 Room-temperature optically detected magnetic resonance of single defects in hexagonal boron nitride *Nature communications* **13** 618

[5] Guo NJ, Li S, Liu W, Yang YZ, Zeng XD, Yu S, Meng Y, Li ZP, Wang ZA, Xie LK, Ge RC, Wang JF, Li Q, Xu JS, Wang YT, Tang JS, Gali A, Li CF, Guo GC 2023 Coherent control of an ultrabright single spin in hexagonal boron nitride at room temperature *Nature communications* **14** 2893

[6] Healey AJ, Scholten SC, Yang T, Scott JA, Abrahams GJ, Robertson IO, Hou XF, Guo YF, Rahman S, Lu Y, Kianinia M, Aharonovich I, Tetienne JP 2023 Quantum microscopy with van der Waals heterostructures *Nature Physics* **19** 87

[7] Huang M, Zhou J, Chen D, Lu H, McLaughlin NJ, Li S, Alghamdi M, Djugba D, Shi J, Wang H, Du CR 2022 Wide field imaging of van der Waals ferromagnet Fe3GeTe2 by spin defects in hexagonal boron nitride *Nature communications* **13** 5369

[8] Lyu X, Tan Q, Wu L, Zhang C, Zhang Z, Mu Z, Zúñiga-Pérez J, Cai H, Gao W 2022 Strain quantum sensing with spin defects in hexagonal boron nitride *Nano Letters* **22** 6553

[9] Gao X, Vaidya S, Li K, Ju P, Jiang B, Xu Z, Allcca AE, Shen K, Taniguchi T, Watanabe K, Bhave SA, Chen YP, Ping Y, Li T 2022 Nuclear spin polarization and control in hexagonal boron nitride *Nature Materials* **21** 1024

[10] Zhou J, Lu H, Chen D, Huang M, Yan GQ, Al-Matouq F, Chang J, Djugba D, Jiang Z, Wang H, Du CR 2024 Sensing spin wave excitations by spin defects in few-layer-thick hexagonal boron nitride *Science Advances* **10** eadk8495.

[11] Scholten SC, Singh P, Healey AJ, Robertson IO, Haim G, Tan C, Broadway DA, Wang L, Abe H, Ohshima T, Kianinia M Reineck P, Aharonovich I, Tetienne JP 2024 Multi-species optically addressable spin defects in a van der Waals material *Nature Communications* **15** 6727

[12] Gong R, Du X, Janzen E, Liu V, Liu Z, He G, Ye B, Li T, Yao NY, Edgar JH, Henriksen EA, Zu C 2024 Isotope engineering for spin defects in van der Waals materials *Nature Communications* **15** 104

[13] Haykal A, Tanos R, Minotto N, Durand A, Fabre F, Li J, Edgar JH, Ivady V, Gali A, Michel T, Dréau A, Gil B, Cassabois G, Jacques V 2022 Decoherence of $V_B^-$ spin defects in monoisotopic hexagonal boron nitride *Nature communications* **13** 4347

[14] Ramsay AJ, Hekmati R, Patrickson CJ, Baber S, Arvidsson-Shukur DR, Bennett AJ, Luxmoore IJ 2023 Coherence protection of spin qubits in hexagonal boron nitride *Nature Communications* **14** 461

[15]     Rizzato R, Schalk M, Mohr S, Hermann JC, Leibold JP, Bruckmaier F, Salvitti G, Qian C, Ji P, Astakhov GV, Kentsch U., Helm M, Stier AV, Finley JJ, Bucher DB 2023 Extending the coherence of spin defects in hBN enables advanced qubit control and quantum sensing *Nature Communications* **14** 5089

[16]     Lee Y, Hu Y, Lang X, Kim D, Li K, Ping Y, Fu KM, Cho K 2022 Spin-defect qubits in two-dimensional transition metal dichalcogenides operating at telecom wavelengths *Nature Communications* **13** 7501

[17]     Tsai JY, Pan J, Lin H, Bansil A, Yan Q 2022 Antisite defect qubits in monolayer transition metal dichalcogenides *Nature communications* **13**, 492.

[18]     Gao X, Vaidya S, Li K, Ge Z, Dikshit S, Zhang S, Ju P, Shen K, Jin Y, Ping Y, Li T 2025 Single nuclear spin detection and control in a van der Waals material. *Nature* **643** 943

[19]     Gao X, Vaidya S, Dikshit S, Ju P, Shen K, Jin Y, Zhang S, Li T 2024 Nanotube spin defects for omnidirectional magnetic field sensing *Nature Communications* **15** 7697

[20]     Liu W, Li S, Guo NJ, Zeng XD, Xie LK, Liu JY, Ma YH, Wu YQ, Wang YT, Wang ZA, Ren JM, Ao C, Xu JS, Tang JS, Gali A, Li CF, Guo GC 2025 Experimental observation of spin defects in van der Waals material $GeS_2$ *Nano Letters* **25** 16330

[21]     Vaidya S, Gao X, Dikshit S, Fang Z, Llacsahuanga Alcca AE, Chen YP, Yan Q, Li T. 2025 Coherent Spins in van der Waals Semiconductor $GeS_2$ at Ambient Conditions. *Nano Letters* **25**, 14356

# 3. Quantum emitters based on 2D systems

**Yue Luo[1], Stefan Strauf[2]**

[1] School of Electronic Science and Engineering, Southeast University, Nanjing, 211189, China
[2]Department of Physics, Stevens Institute of Technology, Hoboken, New Jersey 07030, United States

E-mail: yueluo@seu.edu.cn

**Status**

Quantum emitters derived from two-dimensional (2D) materials were discovered a decade ago and have rapidly emerged as a promising class of solid-state single-photon sources, attracting significant attention in the context of quantum information science[1, 2]. Advances in 2D materials research have revealed that a broad range of 2D materials can host stable, optically addressable quantum emitters with atomic-scale precision. These discoveries present new opportunities for the development of scalable and integrable quantum technologies and add to the rapid progress involving quantum light sources and spin qubits based on self-assembled quantum dots (0D), single-walled carbon nanotubes (1D), or bulk hosts such as diamond, Si, SiGe, or SiC (3D).

Transition metal dichalcogenides (TMDs), including $WSe_2$, $MoSe_2$, and $MoS_2$ monolayers[1, 3], and hexagonal boron nitride (hBN) [4] have emerged as the most prominent 2D hosts for quantum emitters. In TMDs, quantum emitters can form at localized strain sites that create quantum dot-like 0D confinement potentials for excitons, exhibiting high-purity single-photon emission typically at cryogenic temperatures, with tunability achievable through strain engineering, surface-acoustic-waves, or via dielectric environment control. In contrast, the wide bandgap of hBN enables room-temperature operation, with intrinsic defects such as boron and nitrogen vacancies pairing up with impurities such as carbon[5] or oxygen[6], serving as bright, photostable single-photon sources emitting from deep ultraviolet to near-infrared wavelengths. Many research teams are actively investigating methods for deterministic control over emitter location, emission wavelength, polarization and coherence properties. Engineering approaches such as ion implantation, laser writing, twisting, and nanoscale strain patterning are being developed to enhance spatial precision and reproducibility.

Despite significant progress, challenges remain in achieving scalable integration of 2D quantum emitters into photonic and electronic platforms. Spectral instability, low quantum efficiency in some systems, lack of single-photon indistinguishability or entanglement, and difficulties in deterministic fabrication present ongoing hurdles. Nevertheless, the intrinsic advantages of 2D materials, including atomic thickness, compatibility with heterogeneous integration, and widely tunable optical properties, continue to drive strong interest in this area. As fabrication techniques and theoretical understanding of microscopic mechanisms advance, quantum emitters in 2D materials are poised to play a central role in next-generation quantum photonic technologies.

**Current and future challenges**

The discovery of quantum emitters in 2D materials has opened new avenues for quantum photonic technologies, yet several fundamental challenges must be addressed before these systems can achieve widespread implementation. Current limitations fall into three primary categories: emitter stability, fabrication control, and device integration. Spectral instability remains a critical issue, particularly for quantum emitters in TMDs, where acoustic phonon scattering and environmental charge fluctuations lead to spectral diffusion, linewidth broadening, and blinking behaviour. These effects fundamentally limit photon indistinguishability – a crucial requirement for quantum interference applications. In hBN, while emitters demonstrate better stability at room temperature, reproducible control over their electronic structure and suppression of phonon sidebands remains likewise challenging.

A second major challenge lies in achieving deterministic emitter fabrication. Most 2D material quantum emitters form stochastically through either random defect generation or uncontrolled strain variations. While techniques such as electron beam irradiation, atomic force microscope nanoindentation, and substrate stressor engineering have shown promise for controlled emitter creation, they currently lack the precision and scalability needed for practical quantum technologies. The atomically thin nature of these materials presents additional integration challenges, as poor out-of-plane light extraction and weak coupling to photonic structures significantly reduce system efficiency. Recent advances in deterministic coupling via plasmonic nanocavities[7] and strong-coupling via BIC-type metasurfaces [8] offer potential solutions at the prototype level, but require nanometre-precision alignment that complicates large-scale fabrication.

Looking ahead, the field must overcome several key hurdles to realize practical devices. Electrically driven emission, essential for scalable quantum photonic circuits, remains challenging for most 2D material systems. Developing appropriate contact schemes that preserve emitter properties while enabling efficient carrier injection represents a significant materials engineering challenge. At the fundamental level, reducing phonon-induced decoherence and extending spin coherence times at elevated temperatures will be critical for many quantum information applications. Hybrid integration approaches, combining 2D emitters with silicon photonics, superconducting resonators, or topological waveguides, may provide pathways to address these limitations. However, such systems will require advances in material growth, heterogeneous integration techniques and new theoretical frameworks to understand and optimize interfacial effects and reduce dephasing. Overcoming these challenges will demand close collaboration between materials science, nanofabrication, and quantum optics communities to fully exploit the potential of 2D quantum emitters.

**Advances in science and technology to meet challenges**

Recent breakthroughs in nanoscale engineering and hybrid integration are addressing longstanding challenges in 2D material quantum emitters, transforming them from laboratory curiosities a decade ago into viable components for future quantum technologies. Strain engineering has emerged as a particularly powerful approach, with nanopillar arrays, atomic force microscope nanoindentation, and lithographic substrate patterning now enabling deterministic positioning of quantum emitters in TMD monolayers,[7, 9, 10] (Figure 1). These techniques create spatially controlled strain fields that produce quantum dot-like states with tunable emission properties, achieving good uniformity in emitter arrays. Recent advances have also extended quantum emission to other layered systems, such as InSe[11] and MoTe$_2$[12] operating at telecom bands, $\alpha$-MoO$_3$[13] reaching room temperature, or quantum emitters formed in moiré potentials of twisted MoSe$_2$/WSe$_2$ with high energy-tunability[14] and good coherence [15]. Complementary advances in defect engineering of hBN utilize focused ion implantation, laser writing, and electron beam irradiation that can now implant vacancy-related defects with nanometer-scale precision, which is a critical step toward CMOS-compatible quantum device fabrication.

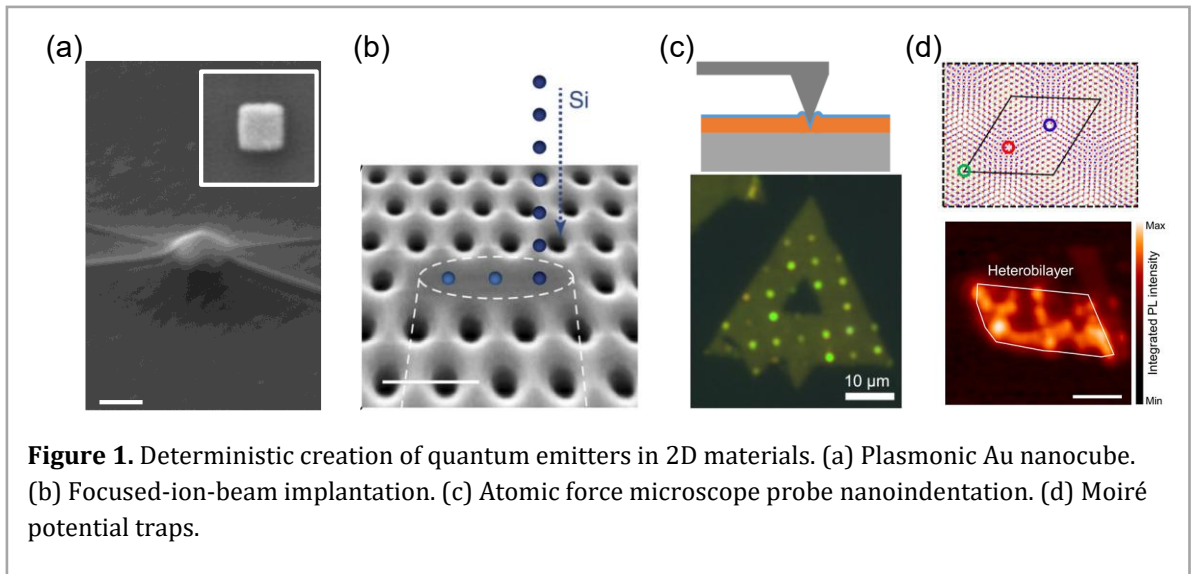

**Figure 1.** Deterministic creation of quantum emitters in 2D materials. (a) Plasmonic Au nanocube. (b) Focused-ion-beam implantation. (c) Atomic force microscope probe nanoindentation. (d) Moiré potential traps.

To combat spectral instability and enhance photon extraction, researchers are developing sophisticated hybrid photonic platforms. Integration with photonic crystal cavities, plasmonic nanostructures, and BIC-type dielectric metasurfaces has demonstrated several key benefits: Purcell-enhanced emission rates, strong-coupling of emitter and mode, reduced spectral diffusion via environmental screening, and improved light extraction efficiency via optimized far-field coupling. These systems now routinely achieve >90% photon collection efficiency when coupled to integrated waveguides or optical fibres. Environmental engineering through hBN encapsulation and oxide passivation has similarly extended emitter coherence times by orders of magnitude, while dielectric tuning enables dynamic control of emission energy and dipole orientation. Perhaps the most important progress can be traced back to the advances towards near defect-free material growth, for example via the flux-growth technique of TMDs that enables now optical emission with near-unity quantum yields[16].

The field is now also witnessing crucial transitions from optical to electrical control, with recent demonstrations of electrically pumped single-photon emission from both hBN defect centres and TMD heterostructures[17]. Light-emitting tunnelling junctions and lateral p-n devices have shown particular promise, though challenges remain in achieving high purity at room temperature. Concurrently, computational approaches are accelerating progress, such as machine learning algorithms trained on first-principles simulations can now predict defect configurations with quantum-relevant electronic states, while inverse design methods are optimizing photonic architectures for specific emitter properties. Together, these advances are establishing a comprehensive toolkit for engineering quantum light sources that combine atomic-scale precision with photonic integration capabilities.

**Concluding remarks**

The rapid progress in engineering quantum emitters in 2D materials highlights their immense potential for scalable quantum photonic technologies. Recent advances in strain and defect engineering have transformed these systems from randomly occurring phenomena to precisely controllable quantum light sources, while hybrid integration with photonic structures has significantly improved their performance and stability. The development of electrically driven devices and computational design tools further bridges the gap between fundamental research and practical applications.

However, critical milestones remain, including achieving room-temperature quantum coherence, perfecting large-scale emitter arrays, and enabling seamless integration with existing quantum hardware. Addressing these challenges will require continued collaboration across materials science, nanophotonics, and quantum engineering. As control over these atomic-scale quantum systems improves, 2D material emitters are poised to play a pivotal role in realizing on-chip quantum networks, sensors, and information processors. Their unique combination of atomic-scale footprint, tunable optical properties, and integration capabilities positions them as a versatile platform for the next generation of quantum technologies.

**Acknowledgements**

Y.L. acknowledges support from the National Natural Science Foundation of China (Grant No.9247710104), Natural Science Foundation of Jiangsu Province (BK20241294) and the Southeast University Interdisciplinary Research Program for Young Scholars. S.S. acknowledges support from the National Science Foundation of the USA (DMR-1809235). The authors also extend their sincere thanks to Dr. Na Liu for her invaluable discussions and insights.

**References**

[1]     Turunen, M, Brotons-Gisbert, M, Dai, Y, Wang, Y, Scerri, E, Bonato, C, Jöns, K D, Sun, Z, and Gerardot, B D 2022 *Nat. Rev. Phys.* 4 219–236

[2]     Kianinia, M, Xu, Z-Q, Toth, M, and Aharonovich, I 2022 *Appl. Phys. Rev.* 9 011306

[3]     Vasconcellos S Michaelis de, Wigger, D, Wurstbauer, U, Holleitner, A W, Bratschitsch, R, and Kuhn, T 2022 *Phys. Status Solidi B* 259 2100566

[4]     Tran, T T, Bray, K, Ford, M J, Toth, M, and Aharonovich, I 2015 *Nat. Nanotechnol.* 11 37–41

[5]     Mendelson, N, Chugh, D, Reimers, J R, Cheng, T S, Gottscholl, A, Long, H, Mellor, C J, Zettl, A, Dyakonov, V, Beton, P H, Novikov, S V, Jagadish, C, Tan, H H, Ford, M J, Toth, M, Bradac, C, and Aharonovich, I 2020 *Nat. Mater.* 20 321–328

[6]     Mohajerani, S S, Chen, S, Alaei, A, Chou, T, Liu, N, Ma, Y, Xiao, L, Lee, S S, Yang, E-H, and Strauf, S 2024 *ACS Photonics* 11 2359–2367

[7]     Luo, Y, Shepard, G D, Ardelean, J V, Rhodes, D A, Kim, B, Barmak, K, Hone, J C, and Strauf, S 2018 *Nat. Nanotechnol.* 13 1137–1142

[8]     Do, T T H, Nonahal, M, Li, C, Valuckas, V, Tan, H H, Kuznetsov, A I, Nguyen, H S, Aharonovich, I, and Ha, S T 2024 *Nat. Commun.* 15 2281

[9]     Palacios-Berraquero, C, Kara, D M, Montblanch, A R-P, Barbone, M, Latawiec, P, Yoon, D, Ott, A K, Loncar, M, Ferrari, A C, and Atatüre, M 2017 *Nat. Commun.* 8 15093

[10]    Branny, A, Kumar, S, Proux, R, and Gerardot, B D 2017 *Nat. Commun.* 8 15053

[11]    Zhao, H, Hus, S M, Chen, J, Yan, X, Lawrie, B J, Jesse, S, Li, A-P, Liang, L, and Htoon, H 2025 *ACS Nano* 19 6911–6917

[12]    Zhao, H, Pettes, M T, Zheng, Y, and Htoon, H 2021 *Nat. Commun.* 12 6753

[13]     Lee, J, Wang, H, Park, K-Y, Huh, S, Kim, D, Yu, M, Kim, C, Thygesen, K S, and Lee, J 2025 *Nano Lett.* 25 1142–1149

[14]     Baek, H, Brotons-Gisbert, M, Koong, Z X, Campbell, A, Rambach, M, Watanabe, K, Taniguchi, T, and Gerardot, B D 2020 *Sci. Adv.* 6 eaba8526

[15]     Wang, H, Kim, H, Dong, D, Shinokita, K, Watanabe, K, Taniguchi, T, and Matsuda, K 2024 *Nat. Commun.* 15 4905

[16]     Kim, B, Luo, Y, Rhodes, D, Bai, Y, Wang, J, Liu, S, Jordan, A, Huang, B, Li, Z, Taniguchi, T, Watanabe, K, Owen, J, Strauf, S, Barmak, K, Zhu, X, and Hone, J 2021 *ACS Nano* 16 140–147

[17]     Yu, M, Lee, J, Watanabe, K, Taniguchi, T, and Lee, J 2024 *ACS Nano* 19 504–511

# 4. Computational Theory of Quantum Defects in 2D materials

**Reda Moukaouine[1,2], Anton Pershin[2,3]\* and Adam Gali[2,3,4]\***

[1] György Hevesy Doctoral School, ELTE Eötvös Loránd University, Institute of Chemistry,
Budapest, Hungary.
[2] HUN-REN Wigner Research Centre for Physics, Institute for Solid State Physics and Optics,
Budapest, Hungary.
[3] Department of Atomic Physics, Budapest University of Technology and Economics,
Budapest, Hungary.
[4] MTA-WFK Lendület "Momentum" Semiconductor Nanostructures Research Group,
Budapest, Hungary.

E-mail: gali.adam@wigner.hun-ren.hu ; pershin.anton@wigner.hun-ren.hu

**Status**

Identifying point defects as a source of spin qubits in bulk semiconductors remains challenging, yet a pragmatic methodology has emerged: hybrid-functional DFT for band-gap correction and level alignment; $\Delta$SCF for low-lying excited states; vibronic modelling and spin-property calculations for comparison with the PL lineshapes and ESR/ODMR data, respectively, see Figure 1 [1–3]. Applied carefully, this toolkit can constrain structural hypotheses and guide measurements. However, its accuracy and transferability to strictly two-dimensional(2D) crystals are not yet established. The 2D landscape already hosts numerous ODMR signals [4,5], but only one qubit with known chemical structure and demonstrated addressable spin, namely the negatively charged boron vacancy $V_B^-$ in hexagonal boron nitride (hBN), has been unambiguously identified [6]. Many other features remain unassigned, owing to methodological shortcomings and limited experimental data.

Several factors confound direct import of bulk workflows to 2D. In monolayers every atom lies at a surface; long-range polarization, substrate and encapsulation screening, local strain, twist angle, and stacking order can modify defect levels enormously [7–9]. Electrostatic treatments specific to 2D (Coulomb truncation, image-charge corrections, dielectric-dependent hybrids) are inconsistently applied and rarely benchmarked. Excited-state physics is equally decisive: strong excitonic effects, spin-orbit coupling, and spin-vibronic interactions (including Jahn–Teller effects) govern optical selection rules, intersystem-crossing (ISC) rates, and thus ODMR contrast, yet are often treated only approximately. New phenomena also emerge in 2D that blur single-center interpretations. Notably, spin polarisation from nominal spin-1/2 systems has been reported and ascribed to donor–acceptor pairs [10,11]; a mechanism not captured by typical single-defect DFT workflows.

Higher-level electronic-structure methods that could arbitrate, such as GW/BSE for quasiparticles and excitons, and post-Hartree–Fock or multireference approaches (EOM-CCSD, CASSCF/NEVPT2, DMRG-based schemes) for strongly correlated defects are not yet routine for open-shell centers in large 2D supercells. Embedding and polarizable-environment formalisms exist [12,13], but lack standardization and error quantification in 2D materials.

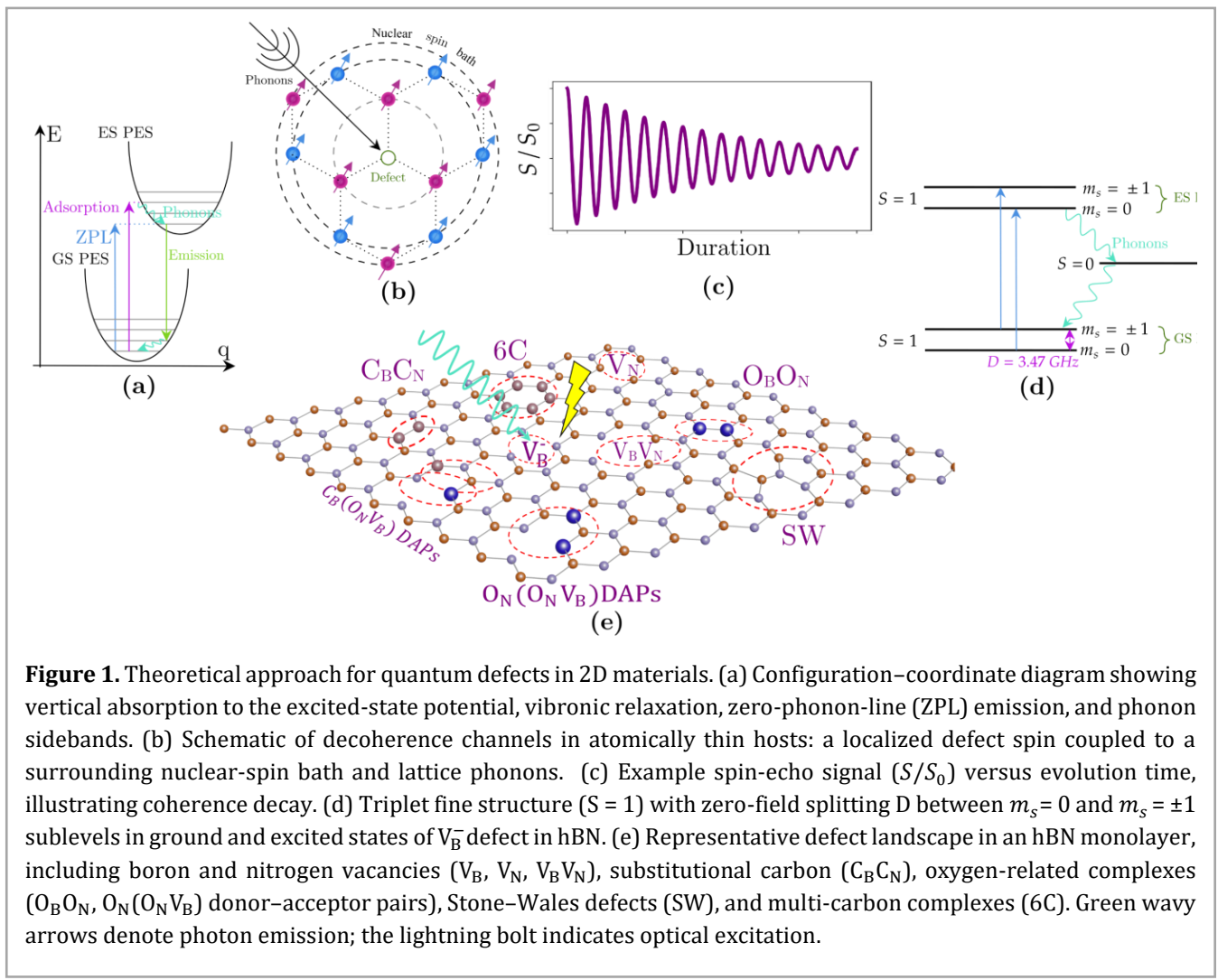

**Figure 1.** Theoretical approach for quantum defects in 2D materials. (a) Configuration–coordinate diagram showing vertical absorption to the excited-state potential, vibronic relaxation, zero-phonon-line (ZPL) emission, and phonon sidebands. (b) Schematic of decoherence channels in atomically thin hosts: a localized defect spin coupled to a surrounding nuclear-spin bath and lattice phonons. (c) Example spin-echo signal ($S/S_0$) versus evolution time, illustrating coherence decay. (d) Triplet fine structure ($S = 1$) with zero-field splitting D between $m_s = 0$ and $m_s = \pm 1$ sublevels in ground and excited states of $V_B^-$ defect in hBN. (e) Representative defect landscape in an hBN monolayer, including boron and nitrogen vacancies ($V_B$, $V_N$, $V_B V_N$), substitutional carbon ($C_B C_N$), oxygen-related complexes ($O_B O_N$, $O_N(O_N V_B)$ donor–acceptor pairs), Stone–Wales defects (SW), and multi-carbon complexes (6C). Green wavy arrows denote photon emission; the lightning bolt indicates optical excitation.

## Current and future challenges

Despite rapid advances, a predictive framework for quantum defects in two-dimensional materials remains elusive. The obstacles are not due to a single missing ingredient but rather to a set of intertwined issues spanning electronic structure, excited states, coherence, and environmental effects. Weak dielectric screening in monolayers amplifies finite-size and image-charge errors, complicating reliable charge-transition levels (CTLs). Large supercells, explicit charge corrections, and hybrid-DFT band-edge alignment are essential, while practical correction schemes were proposed [14]. Yet a standardized workflow applicable across different defects and heterostructures is still lacking. Excited-state properties bring additional complexity. In transition-metal dichalcogenides such as $WS_2$, defects exhibit large spin–orbit splitting and vibronic sidebands that demand GW-level accuracy. Realistic descriptions must therefore incorporate spin–orbit coupling and vibronic Hamiltonians to capture level ordering, optical selection rules, and radiative pathways. Moreover, excited states are often correlated by construction, and efficient geometry relaxation of such multi-determinant states remains limited. While a recent extension of $\Delta$SCF provides practical access to singlet multiplets [7], correlated doublets are insufficiently treated. To this end, a promising route is to approximate correlated excited states by combining ground-state energies of single-reference configurations, though a systematic framework is still lacking.

Quantum coherence is shaped by processes spanning multiple scales. While the atomic structure of a defect dictates its local magnetic properties, decoherence arises from coupling to nuclear spins and phonons. In practice, most theoretical treatments assume magnetic noise as the dominant source, but other contributions, including electric-field fluctuations and interactions with nearby electronic spins, remain largely unexplored. In hBN, natural isotopic composition creates a dense nuclear-spin bath that persists even after isotopic purification. Direct simulation of thousands of spins and phonons is computationally prohibitive, so approximate approaches such as cluster correlation expansion are employed, though higher-order treatments are needed for predictive accuracy [15]. Environmental interactions add another dimension. Surface defects in 2D hosts can chemically interact with adsorbed molecules, shifting energy levels and altering photo-dynamics [16]. Temperature effects are also poorly addressed. Most simulations assume 0 K, yet vibrational and entropic contributions strongly influence defect energetics at finite temperature. In 2D systems, phonons are strongly coupled, but the role of this coupling in limiting $T_1$ spin-relaxation times remains unknown. Addressing it requires advanced approaches such as self-consistent phonon theory or ab initio molecular dynamics, both computationally demanding.

**Advances in science and technology to meet challenges**

Progress towards identification of new qubits in 2D materials will require a coordinated effort in measurement, materials, and computation. Experimentally, datasets with genuine discriminatory power are needed, including targeted isotope substitution, systematic strain- and stacking-dependent PL, and, where feasible, ODMR with hyperfine resolution and STS. Exploring hosts with dilute nuclear-spin backgrounds is equally important; a recent work in $\beta$-GeS$_2$ [17] reports coherence times about two orders of magnitude longer than $V_B^-$ in hBN. An alternative avenue is hybrid architectures in which atomically thin layers act as passive spacers for known spin qubits (for example, molecular spins). hBN is well suited here: van der Waals coupling can preserve the molecule's identity while limiting unwanted nuclear-bath contact, reducing assignment ambiguities. On the theory side, "bulk workflows ported to 2D" should be replaced with predictive, environment-aware approaches that include explicit excited-state and spin–phonon treatments using realistic models. For strongly correlated centres, wavefunction solvers are central. Plane-wave CAS-DMRG/DMRG-SCF augmented by NEVPT2 are needed to handle extended active spaces and should deliver the key observables, not only energies but also state forces (for ZPLs and geometry changes), and spin properties that feed decoherence models for computing $T_2$ time. When quantum computers become widely available, quantum protocols for configuration-interaction (CI) methods may offer a practical path where classical costs explode: selected-CI with quantum subspace expansion, EOM-VQE/ADAPT-VQE, or imaginary-time evolution are increasingly demonstrated in small molecules or defect centers [18–20], yet to be extended to analytic forces and intersystem-crossing rates within periodic, plane-wave frameworks.

Brute force over the combinatorial defect and host spaces is infeasible, so physics-informed machine learning may also act as an intelligent front end to many-body solvers. Crucially, it should predict not only formation and reaction energetics but also properties computed by robust electronic-structure methods, ideally, CTLs, ZPLs, Huang–Rhys factors and oscillator strengths, spin tensors, and ISC rates. Active-learning loops must include charged-defect and substrate/encapsulation descriptors to be capable of relaxing large charged supercells and heterostructures efficiently.

**Concluding remarks**

Bridging theory and experiment for spin defects in 2D materials now depends on methods that go decisively beyond (semi-local) DFT. Predictive control requires quantitatively correct excited states, spin-orbit and vibronic couplings, and intersystem crossing rates in realistic stacks. Many-body approaches should become routine, either using GW/BSE for level- and optical gap- alignment or post-Hartree-Fock methods such as EOM-CCSD for singlet-triplet manifolds and multireference schemes for strongly correlated centers. In either case, they should be extended to incorporate effects of phonons, including non-adiabatic dynamics with explicit spin-vibronic couplings for the ODMR contrast. The environment matters in 2D, so polarizable and quantum embedding approaches (DMET/SEET, CC-in-DFT, GW-in-DFT) also hold promise for capturing substrate and encapsulation effects. Method development should standardize finite-size- and charge-corrections for monolayers and enable automated workflows that return zero-phonon line, zero-field splitting, hyperfine tensors, and ISC rates under strain and bias. Overall, expanding post-HF and many-body research tools from individual case studies to calibrated high-throughput pipelines will turn theoretical propositions into testable predictions and enable true defect-device co-engineering.

**Acknowledgements**

Support by the Ministry of Culture and Innovation and the National Research, Development and Innovation Office within the Quantum Information National Laboratory of Hungary (Grant No. 2022-2.1.1-NL-2022-00004) as well as the European Commission for the projects QuSPARC (Grant No. 101186889) and SPINUS (Grant No. 101135699) are much appreciated. A.G. acknowledges the high-performance computational resources provided by KIFÜ (Governmental Agency for IT Development of Hungary). A.P. acknowledges the financial support of Janos Bolyai Research Fellowship of the Hungarian Academy of Sciences. R.M. is thankful for the support of the Stipendium Hungaricum scholarship.

**References**

[1] Gali Á 2019 Ab initio theory of the nitrogen-vacancy center in diamond **8** 1907–43

[2] Seo H, Ivády V and Ping Y 2024 First-principles computational methods for quantum defects in two-dimensional materials: A perspective *Applied Physics Letters* **125** 140501

[3] Freysoldt C, Grabowski B, Hickel T, Neugebauer J, Kresse G, Janotti A and Van De Walle C G 2014 First-principles calculations for point defects in solids *Rev. Mod. Phys.* **86** 253–305

[4] Gao X, Vaidya S, Li K, Ge Z, Dikshit S, Zhang S, Ju P, Shen K, Jin Y, Ping Y and Li T 2025 Single nuclear spin detection and control in a van der Waals material *Nature* **643** 943–9

[5] Stern H L, M. Gilardoni C, Gu Q, Eizagirre Barker S, Powell O F J, Deng X, Fraser S A, Follet L, Li C, Ramsay A J, Tan H H, Aharonovich I and Atatüre M 2024 A quantum coherent spin in hexagonal boron nitride at ambient conditions *Nature Materials* **23** 1379–85

[6] Ivády V, Barcza G, Thiering G, Li S, Hamdi H, Chou J-P, Legeza Ö and Gali A 2020 Ab initio theory of the negatively charged boron vacancy qubit in hexagonal boron nitride *npj Computational Materials* **6** 41

[7] Iwański J, Korona K P, Tokarczyk M, Kowalski G, Dąbrowska A K, Tatarczak P, Rogala I, Bilska M, Wójcik M, Kret S, Reszka A, Kowalski B J, Li S, Pershin A, Gali A, Binder J and Wysmołek A 2024 Revealing polytypism in 2D boron nitride with UV photoluminescence *npj 2D Materials and Applications* **8** 72

[8] Li S, Pershin A, Li P and Gali A 2024 Exceptionally strong coupling of defect emission in hexagonal boron nitride to stacking sequences *npj 2D Materials and Applications* **8** 16

[9] Plo J, Pershin A, Li S, Poirier T, Janzen E, Schutte H, Tian M, Wynn M, Bernard S, Rousseau A, Ibanez A, Valvin P, Desrat W, Michel T, Jacques V, Gil B, Kaminska A, Wan N, Edgar J H, Gali A and Cassabois G 2025 Isotope Substitution and

Polytype Control for Point Defects Identification: The Case of the Ultraviolet Color Center in Hexagonal Boron Nitride *Phys. Rev. X* **15** 021045

[10] Li S, Pershin A and Gali A 2025 Quantum emission from coupled spin pairs in hexagonal boron nitride *Nature Communications* **16** 5842

[11] Singh P, Robertson I O, Scholten S C, Healey A J, Abe H, Ohshima T, Tan H H, Kianinia M, Aharonovich I, Broadway D A, Reineck P and Tetienne J 2025 Violet to Near-Infrared Optical Addressing of Spin Pairs in Hexagonal Boron Nitride *Advanced Materials* **37** 2414846

[12] Chen S, Yu V W, Jin Y, Govoni M and Galli G 2025 Advances in Quantum Defect Embedding Theory *J. Chem. Theory Comput.* **21** 7797–812

[13] Winter M, Bousquet M H E, Jacquemin D, Duchemin I and Blase X 2021 Photoluminescent properties of the carbon-dimer defect in hexagonal boron-nitride: A many-body finite-size cluster approach *Phys. Rev. Materials* **5** 095201

[14] Komsa H-P, Berseneva N, Krasheninnikov A V and Nieminen R M 2014 Charged Point Defects in the Flatland: Accurate Formation Energy Calculations in Two-Dimensional Materials *Phys. Rev. X* **4** 031044

[15] Tárkányi A and Ivády V 2025 Understanding Decoherence of the Boron Vacancy Center in Hexagonal Boron Nitride *Adv. Funct. Mater. (2025): e11300*

[16] Slassi A, Moukaouine R, Cornil J and Pershin A 2025 Controlling Charge Carrier Lifetime in Defective $WS_2$ Monolayer through Interface Engineering: a Time-Domain Ab Initio Study *J. Phys. Chem. Lett.* **16** 1931–8

[17] Liu W, Li S, Guo N-J, Zeng X-D, Xie L-K, Liu J-Y, Ma Y-H, Wu Y-Q, Wang Y-T, Wang Z-A, Ren J-M, Ao C, Xu J-S, Tang J-S, Gali A, Li C-F and Guo G-C 2024 Experimental observation of spin defects in van der Waals material $GeS_2$ *Nano Lett. 2025, 25, 46, 16330–16339*

[18] Huang B, Govoni M and Galli G 2022 Simulating the Electronic Structure of Spin Defects on Quantum Computers *PRX Quantum* **3** 010339

[19] Feniou C, Hassan M, Traoré D, Giner E, Maday Y and Piquemal J-P 2023 Overlap-ADAPT-VQE: practical quantum chemistry on quantum computers via overlap-guided compact Ansätze *Communications Physics* **6** 192

[20] Bierman J, Li Y and Lu J 2023 Improving the Accuracy of Variational Quantum Eigensolvers with Fewer Qubits Using Orbital Optimization *J. Chem. Theory Comput.* **19** 790–8

# 5. Data-Driven Discovery of Quantum Defects in 2D Materials

**Zhenyao Fang[1] and Qimin Yan[1]**

[1] Department of Physics, Northeastern University, Boston, USA

E-mail: z.fang@northeastern.edu, q.yan@northeastern.edu

**Status**

Quantum defects (QDs) have emerged as a key focus in the field of quantum technologies recent years[1]. With the advancement of first-principles computational methods, the search for optimal QDs is significantly accelerated by a data-driven approach, which leverages high-throughput computations based on density functional theory (DFT) to calculate defect properties and to screen candidate host materials and defect types. Among the vast space of crystalline materials, two-dimensional (2D) materials have gained much attention due to their advantages in experimental synthesis and control [2,3]. Unlike bulk materials where the defects are usually embedded within the structure, defects in 2D materials are exposed at the surface, allowing for easier detection and manipulation in experiments. Besides, the reduced dielectric screening in 2D materials allows for the interaction between QDs and external fields, which is essential for quantum emitter and quantum sensor applications[4–6].

To identify an ideal QD through high-throughput screening, the following criteria are usually considered. Firstly, the formation energy of the target defect determines its likelihood of formation in the host material. An optimal formation energy is necessary to avoid insufficient defect concentration, which may limit its response to external fields, or excessive concentration that may lead to defect clustering and decoherence [7]. Secondly, QDs must host well-localized (or isolated) electronic states within the bandgap that enable photon emission and spin manipulation [8]. Finally, factors such as spin decoherence time, hyperfine coupling, and zero-field splitting also play an important role in determining the efficacy of QDs in sensing and light emission [9,10].

Recent efforts in data-driven screening based on the above criteria have expanded the catalogue of QDs in 2D materials [4,6]. For example, Tsai *et al.* [11] investigated the antisite defects in 2D transition metal dichalcogenides and identified several anion-antisite defects as promising qubit candidates. Additionally, they examined the optical transition loops between the triplet and singlet defect states in $WS_2$, providing insights for experimental manipulation of those qubits. Another work performed by Thomas *et al.* [12] studied more than 700 charged substitutional defects within $WS_2$. Among them, the $Co_S$ defect appeared as a candidate QD with optimal optical excitation energy and were demonstrated experimentally feasible. Furthermore, Bertoldo *et al.* [13] performed high-throughput calculations on single and double defects across ten synthesized 2D host materials. Their screening initially identified around 600 defects with a triplet ground state and then refined to 39 ideal QDs based on their formation energies, hyperfine coupling, and zero-field splitting tensors. These work established data-driven high-throughput calculations as a key approach for discovering novel QDs and optimizing their physical properties for applications in quantum information processing.

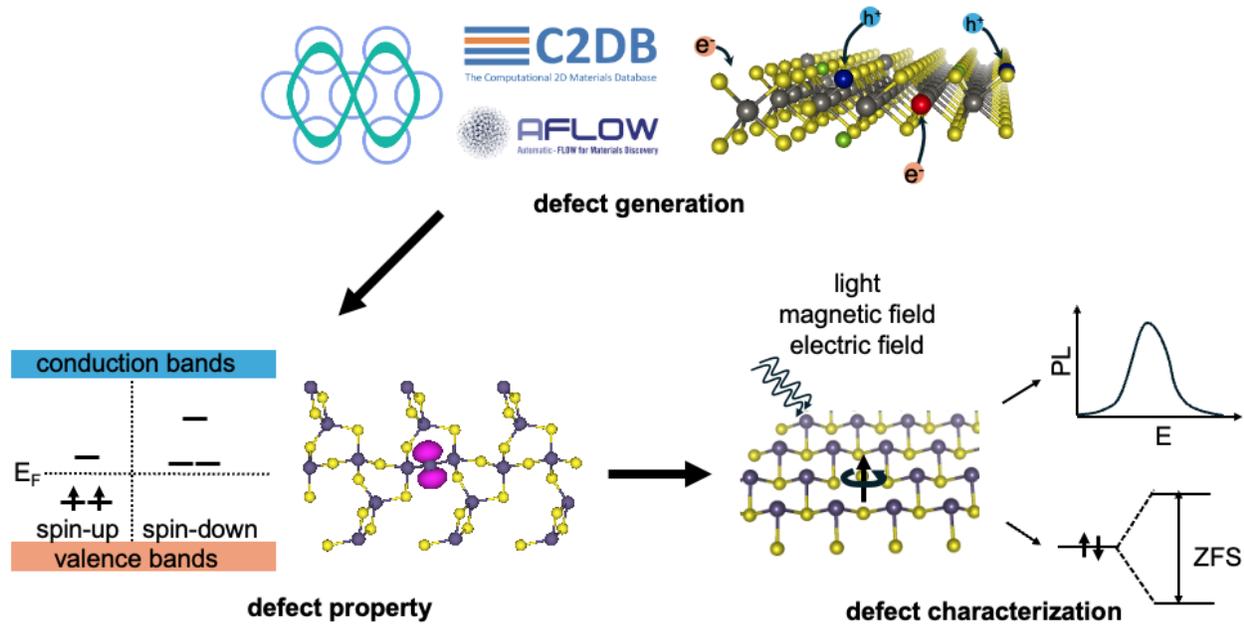

**Figure 1.** High-throughput computational workflow to identify QD candidates from existing databases, calculate defect properties, and connecting to experimentally measurable phenomena.

## Current and future challenges

While recent high-throughput studies have demonstrated the potential of data-driven approaches, several challenges remain in discovering ideal QDs for quantum technology. A primary challenge is the vast configuration space; each host material may contain multiple symmetry-inequivalent atomic sites, each of which can accommodate various types of defects in multiple charge states. Therefore, efficient workflows for computing defect properties and organizing structured databases, essentially a defect genome project, are essential.

Furthermore, current first-principles methods face limitations in accurately capturing the many-body effects in defective systems, such as spin-spin interactions and the coupling between defects and electromagnetic fields. More details on computational methods will be discussed in a later section. Experimental observables, such as photoluminescence spectra and zero-phonon lines, serve as fingerprints for QDs, but connecting these measurements to theoretical predictions remains a major challenge, due to the limited accuracy of theoretical models and the complexity of experimental conditions. In particular, the performance of QDs is affected by dynamical effects, such as defect migration, clustering, and entropy-driven behavior at finite temperatures[14] . While machine learning (ML) approaches based on first-principles calculations were proposed to capture those effects, progress is limited by the scarcity of high-quality training data, making this an open area for future investigation.

## Advances in science and technology to meet challenges

As the exploration of QDs expands, advancements in computational methodologies are essential for addressing the complexity of defect calculations and screening. Efficient workflows have been developed to automate high-throughput calculations, enabling systematic evaluation of fundamental defect properties such as formation energies and charge transition levels[15,16] .

On the other hand, ML methods emerge as powerful tools to address the limitations of first-principles calculations when modelling dynamic properties. ML approaches, such as graph neural network methods trained on high-fidelity calculations, can easily capture the local chemical environment of defects and make predictions on the static defect properties[17,18] . Furthermore, in combination with Monte Carlo simulations, these ML models can also predict the disordering properties, such as the configurational entropy of defects, the ground state defect configuration, and the defect migration pathways[19] . The combination of automated defect workflows, advanced theoretical models, and machine learning techniques will provide a more comprehensive data-driven framework for identifying and characterizing QDs.

**Concluding remarks**

The field of data-driven QD discovery, through at its early stages, has demonstrated significant potential in identifying promising QD candidates for quantum technologies. Using advanced first-principles calculation methods, including the DFT and ML methods, high-throughput screening can enable systematic exploration of QDs in 2D materials and effectively propose experimentally accessible host material and defect candidates. These developments provide a strong foundation for future studies on QDs as quantum sensors and emitters.

Despite these advancements, several open questions remain. A key challenge is to integrate computational techniques of different levels into a unified workflow for QD discovery that can capture the complex interactions in defective systems, such as many-body effects and dynamical effects. Additionally, it is crucial to improve the connection between theoretical predictions and experimental validation for refining the defect models for practical applications. Addressing those challenges will require further development of automated computational workflows, ML models, and collaborative efforts between theory and experiment, and will benefit future discovery and realization of QDs for next-generation quantum technologies.

**Acknowledgements**

This work is supported by the National Science Foundation under Grant No. DMR-2314050.

**References**

[1] Wolfowicz G, Heremans F J, Anderson C P, Kanai S, Seo H, Gali A, Galli G and Awschalom D D 2021 Quantum guidelines for solid-state spin defects *Nat. Rev. Mater.* **6** 906–25

[2] Guo Y, Li J, Dou R, Ye H and Gu C 2024 Quantum defects in two-dimensional van der Waals materials *Fundam. Res.*

[3] Seo H, Ivády V and Ping Y 2024 First-principles computational methods for quantum defects in two-dimensional materials: A perspective *Appl. Phys. Lett.* **125** 140501

[4] Fang H-H, Wang X-J, Marie X and Sun H-B 2024 Quantum sensing with optically accessible spin defects in van der Waals layered materials *Light: Sci. Appl.* **13** 303

[5] Zhang G, Cheng Y, Chou J-P and Gali A 2020 Material platforms for defect qubits and single-photon emitters *Appl. Phys. Rev.* **7** 031308

[6] Lee Y, Hu Y, Lang X, Kim D, Li K, Ping Y, Fu K-M C and Cho K 2022 Spin-defect qubits in two-dimensional transition metal dichalcogenides operating at telecom wavelengths *Nat. Commun.* **13** 7501

[7] Freysoldt C, Grabowski B, Hickel T, Neugebauer J, Kresse G, Janotti A and Walle C G V de 2014 First-principles calculations for point defects in solids *Rev. Mod. Phys.* **86** 253–305

[8] Ping Y and Smart T J 2021 Computational design of quantum defects in two-dimensional materials *Nat. Comput. Sci.* **1** 646–54

[9] Bulancea-Lindvall O, Son N T, Abrikosov I A and Ivády V 2021 Dipolar spin relaxation of divacancy qubits in silicon carbide *npj Comput. Mater.* **7** 213

[10] Gali A 2009 Theory of the neutral nitrogen-vacancy center in diamond and its application to the realization of a qubit *Phys. Rev. B* **79** 235210

[11] Tsai J-Y, Pan J, Lin H, Bansil A and Yan Q 2022 Antisite defect qubits in monolayer transition metal dichalcogenides *Nat. Commun.* **13** 492

[12] Thomas J C, Chen W, Xiong Y, Barker B A, Zhou J, Chen W, Rossi A, Kelly N, Yu Z, Zhou D, Kumari S, Barnard E S, Robinson J A, Terrones M, Schwartzberg A, Ogletree D F, Rotenberg E, Noack M M, Griffin S, Raja A, Strubbe D A, Rignanese G-M, Weber-Bargioni A and Hautier G 2024 A substitutional quantum defect in WS2 discovered by high-throughput computational screening and fabricated by site-selective STM manipulation *Nat. Commun.* **15** 3556

[13] Bertoldo F, Ali S, Manti S and Thygesen K S 2022 Quantum point defects in 2D materials - the QPOD database *npj Comput. Mater.* **8** 56

[14] Leem Y-C, Fang Z, Lee Y-K, Kim N-Y, Kakekhani A, Liu W, Cho S-P, Kim C, Wang Y, Ji Z, Patra A, Kronik L, Rappe A M, Yim S-Y and Agarwal R 2024 Optically Triggered Emergent Mesostructures in Monolayer WS2 *Nano Lett.* **24** 5436–43

[15] Kumagai Y and Oba F 2014 Electrostatics-based finite-size corrections for first-principles point defect calculations *Phys. Rev. B* **89** 195205

[16] Freysoldt C, Mishra A, Ashton M and Neugebauer J 2020 Generalized dipole correction for charged surfaces in the repeated-slab approach *Phys Rev B* **102** 045403

[17] Xie T and Grossman J C 2018 Crystal Graph Convolutional Neural Networks for an Accurate and Interpretable Prediction of Material Properties *Phys Rev Lett* **120** 145301

[18] Fang Z and Yan Q 2025 Leveraging Persistent Homology Features for Accurate Defect Formation Energy Predictions via Graph Neural Networks *Chem. Mater.* **37** 1531–40

[19] Fang Z and Yan Q 2024 Towards accurate prediction of configurational disorder properties in materials using graph neural networks *npj Comput. Mater.* **10** 91

# 5. Spin valve devices based on graphene

**Harvey Stanfield[1] and Ivan J. Vera-Marun[1]**

[1] Department of Physics and Astronomy, The University of Manchester, Manchester, United Kingdom

E-mail: ivan.veramarun@manchester.ac.uk

**Status**

Graphene's unique properties have made it the prototypical platform for spin-transport devices. Its two-dimensional (2D) structure, with high carrier mobility and minimal intrinsic spin–orbit coupling (SOC), enables long spin relaxation lengths even at room temperature [1]. A spin valve is a device typically consisting of two ferromagnetic contacts separated by a non-magnetic spacer that uses the relative alignment of the magnetic layers to control electronic spin transport. In their basic two-terminal configuration, spin valve devices exploit the magnetoresistance arising from the spin-dependent transport at the contact-spacer junctions. Graphene has been shown to act as an excellent spacer for carrying spin information and can be further enhanced when exploiting the valley pseudospin degree of freedom. Early graphene-based spin valve experiments that incorporated magnetic electrodes produced distinct magnetoresistance changes when switching between parallel and antiparallel configurations [2]. Utilising modern methods, recent graphene-based spin valves have shown spin relaxation lengths on the order of >10 μm [3].

Spin valve systems enable the separation of charge and spin degrees of freedom, giving direct access to extracting spin transport parameters when utilising the so-called non-local measurement configuration. The latter is a key factor for their large presence in the field of spintronics. Two of the key parameters that are the focus of spin valve development are spin injection/detection efficiency, and spin relaxation length [1]. Spin injection/detection efficiency quantifies the degree to which a spin current is generated within/detected from the spin-carrying channel. Spin relaxation length is the defining length in which the said spin current decays within the channel.

The development of graphene-based spin valve devices is motivated by both fundamental and technological aspects. The Datta-Das spin transistor offers an architecture where the state switching utilises an applied electric field, as opposed to the magnetic field used by standard spin valve devices. Despite this difference, the Datta-Das spin transistor would benefit greatly from advancements in spin coherence and injection technologies, such as those demonstrated in spin valve devices [4]. Spin valves have also been shown to be a potential avenue for quantum dot technology [5]. To achieve the required confinement for quantum technologies via electrostatic effects bilayer graphene (BLG) is also highly relevant, given the presence of a bandgap.

**Current and future challenges**

There are many means by which a spin current (pure or polarised) can be generated within a graphene channel, such as electrical, thermal and optical [1,6,7]. The most commonly used method is electrical, therefore is the focus of this discussion. Traditionally, this is done by passing an electrical current through a ferromagnetic contact into graphene, generating a spin-polarised current. Ohmic contacts suffer from conductivity mismatch at the contact-graphene interface, which gives rise to spin backscattering, lowering the spin-injection efficiency. To overcome conductivity mismatch, it is common to use tunnel barriers that increase interfacial resistance [8]. The standard method of creating tunnel contacts utilises the growth of oxides (such as MgO) at the contact region. More recently, to enhance injection efficiency and facilitate scalability, tunnel barriers using 2D materials (such as hexagonal boron nitride (hBN)) are becoming more commonly used.

A recent study utilised van der Waals ferromagnetic contacts (indium and cobalt) deposited directly on top of the graphene channel. While maintaining clean interfaces, these contacts form a 2–4 Å vacuum gap that effectively functions as a van der Waal barrier while preserving ohmic transport characteristics [9]. Van der Waal contacts also lead to the possibility of all 2D spin valves, with further research demonstrating the proximity interactions of vdW ferromagnets/graphene heterostructures, even at room temperature [10]. One-dimensional (1D) contacts, an architecture in which the contact interface is restricted to the edge of the graphene channel, further reduce contact doping in the channel, thus allowing efficient non-invasive spin injection [3]. These three methods of overcoming conductivity mismatch are shown in Figure (1).

Although the nature of spin relaxation in graphene is still the subject of study, it is understood that SOC (intrinsic and extrinsic), along with scattering events, such as those caused by impurities or structural imperfections, are the main factors that reduce spin relaxation lengths and spin lifetimes within graphene channels. The encapsulation of graphene in materials such as hBN is a standard method to maintain the high quality of graphene and reduce degradation, which has been shown to increase the spin relaxation length within graphene channels [3]. The use of tunnel contacts, while helping overcome conductivity mismatch, also isolates the graphene channel from contact-induced spin relaxation [8]. Further to this, utilising advanced fabrication techniques, such as bottom-up fabrication, has been shown to increase the spin lifetime within high-quality single-layer graphene (SLG) to exceed 10 ns [11]. For bilayer graphene, the out-of-plane spin lifetime has been shown to be substantially changed through the induction of spin-valley coupling [12].

**Advances in science and technology to meet challenges**

Alternative methods for injecting spin currents, without the need of applying an electrical current via a magnetic electrode, are important. One method is via the spin Hall effect, achieved by inducing spin-orbit coupling (SOC) within the graphene channel. Since SOC tends to diminish graphene's ability to preserve spin currents, such an architecture typically consists of regions of graphene with high SOC to generate spin currents and regions with low SOC to propagate those currents. A recent example of this approach demonstrated how graphene regions in proximity to $Cr_2Ge_2Te_6$ had induced SOC and magnetic exchange coupling [13]. This allowed for the creation of a pure spin current via the spin Hall effect within a region of the graphene, which could then be carried by pristine graphene and finally detected via a ferromagnetic contact, resulting in a hybrid spin valve device, as shown in Figure (1). Novel proximity effects involving magnetic insulators have also been demonstrated to enhance traditional spin injection methods, allowing robust and tunable spin splitting in graphene [14]. These insights into proximity-induced effects in graphene pave the way for exploring how induced magnetism can complement SOC mechanisms.

For SOC induced within the graphene channel, a combination of electrostatic tuning of the fermi level of graphene and Rashba-type spin orbit interactions has been demonstrated to allow non-volatile control of the spin-to-charge conversion at room temperature [15]. This was induced by proximity effects with a ferromagnetic insulator (such as YIG). Recent experimental work has also demonstrated how the twist angle between graphene and a transition metal dichalcogenide (TMD) provides an alternative route for tunable charge-to-spin conversion in graphene via novel radial spin textures [16].

In terms of making graphene magnetic, experiments have demonstrated how hydrogen adatoms and lattice vacancies are shown to generate localised magnetic moments. When probed with pure spin currents, the exchange coupling between the conduction electrons and the magnetic moments was evidenced, demonstrating the presence of induced paramagnetism in the graphene channel [17]. In addition, the doping of isolated cobalt atoms, via utilising coordinated nitrogen species, has shown promise in generation of a magnetic ordering within the cobalt-doped graphene channels [18]. As with the other range of effects covered, proximity effects can also induce magnetism in graphene, in which this induced magnetism can be used to generate spin currents via electrical and thermal means [6].

## Concluding remarks

The works discussed above highlight the variety of effects that can be exploited to enhance the functionality of graphene in spin valve architectures. The precise engineering of proximity phenomena in graphene-based heterostructures, including spin–orbit coupling, magnetic exchange, electrostatic gating and twist angles, has opened novel pathways to realise robust quantum-coherent spin transport channels [19]. This potential is underscored by recent efforts to explore the emergence of quantum spin Hall states in magnetic graphene, thereby confirming the effective interplay of proximity-induced spin–orbit interactions [20]. These advances not only promise to improve gate-tunable spin valves but also pave the way for evolving architectures that harness both spin and valley pseudospin degrees of freedom. Theoretical predictions indicate that graphene-based multilayers, when integrated with TMDs, can enable dynamically tunable spin and/or valley polarisation via gate voltages [21]. This tunability is crucial for the incorporation of graphene-based spin valves in spin logic devices, while further enhanced control over spin dynamics could produce energy-efficient and scalable spin FETs [22], as motivated by the Datta-Das spin transistor. Demonstrations of quantum transport components, such as quantum dot spin injectors previously demonstrated on semiconductor platforms [5] and more recently spin-polarised quantum point contacts in graphene-based spin valves [23], highlight design principles that may be used to advance graphene-based quantum spintronics.

## Acknowledgements

UK participants in Horizon Europe Project "2D Heterostructure Non-volatile Spin Memory Technology" (2DSPIN-TECH) are supported by UKRI grant number [10101734] (The University of Manchester).

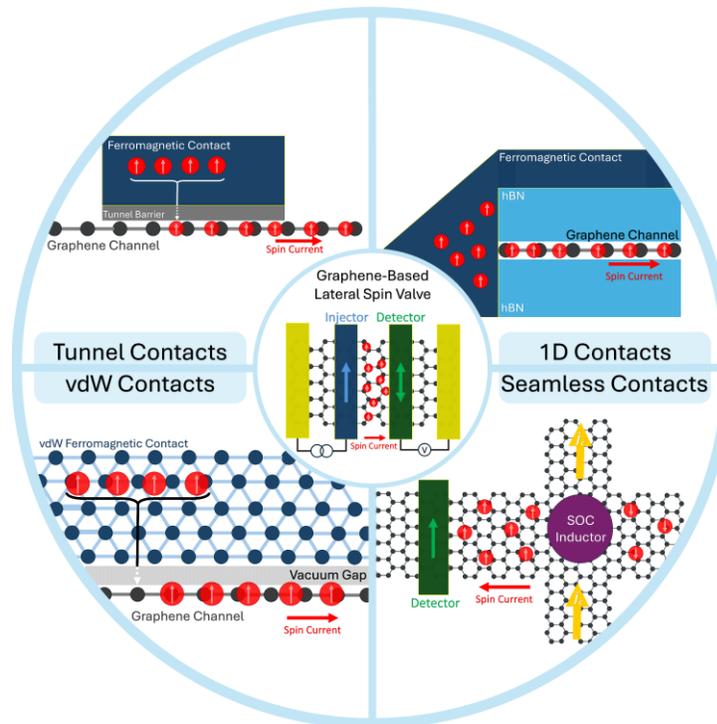

**Figure 1.** Electrical injection methods of spin current into a graphene channel. Central inset displays the basic architecture of a four-terminal spin valve with a graphene channel. The top left, top right and bottom left quadrants show three methods of increasing interfacial resistance to overcome conductivity mismatch: tunnel junction contacts, 1D contacts, and van der Waal (vdW) contacts respectively. The bottom right quadrant shows the use of proximity effects to induce quantum spin Hall effect to source a spin current within the graphene channel.

### References

[1] Nikolaos Tombros, Csaba Jozsa, Mihaita Popinciuc, H. Jonkman, and Bart Wees. Electronic spin transport and spin precession in single graphene layers at room temperature. Nature, 448:571–4, 09 2007.

[2] E. Hill, A. K. Geim, K. S. Novoselov, P. Blake, and F. Schedin. Graphene based spin valve devices. In 2006 IEEE International Magnetics Conference (INTERMAG), pages 385–385, 2006.

[3] Victor Guarochico-Moreira, Jose Sambricio, Khalid Omari, Christopher Anderson, Denis Bandurin, Jesus Toscano, Noel Cordero, Kenji Watanabe, Takashi Taniguchi, Irina Grigorieva, and Ivan Vera-Marun. Tunable spin injection in high-quality graphene with one-dimensional contacts. Nano Letters, 22, 01 2022.

[4] Supriyo Datta and Biswajit Das. Electronic analog of the electro-optic modulator. Applied Physics Letters, 56(7):665–667, 02 1990.

[5] Arunav Bordoloi, Valentina Zannier, Lucia Sorba, Christian Schoenenberger, and Andreas Baumgartner. A double quantum dot spin valve. Communications Physics, 3, 08 2020.

[6] Talieh S. Ghiasi, Alexey A. Kaverzin, Avalon H. Dismukes, Dennis K. de Wal, Xavier Roy, and Bart J. van Wees. Electrical and thermal generation of spin currents by magnetic bilayer graphene. Nature Nanotechnology, 16(7):788–794, Jul 2021.

[7] Ahmet Avsar, Dmitrii Unuchek, Jiawei Liu, Oriol Lopez-Sanchez, Kenji Watanabe, Takashi Taniguchi, Barbaros Ozyilmaz, and Andras Kis. Optospintronics in graphene via proximity coupling. ACS Nano, 11, 05 2017.

[8] Mallikarjuna Gurram, Siddhartha Omar, and Bart Wees. Bias induced up to 100% spin-injection and detection polarizations in ferromagnet/bilayer-hBN/graphene/hBN heterostructures Nature Communications, 8, 04 2017.

[9] Soumya Sarkar, Saeyoung Oh, Peter Newton, Yang Li, Yiru Zhu, Maheera Ghani, Han Yan, Hu Young Jeong, Yan Wang, and Manish Chhowalla. Spin injection in graphene using ferromagnetic van der waals contacts of indium and cobalt. Nature Electronics, 8:215–221, 01 2025.

[10] Bing Zhao, Roselle Ngaloy, Sukanya Ghosh, Soheil Ershadrad, Rahul Gupta, Khadiza Ali, Anamul Md. Hoque, Bogdan Karpiak, Dmitrii Khokhriakov, Craig Polley, Balasubramanian Thiagarajan, Alexei Kalaboukhov, Peter Svedlindh, Biplab Sanyal, and Saroj P. Dash. A room-temperature spin-valve with van der waals ferromagnet fe5gete2/graphene heterostructure. Advanced Materials, 35(16):2209113, 2023.

[11] Marc Drögeler, Christopher Franzen, Frank Volmer, Tobias Pohlmann, Luca Banszerus, Maik Wolter, Kenji Watanabe, Takashi Taniguchi, Christoph Stampfer, and Bernd Beschoten. Spin lifetimes exceeding 12 ns in graphene nonlocal spin valve devices. Nano Letters, 16(6):3533 3539, 2016. PMID: 27210240.

[12] Johannes Christian Leutenantsmeyer, Josep Ingla- Aynés, Jaroslav Fabian, and Bart J. van Wees. Observation of spin-valley-coupling-induced large spin-lifetime anisotropy in bilayer graphene. Phys. Rev. Lett., 121:127702, Sep 2018. 4

[13] Haozhe Yang, Marco Gobbi, Franz Herling, Van Tuong Pham, Francesco Calavalle, Beatriz Martín-García, Fert Albert, Luis Hueso, and Felix Casanova. A seamless graphene spin valve based on proximity to van der waals magnet cr2ge2te6. Nature Electronics, 8:15–23, 11 2024.

[14] Junxiong Hu, Yulei Han, Xiao Chi, Ganesh Ji Omar, Mohammed Mohammed Esmail Al Ezzi, Jian Gou, Xiaojiang Yu, Rusydi Andrivo, Kenji Watanabe, Takashi Taniguchi, Andrew Thye Shen Wee, Zhenhua Qiao, and A. Ariando. Tunable spin-polarized states in graphene on a ferrimagnetic oxide insulator. Advanced Materials, 36, 2 2024.

[15] Jonghyeon Choi, Jungmin Park, Seunghyeon Noh, Jaebyeong Lee, Seunghyun Lee, Daeseong Choe, Hyeonjung Jung, Junhyeon Jo, Inseon Oh, Juwon Han, Soon-Yong Kwon, Chang Ahn, Byoung-Chul Min, Hosub Jin, Choong Kim, Kyoung-Whan Kim, and Jung-Woo Yoo. Non-volatile fermi level tuning for the control of spin-charge conversion at room temperature. Nature Communications, 15, 10 2024.

[16] Haozhe Yang, Beatriz Martín-García, Jozef Kimák, Eva Schmoranzerová, Eoin Dolan, Zhen dong Chi, Marco Gobbi, Petr Němec, Luis Hueso, and Felix Casanova. Twist-angle-tunable spin texture in wse2/graphene van der waals heterostructures. Nature Materials, 23:1502–1508, 08 2024.

[17] Kathleen M. McCreary, Adrian G. Swartz, Wei Han, Jaroslav Fabian, and Roland K. Kawakami. Magnetic moment formation in graphene detected by scattering of pure spin currents. Phys. Rev. Lett., 109:186604, Nov 2012.

[18] Wei Hu, Chao Wang, Hao Tan, Hengli Duan, Guinan Li, Na Li, Qianqian Ji, Ying Lu, Yao Wang, Zhihu Sun, Fengchun Hu, and Wensheng Yan. Embedding atomic cobalt into graphene lattices to activate room-temperature ferromagnetism. Nature Communications, 12:1854, 3 2021.

[19] Klaus Zollner and Jaroslav Fabian. Proximity effects, topological states, and correlated physics in graphene heterostructures. 2D Materials, 12(1):013004, January 2025.

[20] Talieh S Ghiasi, Davit Petrosyan, Josep Ingla-Aynés, Tristan Bras, Kenji Watanabe, Takashi Taniguchi, Samuel Mañnas-Valero, Eugenio Coronado, Klaus Zollner, Jaroslav Fabian, Philip Kim, and Herre S. J. van der Zant. *Nat Commun* 16, 5336 (2025).

[21] Taige Wang, Marc Vila, Michael P Zaletel, and Shubhayu Chatterjee. Electrical control of spin and valley in spin-orbit coupled graphene multilayers. Physical Review Letters, 132:116504, 3 2024.

[22] Josep Ingla-Aynés , Franz Herling, Jaroslav Fabian, Luis E Hueso, and Fèlix Casanova. Electrical control of valley-zeeman spin-orbit-coupling–induced spin precession at room temperature. Physical Review Letters, 127:047202, 7 2021.

[23] Daniel Burrow, Jesus Toscano, Victor  Guarochico-Moreira, Khalid Omari, Irina Grigorieva, Thomas Thomson, and Ivan Vera-Marun. Spin polarised quantised transport via one dimensional nanowire-graphene contacts. Communications Materials, 6, 02 2025.

# 7. Quantum Magnonics in 2D Materials

**Michael Newburger[1], Simranjeet Singh[2] and Tiancong Zhu[3,4,5]**

[1] Materials and Manufacturing Directorate, Air Force Research Laboratory, Wright-Patterson AFB, OH, 45433, USA
[2] Department of Physics, Carnegie Mellon University, Pittsburgh, PA, 15213, USA
[3] Department of Physics and Astronomy, Purdue University, West Lafayette, IN, 47907, USA
[4] Birck Nanotechnology Center, Purdue University, West Lafayette, IN, 47907, USA
[5] Purdue Quantum Science and Engineering Institute, Purdue University, West Lafayette, IN, 47907, USA

E-mail: michael.newburger.1@us.af.mil, simranjs@andrew.cmu.edu, zhu1242@purdue.edu

**Status**

Quantum magnonics explores the quantum states of magnons—the quanta of collective spin excitations—their entanglement, and their interactions with other quantum platforms [1], with the goal of bridging magnon spintronics and quantum information science. This interdisciplinary field has emerged at the convergence of two rapidly evolving domains. On one hand, spintronics has advanced into the nanoscale regime, leveraging quasiparticles such as magnons to develop high-density, energy-efficient information technologies. At these scales, quantum effects become increasingly significant, necessitating the consideration of quantum behaviour in information processing and storage. On the other hand, the rapid progress in quantum technologies has driven the search for hybrid solid-state systems that combine complementary functionalities, enabling complex architectures and integration with existing systems. Magnons, with their inherent spin signatures, solid-state nature, long lifetimes, and low-dissipation transport for information transmission, present a promising platform for such hybrid quantum systems.

Recent discoveries of intrinsic long-range magnetic order in 2D van der Waals (vdW) materials, such as $CrI_3$, $Cr_2Ge_2Te_6$, and $MnPS_3$, have opened new opportunities for advancing quantum magnonics [2]. Compared to their three-dimensional counterparts, 2D magnets offer several compelling advantages. Their magnetic properties, including anisotropy, exchange coupling, and spin orbit coupling, are highly tunable via external stimuli such as electric fields, electrostatic gating, and mechanical strain, offering a platform to engineering tailored magnonic behavior [3]. The vdW nature of these materials enables facile integration with other quantum systems, allowing versatile design of hybrid quantum magnonic platforms [4]. Furthermore, their reduced dimensionality also gives rise to emergent magnonic phenomena, including exotic modes and non-trivial topological states that are inaccessible in the bulk systems, providing fertile ground for exploring new quantum effects [5].

Despite being an emergent field, research on 2D materials has shown promising advancement for quantum magnonics. For instance, low magnetic damping and high spin mixing conductance were observed in few-layer vdW magnet $Cr_2Ge_2Te_6$ [6], and propagating spin waves has also been directly observed in vdW ferromagnet $Fe_5GeTe_2$ [7] (Figure 1). These demonstrate the potential of using coherent magnons in 2D vdW materials for efficient information transmission. Researchers have also taken steps into understanding the coupling between 2D magnons with spin qubits[8], cavity photons[9], phonons[10], excitons[11], and other quantum systems. Meanwhile, new schemes such as manipulating one or more spin qubits with magnons, magnon-fluxon interconnection, and Bose-

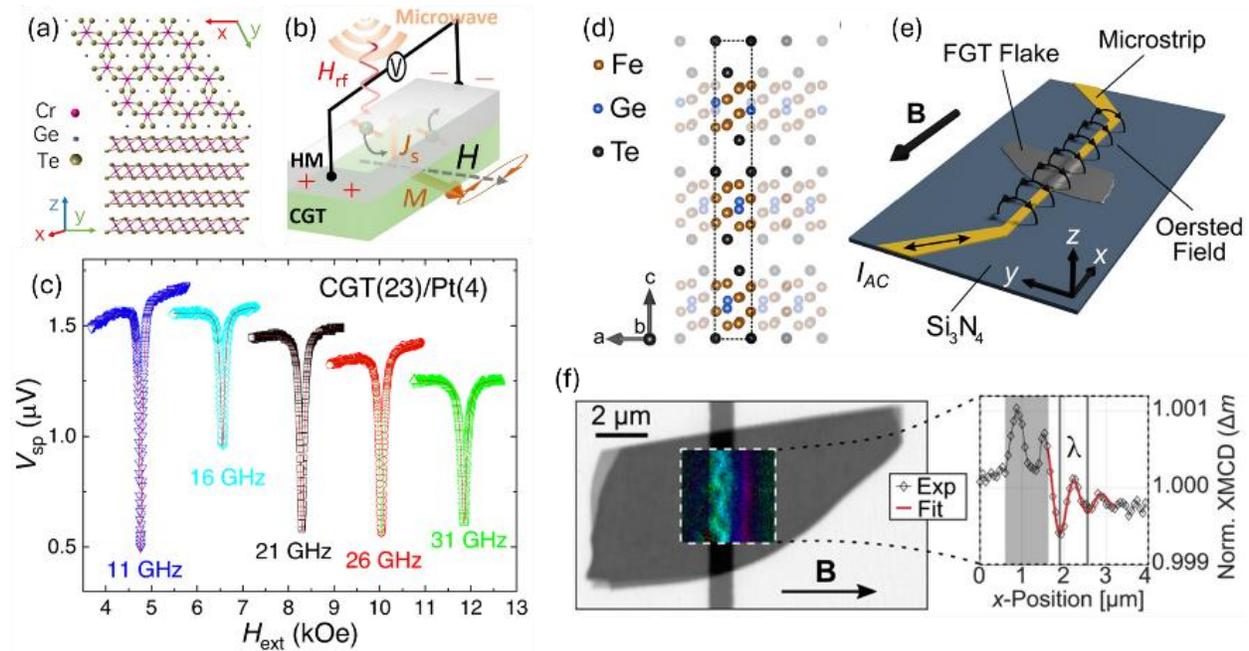

**Figure 1.** (a) Atomic structure of vdW magnet $Cr_2Ge_2Te_6$. (b) Schematic illustration of the spin-pumping experimental set-up. (c) A series of ferromagnetic resonance peaks from $Cr_2Ge_2Te_6$/Pt device under a wide range of RF signal frequencies (8-40 GHz). (a-c) adapted from [6] under CC-BY 4.0. (d) Atomic structure of vdW magnet $Fe_5GeTe_2$ (e) Illustration of sample geometry for imaging spin-waves with time-resolved scanning transmission X-ray microscopy. (f) Phase-resolved map and cross-section of the spin-wave dynamics in $Fe_5GeTe_2$. (d-f) adapted from [7] under CC-BY 4.0.

Einstein condensation of magnons have been proposed and tested [1], while their realization with 2D vdW materials remains an exciting direction.

**Current and future challenges**

Quantum magnonics require reliable generation, control, and detection of coherent magnons at a desired frequency and wave vector, which is challenging for 2D materials. Magnon generation with conventional inductive methods using microwaves suffers from a lack of magnetic volume in 2D materials due to reduced sample thickness and size, while detection also requires exceptionally high-sensitivity probes to pick up the small magnetic field response [4]. The situation is further complicated by the dependency of magnetization on the number of layers in these materials. To date, experiments have mainly focused on uniform ferromagnetic resonance (FMR) modes in thicker flakes, while demonstration of propagating finite-wavelength spin waves remains rare and challenging [7].

An alternative solution is to create hybrid quantum devices that couple magnons with other degrees of freedom including phonons, photons, or polarons, and detect coherent magnons via spectroscopic methods such as Raman spectroscopy and Brillouin light scattering. Indeed, 2D magnets often support strong magnon-phonon and magnon-plasmon interactions. For example, semiconducting vdW magnets exhibit spin-exciton coupling that enables optical pumping and readout of magnons [11], and strong magnon-polariton coupling in CrSBr provides enhanced optical control of magnetic states [9]. These examples show promise, but the creation of reproducible hybrid magnonic interfaces is still under active development. Coupling magnons to spin qubits is another intriguing avenue. Recent first-principles work predicts that a carefully chosen molecular spin qubit on a CrSBr surface could exchange-couple with 2D magnons and allow quantum control of spin waves [8], but realizing such an ultrafast, low-decoherence magnon–qubit interface will require exquisite control of the chemistry and the interfacial exchange interaction.

Material synthesis is fundamental to 2D quantum magnonics, yet scalable growth of high quality, large area 2D magnetic crystals is rare, making reliable magnonic devices hard to produce. A majority of 2D magnet studies to date still rely on mechanical exfoliation from bulk crystals, yielding small flakes with varied thickness. Many 2D magnets also degrade rapidly in ambient conditions, posing issues for stability. Particularly, hybrid quantum magnonic systems require scalable synthesis of designed heterostructures with clean interfaces. Growing large-area films with chemical vapor deposition (CVD) and molecular beam epitaxy (MBE) has recently become feasible [12], but extending these bottom-up methods to many 2D magnets for scalable devices remains a challenge. The high crystallinity nature of 2D magnets also poses strict requirements for eliminating lattice mismatch, minimizing interfacial diffusion, and ensuring low defect density and exact stoichiometry during deposition. Achieving large area, high quality films and heterostructures with clean interfaces by-design is still far from routine.

**Advances in science and technology to meet challenges**

Quantum sensing has emerged as a new way to detect magnons in 2D materials with ultimate resolution and sensitivity. Optical detection of magnetic resonance using quantum spin sensors has allowed probing of local spin dynamics and magnon propagation with high spatial and temporal resolution [13–15] (Figure 2). In particular, spin centers in hexagonal boron nitride (hBN) exhibit superior potential for sensing 2D magnonics due to their seamless integration with other 2D systems, sole defect axis perpendicular to the 2D plane, wide range of operating temperature, long spin relaxation time, and amenability to single photon emission from individual defects [16]. Importantly, the coupling strength of these sensors can be enhanced due to proximity of the defect to the magnetic material. Thus, defects in few-layer hBN may enable detection of weak stray fields from the propagation of magnons in 2D magnets down to single-atom thickness.

For the development of hybrid quantum systems, it is also critical to study the fundamental electric, magnetic, optical, and strain coupling with 2D magnets, and the degree to which they can be modified through design of heterostructures and alloys. While preliminary efforts have demonstrated new phenomena such as Skyrmions in vdW heterostructures [12], the dynamics of these systems, and the degree to which magnon dispersion can be engineered, remain largely unexplored. Expanded optical characterization efforts, including time and spatially resolved magneto-optic Kerr effect and Brillouin light scattering, appear slated to greatly impact this understanding by directly measuring coupled magnetic excitations in these materials. How patterned super-structures such as magnonic crystals could be used to modify the properties of magnons is also of interest. A collaborative approach between theory, computational simulation, and experiment is highly desired for those directions.

Searching for new 2D magnetic materials and developing novel synthesis methods is also highly desirable in the development of 2D quantum magnonics. Apart from traditional approaches, the use of machine learning and artificial intelligence for high-throughput material screening [17] and autonomous growth [18] has emerged as an attractive route. Applying these methods could boost material discoveries and create designer 2D magnets for quantum magnonics. On the other hand, new synthesis methods such as remote epitaxy [19] and full-film dry transfer of MBE-grown van der Waals materials [20] might provide new avenues to circumventing compatibility issues fabricating wafer-scale vdW heterostructures, combining the advantages of vdW stacking and scalable bottom-up synthesis to enable rapid compatibility with transmon qubits, point defects, and other existing devices.

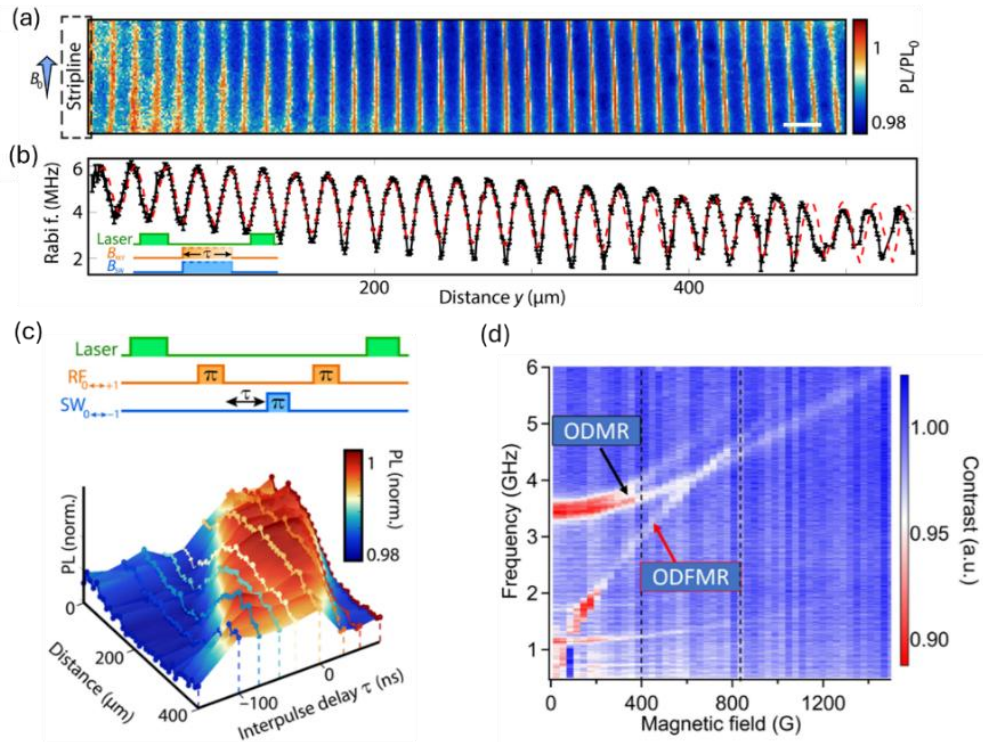

**Figure 2.** (a-c) Spatial and temporal image of magnon transport in YIG measured with NV center. (a) Spatial ESR contrast of NV photoluminescence excited in YIG by a microwave current in a stripline. (b) Rabi frequency versus distance from the strapline. Inset: measurement sequence. (c) Spin-wave dispersion in the space and time domain, where the slope of the red area indicates a spin-wave group velocity of 3.6 km/s. (a-c) adapted from [13] under <u>CC BY-NC 4.0</u>. (d) Optical response of a hybrid YIG/ $V_{B^-}$ showing optically detected ferromagnetic resonance measured by $V_{B^-}$ centers in hBN. (d) Reprinted with permission from [14], license number RNP/25/MAY/091021.

**Concluding remarks**

Quantum magnonics is a rapidly advancing field, and its extension to 2D materials remains in its early stages. Nonetheless, proof-of-principle experiments have demonstrated their feasibility, highlighting the promise of 2D systems with their exceptional tunability, efficient manipulation, and integration potential. At present, the reliable creation, control, and detection of coherent magnons in 2D magnets remain grand challenges which demand a closer integration of theoretical modeling, advanced material synthesis, and cutting-edge measurement techniques. Encouragingly, recent progress in high-sensitivity quantum sensing and bottom-up growth methods has begun to bridge the gap between conceptual frameworks and experimental realization. Novel ideas, such as generating spin superfluid transport through Bose-Einstein-condensation-like instability and spin transport via topologically protected magnon edge states, are also intriguing to be pursued. Looking ahead, the integration of 2D magnets into designed hybrid quantum systems will offer an exciting pathway to uncover novel magnonic phenomena and unlock new functionalities for quantum technologies.

**Acknowledgements**

M.N. acknowledges support from the Air Force Office of Scientific Research (FA955023RXCOR001). S.S. acknowledges the financial support from National Science Foundation (NSF) through Grant No. DMR-221051 and NSF-CAREER Award through Grant No. ECCS-2339723. T. Z. acknowledges support from Center for Quantum Technologies under the Industry-University Cooperative Research Center Program at the US National Science Foundation under Grant Nos. 2224960 (Purdue University), 2224928 (Indiana University), and 2224985 (University of Notre Dame).

**References**

[1]      Yuan H Y, Cao Y, Kamra A, Duine R A and Yan P 2022 Quantum magnonics: When magnon spintronics meets quantum information science *Phys. Rep.* **965** 1–74

[2]      Gong C and Zhang X 2019 Two-dimensional magnetic crystals and emergent heterostructure devices *Science* **363** eaav4450

[3]      Ahn Y, Guo X, Son S, Sun Z and Zhao L 2024 Progress and prospects in two-dimensional magnetism of van der Waals materials *Prog. Quantum Electron.* **93** 100498

[4]      Mañas-Valero S, Sar T van der, Duine R A and Wees B van 2025 Fundamentals and applications of van der Waals magnets in magnon spintronics *Newton* **1**

[5]      Zhuo F, Kang J, Manchon A and Cheng Z 2025 Topological Phases in Magnonics *Adv. Phys. Res.* **4** 2300054

[6]      Xu H, Jia K, Huang Y, Meng F, Zhang Q, Zhang Y, Cheng C, Lan G, Dong J, Wei J, Feng J, He C, Yuan Z, Zhu M, He W, Wan C, Wei H, Wang S, Shao Q, Gu L, Coey M, Shi Y, Zhang G, Han X and Yu G 2023 Electrical detection of spin pumping in van der Waals ferromagnetic Cr2Ge2Te6 with low magnetic damping *Nat. Commun.* **14** 3824

[7]      Schulz F, Litzius K, Powalla L, Birch M T, Gallardo R A, Satheesh S, Weigand M, Scholz T, Lotsch B V, Schütz G, Burghard M and Wintz S 2023 Direct Observation of Propagating Spin Waves in the 2D van der Waals Ferromagnet Fe5GeTe2 *Nano Lett.* **23** 10126–31

[8]      Dey S, Rivero-Carracedo G, Shumilin A, Gonzalez-Ballestero C and Baldoví J J 2025 Coupling molecular spin qubits with 2D magnets for coherent magnon manipulation

[9]      Dirnberger F, Quan J, Bushati R, Diederich G M, Florian M, Klein J, Mosina K, Sofer Z, Xu X, Kamra A, García-Vidal F J, Alù A and Menon V M 2023 Magneto-optics in a van der Waals magnet tuned by self-hybridized polaritons *Nature* **620** 533–7

[10]     Lyons T P, Puebla J, Yamamoto K, Deacon R S, Hwang Y, Ishibashi K, Maekawa S and Otani Y 2023 Acoustically Driven Coherent Magnon-Phonon Coupling in a Layered Antiferromagnet *Phys. Rev. Lett.* **131** 196701

[11]     Bae Y J, Wang J, Scheie A, Xu J, Chica D G, Diederich G M, Cenker J, Ziebel M E, Bai Y, Ren H, Dean C R, Delor M, Xu X, Roy X, Kent A D and Zhu X 2022 Exciton-coupled coherent magnons in a 2D semiconductor *Nature* **609** 282–6

[12]     Zhang B, Lu P, Tabrizian R, Feng P X-L and Wu Y 2024 2D Magnetic heterostructures: spintronics and quantum future *Npj Spintron.* **2** 1–10

[13]     Bertelli I, Carmiggelt J J, Yu T, Simon B G, Pothoven C C, Bauer G E W, Blanter Y M, Aarts J and van der Sar T 2020 Magnetic resonance imaging of spin-wave transport and interference in a magnetic insulator *Sci. Adv.* **6** eabd3556

[14]     Das S, Melendez A L, Kao I-H, García-Monge J A, Russell D, Li J, Watanabe K, Taniguchi T, Edgar J H, Katoch J, Yang F, Hammel P C and Singh S 2024 Quantum Sensing of Spin Dynamics Using Boron-Vacancy Centers in Hexagonal Boron Nitride *Phys. Rev. Lett.* **133** 166704

[15]     Zhou J, Lu H, Chen D, Huang M, Yan G Q, Al-matouq F, Chang J, Djugba D, Jiang Z, Wang H and Du C R 2024 Sensing spin wave excitations by spin defects in few-layer-thick hexagonal boron nitride *Sci. Adv.* **10** eadk8495

[16]     Vaidya S, Gao X, Dikshit S, Aharonovich I and Li T 2023 Quantum sensing and imaging with spin defects in hexagonal boron nitride *Adv. Phys. X* **8** 2206049

[17]     Torelli D, Moustafa H, Jacobsen K W and Olsen T 2020 High-throughput computational screening for two-dimensional magnetic materials based on experimental databases of three-dimensional compounds *Npj Comput. Mater.* **6** 1–12

[18]     Sabattini L, Coriolano A, Casert C, Forti S, Barnard E S, Beltram F, Pontil M, Whitelam S, Coletti C and Rossi A 2025 Towards AI-driven autonomous growth of 2D materials based on a graphene case study *Commun. Phys.* **8** 1–8

[19]     Kim H, Chang C S, Lee S, Jiang J, Jeong J, Park M, Meng Y, Ji J, Kwon Y, Sun X, Kong W, Kum H S, Bae S-H, Lee K, Hong Y J, Shi J and Kim J 2022 Remote epitaxy *Nat. Rev. Methods Primer* **2** 1–21

[20]     Li Z, Zhou W, Swann M, Vorona V, Scott H and Kawakami R K 2025 Full-film dry transfer of MBE-grown van der Waals materials *2D Mater.* **12** 035003

# 8. Nonlinear Quantum Photonics with 2D Materials

**Mauro Brotons-Gisbert[1], Klaus D. Jöns[2] and Brian D. Gerardot[1]**

[1] Institute of Photonics and Quantum Sciences, SUPA, Heriot-Watt University, EH14 4AS, United Kingdom
[2] Institute for Photonic Quantum Systems (PhoQS), Center for Optoelectronics and Photonics Paderborn (CeOPP) and Department of Physics, Paderborn University, 33098 Paderborn, Germany

E-mail: b.d.gerardot@hw.ac.uk

**Status**

Quantum photonics leverages the principles of quantum mechanics to manipulate light at the single-photon level, enabling applications in quantum computing, communication, and sensing. At the heart of such quantum photonic technologies lies the generation and manipulation of entangled photon pairs. Technologically, entangled photon pairs are primarily achieved through spontaneous parametric down-conversion (SPDC), a nonlinear optical process where a photon from a high-energy pump beam is converted into two lower-energy entangled photons within a nonlinear medium. This phenomenon forms the cornerstone of quantum optics, providing a vital resource for various quantum protocols.

Traditional SPDC relies on bulk nonlinear optical materials like beta-barium borate or lithium niobate, which, despite their effectiveness, often necessitate large-scale setups and exhibit limited tunability and integration capabilities. These constraints pose significant challenges in scaling quantum photonics to practical, deployable technologies. The advent of two-dimensional (2D) materials offers the potential for a transformative shift in the field. Due to the intrinsic broken inversion symmetry of many 2D crystals, they exhibit extraordinarily high optical nonlinearities. Combined with ultra-thin profiles and a high index of refraction, this can facilitate enhanced light-matter interactions at the nanoscale. Their remarkable nonlinear optical properties can be orders of magnitude greater than those of traditional bulk materials, making them ideal for miniaturized and highly efficient entangled photon sources which simultaneously offer facile integration with photonic circuits. Moreover, advancements in the fabrication techniques of 2D materials, allowing precise control over layer number and stacking configurations, have unlocked new ways to tailor their optical responses. This capability enables the customization of quantum light sources to specific needs, addressing efficiency, tunability, and integration limitations currently faced by conventional photon pair sources.

Looking forward, the role of 2D materials in nonlinear quantum photonics is poised to expand significantly. By enhancing the generation and manipulation of entangled photons, 2D nonlinear quantum photonics will enable novel chip-scale devices which underpin future advancements in quantum computing, communication, and sensing.

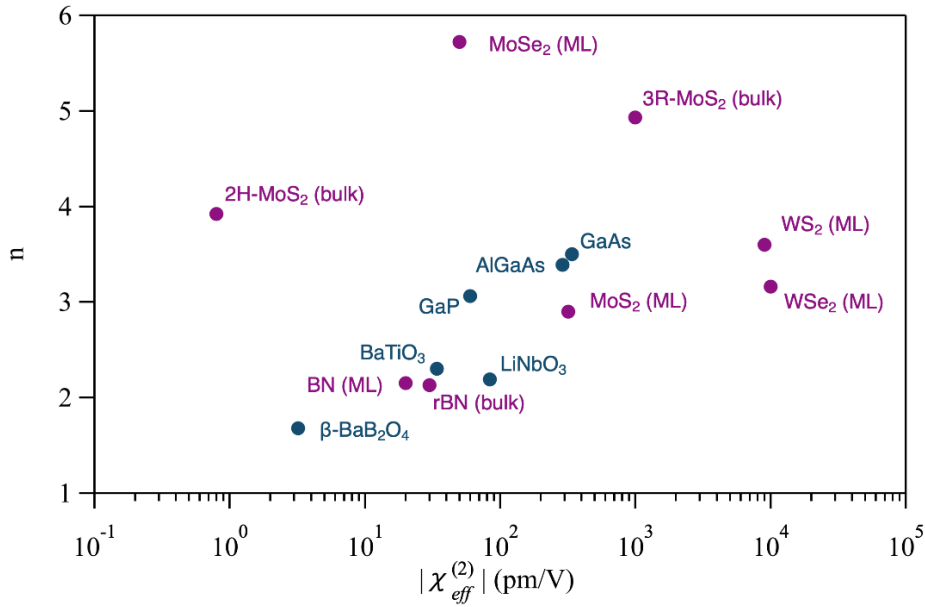

**Figure 1**. Comparison of the maximum effective bulk-like second-order susceptibility $|\chi_{eff}^{(2)}|$ component versus refractive index n for several layered van der Waals materials and conventional nonlinear bulk crystals. The colour labelling indicates the SHG wavelength at which the corresponding $|\chi_{eff}^{(2)}|$ values were measured. Optical data for various materials are taken from: bulk 2H-MoS$_2$ (405 nm) [1, 2], ML MoS$_2$ (405 nm) [2, 3], 3R-MoS$_2$ (532 nm) [1, 4], ML WSe$_2$ (408 nm) [5, 6], ML WS$_2$ (416 nm) [6, 7], ML MoSe$_2$ (800 nm) [8], ML hBN (405 nm) [1, 2], bulk rBN (532 nm) [1,9], GaAs (405 nm), AlGaAs (785 nm), GaP (608 nm), BaTiO$_3$ (532 nm), LiNbO$_3$ (532 nm), β-BaB$_2$O$_3$ (532 nm) [10].

**Current and future challenges**

Realizing the full potential of nonlinear quantum photonics with 2D materials hinges on overcoming several scientific and technological challenges. One major limitation is the centrosymmetric nature of the hexagonal crystal structure in most commonly available bulk van der Waals materials, which results in vanishing effective second-order nonlinear susceptibilities ($\chi_{eff}^{(2)}$). This fundamentally hinders their direct integration into nonlinear photonic applications. Recent demonstrations of SPDC in rhombohedral polytypes of MoS$_2$[11], WS$_2$[12], BN [13], as well as both natural and twisted NbOCl$_2$ [14,15] show the promise of exploiting crystal symmetry and stacking order, but scalable **high-purity synthesis** routes of such non-centrosymmetric crystal phases are still underdeveloped.

Enhancing the conversion efficiency of SPDC in atomically thin media is another crucial requirement for advancing the application of 2D materials in nonlinear quantum photonics. Despite exhibiting giant, effective bulk-like second-order nonlinear susceptibilities [2,8,16], the nanometer-scale thickness inherently limits the interaction length, and thus the photon-pair generation efficiency. Thus, advancing this field requires **enhanced light–matter interaction**. Potential strategies include employing intrinsically non-centrosymmetric van der Waals crystals with micro-scale (rather than nanoscale) thicknesses, minimizing losses due to excitonic absorption (which often accompanies high χ2 values), increasing the optical nonlinearity via the broken inversion symmetry arising from twisted interfaces [17], and integrating 2D materials into resonant cavities, waveguides, or metasurfaces.

Further, engineering **phase matching** in nonlinear 2D materials is a key opportunity in an atomically thin platform. Unlike bulk crystals, 2D materials lack dispersion-induced phase mismatch, enabling ultra-broadband nonlinear interactions at the ultimate thickness limit [18, 19]. However, as the thickness of the 2D media scales to increase the light-matter interaction efficiency, "tricks" first pioneered in bulk nonlinear optics such as periodic poling can be implemented for 2D materials. Initial progress has been made toward phase-matched [20] as well as periodically poled van der Waals structures [9, 21]. Notably, periodically poled $3R\text{-}MoS_2$ has demonstrated enhanced SPDC efficiency and improved coincidence-to-accidental ratios at telecom wavelengths [21]. However, its performance still lags behind that of conventional nonlinear crystals. Further improvements in SPDC efficiency are expected with an **increased number of poled periods** beyond the current limit of three.

Technologically, **on-chip integration** of 2D materials with scalable photonic platforms remains another critical frontier. Efficient coupling of pump and generated photons into and out of 2D layers—ideally on-chip—requires the development of advanced nanophotonic interfaces, including gratings, waveguides, cavities or resonators, and perhaps metasurfaces.

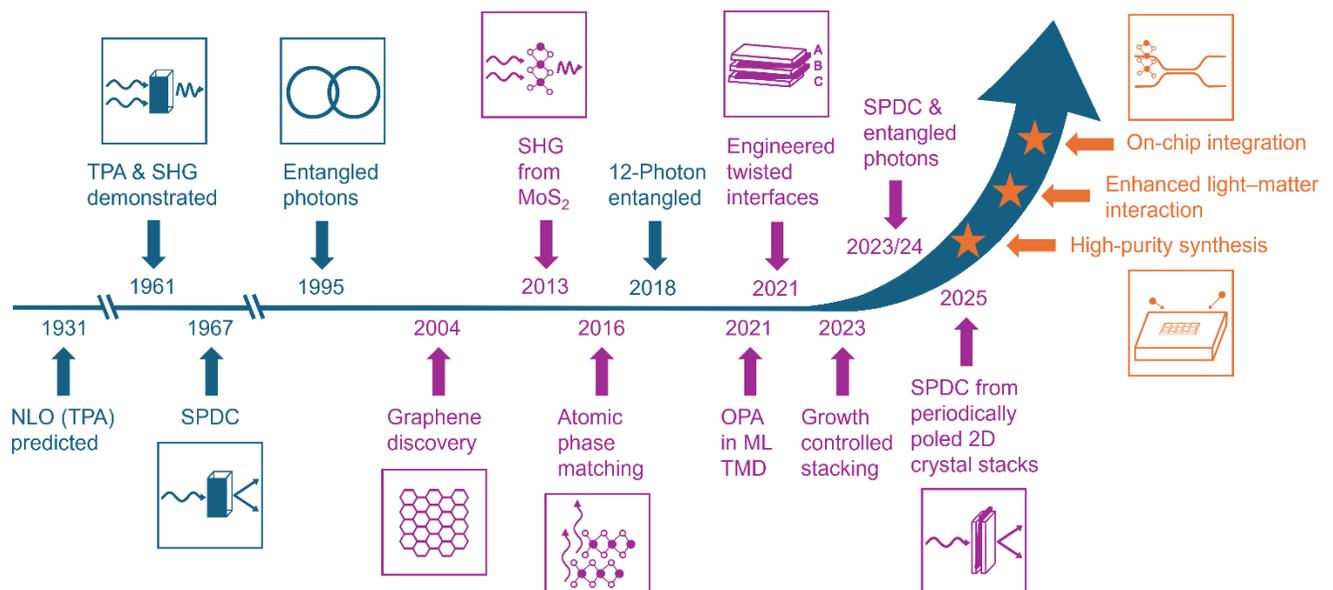

**Figure 2**. Timeline of the most relevant achievements and developments in the field of nonlinear optics for quantum optics and using conventional bulk (turquoise) and 2D (magenta) materials. The fast progress and short timeframe in-between of discoveries using 2D materials helps put these achievements in perspective. Notably, it only took 11 years from the demonstration of SHG from 2D materials to achieve entangled photon generation, compared to 34 years in the field of nonlinear optics using bulk crystals. The stars (orange) highlight the three next possible landmarks which will push the field forward.

Finally, to enable quantum applications, the generated photon pairs must meet stringent requirements in terms of purity, brightness, indistinguishability, and entanglement fidelity. This calls for novel approaches to engineer the nonlinearity profiles of the 2D-based crystals.

**Advances in science and technology to meet challenges**

Advancing nonlinear 2D materials for photonic quantum technologies requires several fundamental and technological breakthroughs. From a materials science perspective, the field demands reliable, **high-purity synthesis** methods for large-area, non-centrosymmetric 2D crystals and heterostructures—such as 3R-phase transition metal dichalcogenides (3R-TMDs), rhombohedral boron nitride (rBN), and ferroelectric layered materials. A major challenge lies in uncontrolled crystal stacking and stacking faults, which induce optical phase deviations across different 2D layers, leading to destructive interference of the nonlinear response. This significantly hinders efficient light–matter interaction. Moreover, the chemical stability of these materials, particularly under ambient conditions and high optical powers, must be addressed to ensure long-term device operation.

From a technological standpoint, **greater precision in material placement, alignment, and encapsulation** is essential to preserve optical quality, improve device yield, and ensure performance reproducibility. In addition, the development of automated or wafer-scale transfer techniques is crucial to move beyond proof-of-concept demonstrations. Such advances will enable the fabrication of large-scale stacks of engineered, quasi-phase-matched 2D nonlinear devices with a number of poling periods comparable to those found in conventional periodically poled nonlinear crystals.

From a more fundamental perspective, the ability to apply arbitrary twist angles between crystal domains enables control over the nonlinear optical response through engineered "**twist-phase-matching**". This approach holds promise for enhancing nonlinear conversion efficiency and for tuning the polarization and spectral properties of the generated photons—capabilities that go beyond those offered by conventional periodically poled bulk nonlinear crystals.

Furthermore, the effects of nonlocal electric responses to **spatially structured electromagnetic fields** in van der Waals crystals remain largely unexplored, both theoretically and experimentally. Spatial dispersion effects, assumed to be negligible in bulk nonlinear materials, may become significant in layered materials where the characteristic dimensions of electromagnetic field variations approach or exceed interlayer distances. In such cases, emergent selection rules, altered phase-matching conditions, or spatially varying nonlinear susceptibilities not previously considered may emerge. Advancing our understanding in this area could open new pathways for nonlinear quantum optical applications that transcend the limitations imposed by the intrinsic crystal symmetry of nonlinear materials. For example, harnessing such effects could provide access to otherwise forbidden nonlinear processes, enable dynamic control over quantum light generation via spatial field shaping, or permit tailored nonlinearities through engineered ferroelectric domains, dielectric environments, or strain profiles. Exploring spatially structured nonlinear responses may thus unlock entirely new design principles for compact, tunable, and symmetry-adapted entangled photon sources.

**Concluding remarks**

In conclusion, nonlinear quantum optics with 2D materials is a rapidly evolving field that, despite its youth, has achieved remarkable progress in recent years. The unique optical and electronic properties of 2D systems and the ability to tailor matter on the atomic level offer great promise for next-generation nonlinear optics devices. However, critical challenges remain, particularly in achieving scalability, high-purity large-area fabrication and efficient on-chip integration are required. Harnessing the strong non-linearities in 2D materials will be enabled by stronger light–matter interaction. Coordinated interdisciplinary advances across nonlinear optics, materials science, nanofabrication, and quantum photonics will be required—but the reward may be scalable, room-temperature quantum light sources with unprecedented compactness and functionality.

**Acknowledgements**

This work is supported by the Deutsche Forschungsgemeinschaft (German Research Foundation) through the transregional collaborative research center TRR142/3-2022 (231447078), the ERC grant (LiNQs, 101042672), and the UK Engineering and Physical Sciences Research Council (EP/P029892/1, EP/Y026284/1, EP/Z533208/1). M.B.-G. is supported by a Royal Society University Research Fellowship. B.D.G. is supported by a Chair in Emerging Technology from the Royal Academy of Engineering.

**References**

[1]        Zotev P G, Wang Y, Andres-Penares D, Severs-Millard T, Randerson S, Hu X, Sortino L, Louca C, Brotons-Gisbert M, Huq T, Vezzoli S, Sapienza R, Krauss T F, Gerardot B D and Tartakovskii A I 2023 Van der Waals Materials for Applications in Nanophotonics *Laser Photonics Rev.* **17** 2200957

[2]        Li Y, Rao Y, Mak K F, You Y, Wang S, Dean C R and Heinz T F 2013 Probing Symmetry Properties of Few-Layer MoS2 and h-BN by Optical Second-Harmonic Generation *Nano Lett.* **13** 3329–33

[3]        Islam K M, Synowicki R, Ismael T, Oguntoye I, Grinalds N and Escarra M D 2021 In-Plane and Out-of-Plane Optical Properties of Monolayer, Few-Layer, and Thin-Film MoS2 from 190 to 1700 nm and Their Application in Photonic Device Design *Adv. Photonics Res.* **2** 2000180

[4]        Wagoner G A, Persans P D, Wagenen E A V and Korenowski G M 1998 Second-harmonic generation in molybdenum disulfide *JOSA B* **15** 1017–21

[5]        Gu H, Song B, Fang M, Hong Y, Chen X, Jiang H, Ren W and Liu S 2019 Layer-dependent dielectric and optical properties of centimeter-scale 2D WSe2: evolution from a single layer to few layers *Nanoscale* **11** 22762–71

[6]        Ribeiro-Soares J, Janisch C, Liu Z, Elías A L, Dresselhaus M S, Terrones M, Cançado L G and Jorio A 2015 Second Harmonic Generation in WSe2 *2D Mater.* **2** 045015

[7]        Ermolaev G A, Yakubovsky D I, Stebunov Y V, Arsenin A V and Volkov V S 2019 Spectral ellipsometry of monolayer transition metal dichalcogenides: Analysis of excitonic peaks in dispersion *J. Vac. Sci. Technol. B* **38** 014002

[8]        Le C T, Clark D J, Ullah F, Senthilkumar V, Jang J I, Sim Y, Seong M-J, Chung K-H, Park H and Kim Y S 2016 Nonlinear optical characteristics of monolayer MoSe2 *Ann. Phys.* **528** 551–9

[9]        Qi J, Ma C, Guo Q, Ma C, Zhang Z, Liu F, Shi X, Wang L, Xue M, Wu M, Gao P, Hong H, Wang X, Wang E, Liu C and Liu K 2024 Stacking-Controlled Growth of rBN Crystalline Films with High Nonlinear Optical Conversion Efficiency up to 1% *Adv. Mater.* **36** 2303122

[10]       Eimerl D, Davis L, Velsko S, Graham E K and Zalkin A 1987 Optical, mechanical, and thermal properties of barium borate *J. Appl. Phys.* **62** 1968–83

[11]       Weissflog M A, Fedotova A, Tang Y, Santos E A, Laudert B, Shinde S, Abtahi F, Afsharnia M, Pérez Pérez I, Ritter S, Qin H, Janousek J, Shradha S, Staude I, Saravi S, Pertsch T, Setzpfandt F, Lu Y and Eilenberger F 2024 A tunable transition metal dichalcogenide entangled photon-pair source *Nat. Commun.* **15** 7600

[12]       Feng J, Wu Y-K, Duan R, Wang J, Chen W, Qin J, Liu Z, Guo G-C, Ren X-F and Qiu C-W 2024 Polarization-entangled photon-pair source with van der Waals 3R-WS2 crystal *eLight* **4** 16

[13]     Liang H, Gu T, Lou Y, Yang C, Ma C, Qi J, Bettiol A A and Wang X 2025 Tunable polarization entangled photon-pair source in rhombohedral boron nitride *Sci. Adv.* **11** eadt3710

[14]     Kallioniemi L, Lyu X, He R, Rasmita A, Duan R, Liu Z and Gao W 2025 Van der Waals engineering for quantum-entangled photon generation *Nat. Photonics* **19** 142–8

[15]     Guo Q, Qi X-Z, Zhang L, Gao M, Hu S, Zhou W, Zang W, Zhao X, Wang J, Yan B, Xu M, Wu Y-K, Eda G, Xiao Z, Yang S A, Gou H, Feng Y P, Guo G-C, Zhou W, Ren X-F, Qiu C-W, Pennycook S J and Wee A T S 2023 Ultrathin quantum light source with van der Waals NbOCl2 crystal *Nature* **613** 53–9

[16]     Malard L M, Alencar T V, Barboza A P M, Mak K F and de Paula A M 2013 Observation of intense second harmonic generation from MoS${}_{2}$ atomic crystals *Phys. Rev. B* **87** 201401

[17]     Yao K, Finney N R, Zhang J, Moore S L, Xian L, Tancogne-Dejean N, Liu F, Ardelean J, Xu X, Halbertal D, Watanabe K, Taniguchi T, Ochoa H, Asenjo-Garcia A, Zhu X, Basov D N, Rubio A, Dean C R, Hone J and Schuck P J 2021 Enhanced tunable second harmonic generation from twistable interfaces and vertical superlattices in boron nitride homostructures *Sci. Adv.* **7** eabe8691

[18]     Zhao M, Ye Z, Suzuki R, Ye Y, Zhu H, Xiao J, Wang Y, Iwasa Y and Zhang X 2016 Atomically phase-matched second-harmonic generation in a 2D crystal *Light Sci. Appl.* **5** e16131–e16131

[19]     Trovatello C, Marini A, Xu X, Lee C, Liu F, Curreli N, Manzoni C, Dal Conte S, Yao K, Ciattoni A, Hone J, Zhu X, Schuck P J and Cerullo G 2021 Optical parametric amplification by monolayer transition metal dichalcogenides *Nat. Photonics* **15** 6–10

[20]     Xu X, Trovatello C, Mooshammer F, Shao Y, Zhang S, Yao K, Basov D N, Cerullo G and Schuck P J 2022 Towards compact phase-matched and waveguided nonlinear optics in atomically layered semiconductors *Nat. Photonics* **16** 698–706

[21]     Trovatello C, Ferrante C, Yang B, Bajo J, Braun B, Peng Z H, Xu X, Jenke P K, Ye A, Delor M, Basov D N, Park J, Walther P, Dean C R, Rozema L A, Marini A, Cerullo G and Schuck P J 2025 Quasi-phase-matched up- and down-conversion in periodically poled layered semiconductors *Nat. Photonics* **19** 291–9

# 9. Cavity-based engineering of 2D quantum materials

**Brian S. Y. Kim[1,2], John R. Schaibley[2], Kyle L. Seyler[3]**

[1] Department of Materials Science & Engineering, University of Arizona, Tucson, Arizona 85721, USA
[2] Department of Physics, University of Arizona, Tucson, Arizona 85721, USA
[3] College of Optical Sciences, University of Arizona, Tucson, Arizona 85719, USA

E-mail: briankim@arizona.edu

**Status**

Controlling light-matter interactions in quantum materials is a challenge at the frontier of quantum optics. This capability enables manipulation of their optical and electronic properties with tremendous potential for applications in quantum information science, quantum sensing, and lasers. However, the intrinsic strength of light-matter coupling is weak, as reflected by the small fine-structure constant $\alpha$ ($\approx 1/137$). A powerful approach to strengthen light-matter interaction is to leverage intense electromagnetic field enhancement within the confined optical mode volume of optical cavities [1]. This cavity-based approach, first conceptualized by Purcell in 1946, involves embedding a quantum emitter (e.g., a two-level system) inside a resonant optical cavity, which modifies the electromagnetic environment and spontaneous emission rate of the emitter [2]. Recently, 2D quantum materials have emerged as an ideal platform to explore cavity-based engineering. Their reduced dimensionality fosters enhanced light-matter interactions, and their ability to be assembled into van der Waals (vdW) heterostructures allows for integration into state-of-the-art optical cavity structures.

Integrating 2D materials with resonant optical cavities has become an important strategy to tune their optical emission properties. There are now many studies demonstrating how coupling to cavity modes can dramatically enhance optical nonlinearities, modify emission rates, and tune spin-valley properties in a wide variety of 2D materials such as monolayer transition metal dichalcogenides (Figure 1a,b). Examples include improvements in second harmonic generation [3], single photon emission rate [4], valley splitting [5], and the creation of efficient nanolasers [6]. Another promising direction involves combining optical cavities with Floquet engineering schemes [7], which use periodic optical driving to transiently modify the ground state of quantum materials. While Floquet engineering offers tremendous potential to realize new phases inaccessible in equilibrium, the main challenge is the high driving field intensity required to sufficiently perturb the ground state of 2D materials. Importantly, coupling to cavity modes provides a viable route to accessing intrinsic Floquet-engineered properties induced by light-matter interaction effects while minimizing deleterious heating effects associated with high field intensities. For example, a recent study showed that excitons coupled to cavity modes in monolayer $WSe_2$ encapsulated within distributed Bragg reflectors exhibit a two-order-of-magnitude reduction in the threshold field intensity required to induce appreciable spin and valley splitting [8].

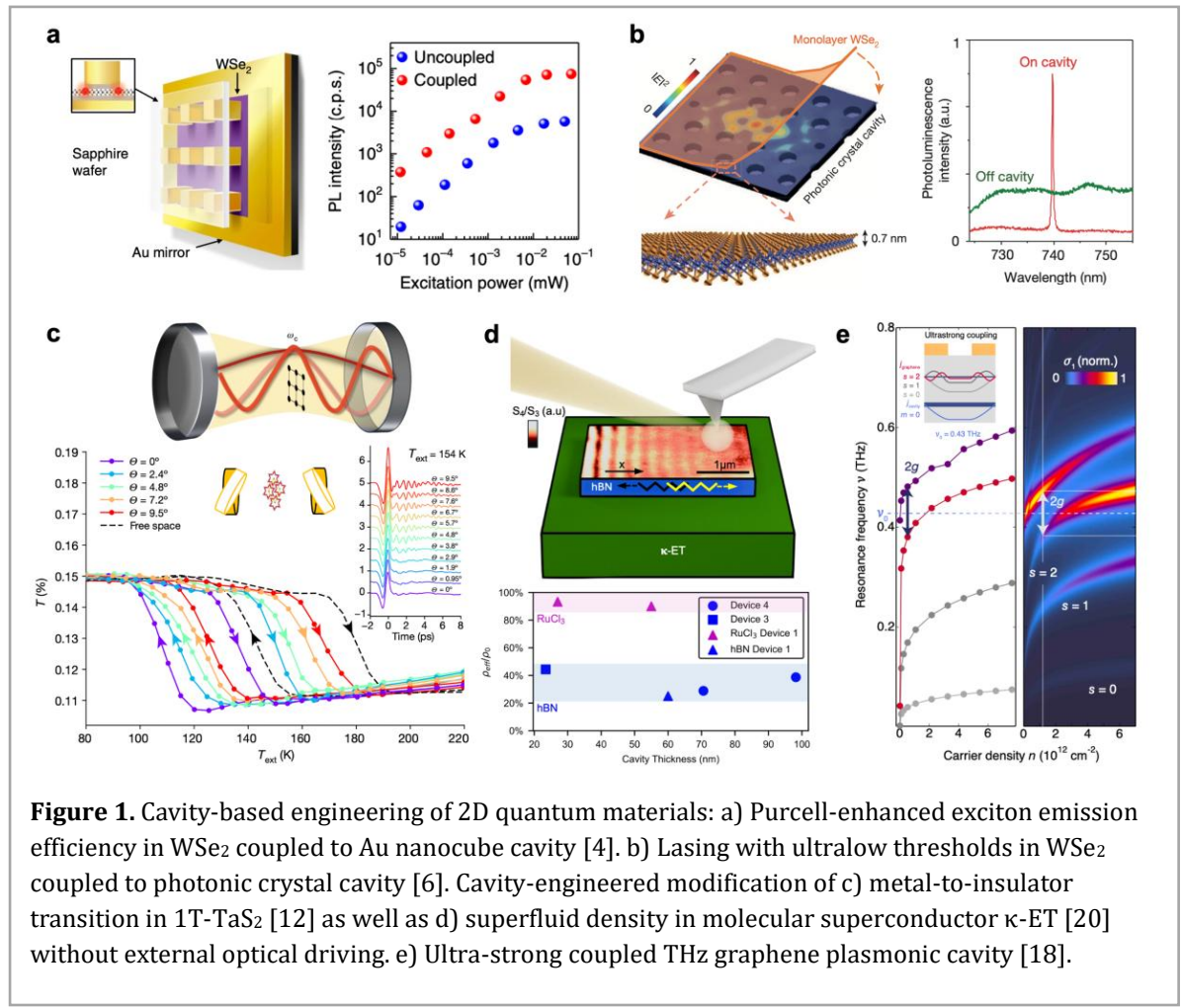

**Figure 1.** Cavity-based engineering of 2D quantum materials: a) Purcell-enhanced exciton emission efficiency in WSe$_2$ coupled to Au nanocube cavity [4]. b) Lasing with ultralow thresholds in WSe$_2$ coupled to photonic crystal cavity [6]. Cavity-engineered modification of c) metal-to-insulator transition in 1T-TaS$_2$ [12] as well as d) superfluid density in molecular superconductor κ-ET [20] without external optical driving. e) Ultra-strong coupled THz graphene plasmonic cavity [18].

## Current and future challenges

The strength of light-matter interaction can be parametrized by the coupling constant $g$ within the quantum electrodynamics framework. Depending on the magnitude of $g$ relative to the overall loss $\gamma$ of the system as well as the bare cavity frequency $\omega$, varying regimes of light-matter interaction can be achieved. This spans from the weak and strong coupling regimes, where $g$ competes with the system's energy loss rate, to the ultra-strong and deep-strong coupling regimes, where $g$ approaches or even exceeds the characteristic cavity frequency $\omega$. Light-matter coupling phenomena demonstrated in 2D materials have primarily focused on weak or strong coupling regimes thus far. These regimes modify the electromagnetic environment or optical density of states to alter the optical emission properties. An exciting frontier is to reach the ultra- or deep-strong coupling regime based on 2D materials, which has only recently been experimentally realized in other select solid-state systems [9]. The ability to access this latter regime would enable exciting nonlinear optical phenomena driven by dominant higher order as well as virtual excitation processes. This includes entangled photon pair generation and Dicke superradiant phase transition in quantum light source arrays, both of which are useful for quantum technologies.

An important open question is to what extent the electronic ground state of a 2D material can be modified simply by coupling to an optical cavity even without any external optical driving. In the ultra-strong coupling regime, quantum vacuum field fluctuations of the virtual cavity photons can become non-negligible and strongly interact with electronic and vibrational excitations in solid-state materials, fundamentally modifying the energy landscape of ionic and electronic degrees of freedom [10]. The use of vacuum field fluctuations in a dark cavity has recently gained much interest due to its untapped potential in directly modifying electron-electron interactions to induce phase transitions or alter the ground state of solid-state materials [11,12]. A rich variety of cavity-modified phases in equilibrium—including superconductivity, fractional quantum Hall effects, and ferroelectricity, among many others—may arise in the absence of external optical driving (Figure 1c) [13]. This dark cavity engineering presents an exciting new opportunity to explore emergent quantum phases and new ground states in 2D materials. Importantly, the approach can also overcome fundamental limitations of light-induced matter states in driven optical cavities, such as undesired materials degradation associated with the thermal effects as well as the short lifetime of the cavity-modified matter states.

**Advances in science and technology to meet challenges**

To unlock the full potential offered by cavity engineering of 2D materials, several technological and experimental gaps need to be addressed. A primary challenge is materials integration: we need scalable methods to fabricate high-finesse optical cavities embedded with 2D materials with atomically precise interfacial quality. Existing approaches typically focus on micron-scale exfoliated 2D materials assembled into optical cavities using transferable solid-state optical mirrors [14]. Such vdW assembly processes typically result in a high density of interfacial bubbles even when performed using the best available assembly protocols, including stacking in a glovebox filled with an inert gas such as argon. Considering the high sensitivity of light-matter coupling to the dielectric environment at the nanoscale, these interfacial imperfections are detrimental to producing highly reproducible and scalable light-matter coupled vdW heterostructures. Systematic efforts are required to isolate and remove sources of contamination and bubbles that typically arise during the vdW stacking processes. Recent studies demonstrate that polymer-free assembly of vdW heterostructures in ultrahigh vacuum (UHV) conditions can enable ultra-high-quality bubble-free interfaces in assembled vdW devices [15]. Further development of large-area transfer techniques based on UHV stacking may provide a viable route to scalable and ultra-high-quality 2D materials-integrated optical cavity systems.

A powerful alternative is to leverage cavities made of polaritons, light-matter hybrid modes with extreme light confinement at the nanoscale and ultrasmall mode volumes that are ubiquitous in 2D materials. These vdW polaritonic cavities can facilitate seamless vdW integration with other 2D material components, effectively overcoming fundamental challenges associated with integrating 2D materials into conventional bulky solid-state optical cavities and mirrors. It is also feasible to fabricate nanoscale vdW polaritonic cavities—such as graphene-based plasmonic cavities [16] and hyperbolic phonon polaritonic cavities in hexagonal boron nitride (hBN) [17]—that exhibit near-intrinsic quality factors, limited primarily by the fundamental materials loss channels. In this respect, a growing number of recent studies demonstrate the use of vdW polaritonic cavities to reach ultra-strong coupling [18] and modify matter states in proximal layered materials [19,20], including without external optical driving (Figure 1d,e). Therefore, we envision that further developments of these ultra-high quality vdW polaritonic cavities can provide a unique platform to explore cavity engineering of 2D materials across varying light-matter coupling regimes on demand.

### Concluding remarks

Cavity-based engineering of 2D quantum materials represents a rapidly evolving frontier at the intersection of quantum optics, condensed matter physics, and materials science. Theoretical predictions have uncovered exotic cavity-induced phenomena in solid-state materials, from vacuum-fluctuation-mediated ground state modifications to novel quantum phases, yet experimental realizations remain in their early stages. Closing this theory-experiment gap is a central challenge. A key priority is realizing scalable, high-fidelity integration of 2D materials with ultra-high-quality cavities, including emerging vdW polaritonic nanocavity platforms, and the ability to continuously and precisely tune light-matter coupling strength across multiple coupling regimes. By merging the exceptional tunability of vdW 2D materials with a vast array of possible cavity designs, the field of cavity-based 2D material engineering is poised to open new frontiers for exploring fundamental light-matter interactions and realizing novel quantum optical technologies.

### Acknowledgements

This work was supported by the University of Arizona's School of Mining Engineering and Mineral Resources; Vertically Integrated Projects Program; and Office of Research and Partnerships under the RII Invited Travel Grant and RII Core Facilities Pilot Program.

### References

[1] Vahala K J 2003 Optical microcavities *Nature* **424** 839

[2] Purcell E M 1946 Spontaneous emission probabilities at radio frequencies *Phys. Rev.* **69** 681

[3] Fryett T K, Seyler K L, Zheng J, Liu C-H, Xu X and Majumdar A 2017 Silicon photonic crystal cavity enhanced second-harmonic generation from monolayer $WSe_2$ *2D Mater.* **4** 015031

[4] Luo Y, Shepard G D, Ardelean J V, Rhodes D A, Kim B, Barmak K, Hone J C and Strauf S 2018 Deterministic coupling of site-controlled quantum emitters in monolayer $WSe_2$ to plasmonic nanocavities *Nature Nanotech.* **13** 1137

[5] Lyons T P, Gillard D J, Leblanc C, Puebla J, Solnyshkov D D, Klompmaker L, Akimov I A, Louca C, Muduli P, Genco A, Bayer M, Otani Y, Malpuech G and Tartakovskii A I 2022 Giant effective Zeeman splitting in a monolayer semiconductor realized by spin-selective strong light–matter coupling *Nature Photon.* **16** 632

[6] Wu S, Buckley S, Schaibley J R, Feng L, Yan J, Mandrus D G, Hatami F, Yao W, Vučković J, Majumdar A and Xu X 2015 Monolayer semiconductor nanocavity lasers with ultralow thresholds *Nature* **520** 69

[7] Oka T and Kitamura S 2019 Floquet engineering of quantum materials *Annu. Rev. Condens. Matter Phys.* **10** 387

[8] Zhou L, Liu B, Liu Y, Lu Y, Li Q, Xie X, Lydick N, Hao R, Liu C, Watanabe K, Taniguchi T, Chou Y-H, Forrest S R and Deng H 2024 Cavity Floquet engineering *Nature Commun.* **15** 7782

[9] Forn-Díaz P, Lamata L, Rico E, Kono J and Solano E 2019 Ultrastrong coupling regimes of light-matter interaction *Rev. Mod. Phys.* **91** 025005

[10] Lu I-T, Shin D, Svendsen M K, Latini S, Hübener H, Ruggenthaler M and Rubio A 2025 Cavity engineering of solid-state materials without external driving *Adv. Opt. Photon.* **17** 441

[11] Appugliese F, Enkner J, Paravicini-Bagliani G L, Beck M, Reichl C, Wegscheider W, Scalari G, Ciuti C and Faist J 2022 Breakdown of topological protection by cavity vacuum fields in the integer quantum Hall effect *Science* **375** 1030

[12] Jarc G, Mathengattil S Y, Montanaro A, Giusti F, Rigoni E M, Sergo R, Fassioli F, Winnerl S, Dal Zilio S, Mihailovic D, Prelovšek P, Eckstein M and Fausti D 2023 Cavity-mediated thermal control of metal-to-insulator transition in 1T-$TaS_2$ *Nature* **622** 487

[13] Schlawin F, Kennes D M and Sentef M A 2022 Cavity quantum materials *Appl. Phys. Rev.* **9** 011312

[14] Paik E Y, Zhang L, Hou S, Zhao H, Chou Y-H, Forrest S R and Deng H 2023 High quality factor microcavity for van der Waals semiconductor polaritons using a transferrable mirror *Adv. Opt. Mater.* **11** 2201440

[15] Wang W, Clark N, Hamer M, Carl A, Tovari E, Sullivan-Allsop S, Tillotson E, Gao Y, de Latour H, Selles F, Howarth J, Castanon E G, Zhou M, Bai H, Li X, Weston A, Watanabe K, Taniguchi T, Mattevi C, Bointon T H, Wiper P V, Strudwick A J, Ponomarenko L A, Kretinin A V, Haigh S J, Summerfield A and Gorbachev R 2023 Clean assembly of van der Waals heterostructures using silicon nitride membranes *Nature Electron.* **6** 981

[16] Kim B S Y, Sternbach A J, Choi M S, Sun Z, Ruta F L, Shao Y, McLeod A S, Xiong L, Dong Y, Chung T S, Rajendran A, Liu S, Nipane A, Chae S H, Zangiabadi A, Xu X, Millis A J, Schuck P J, Dean C R, Hone J and Basov D N 2023 Ambipolar charge-transfer graphene plasmonic cavities *Nature Mater.* **22** 838

[17] Herzig Sheinfux H, Orsini L, Jung M, Torre I, Ceccanti M, Marconi S, Maniyara R, Barcons Ruiz D, Hötger A, Bertini R, Castilla S, Hesp N C H, Janzen E, Holleitner A, Pruneri V, Edgar J H, Shvets G and Koppens F H L 2024 High-quality nanocavities through multimodal confinement of hyperbolic polaritons in hexagonal boron nitride *Nature Mater.* **23** 499

[18] Kipp G, Bretscher H M, Schulte B, Herrmann D, Kusyak K, Day M W, Kesavan S, Matsuyama T, Li X, Langner S M, Hagelstein J, Sturm F, Potts A M, Eckhardt C J, Huang Y, Watanabe K, Taniguchi T, Rubio A, Kennes D M, Sentef M A, Baudin E, Meier G, Michael M H and McIver J W 2025 Cavity electrodynamics of van der Waals heterostructures *Nature Phys.* **21** 1926

[19] Berkowitz M E, Kim B S Y, Ni G, McLeod A S, Lo C F B, Sun Z, Gu G, Watanabe K, Taniguchi T, Millis A J, Hone J C, Fogler M M, Averitt R D and Basov D N 2021 Hyperbolic Cooper-pair polaritons in planar graphene/cuprate plasmonic cavities *Nano Lett.* **21** 308

[20] Keren I, Webb T A, Zhang S, Xu J, Sun D, Kim B S Y, Shin D, Zhang S S, Zhang J, Pereira G, Yao J, Okugawa T, Michael M H, Edgar J H, Wolf S, Julian M, Prasankumar R P, Miyagawa K, Kanoda K, Gu G, Cothrine M, Mandrus D, Buzzi M, Cavalleri A, Dean C R, Kennes D M, Millis A J, Li Q, Sentef M A, Rubio A and Pasupathy A N 2025 arXiv:2505.17378

# 10. Superconducting qubits based on 2D materials

**Jesse Balgley[1], James Hone[1] and Kin Chung Fong[2,3,4]**

[1] Department of Mechanical Engineering, Columbia University, New York, NY 10027, USA
[2] Department of Electrical and Computer Engineering, Northeastern University, Boston, MA 02115, USA
[3] Department of Physics, Northeastern University, Boston, MA 02115, USA
[4] Quantum Materials and Sensing Institute, Burlington, MA 01803, USA

E-mail: k.fong@northeastern.edu

**Status**

Advances from materials research hold significant promise for enhancing both qubit coherence and operational functionality. Current superconducting qubits rely predominantly on amorphous aluminum and aluminum oxide, yet these materials impose intrinsic limitations. A key challenge is the narrow parameter space conventionally used materials offer in terms of critical temperature, superconducting gap, and barrier height, which constrains the operating frequencies and temperatures accessible to these qubits because of thermally activated quasiparticles. In addition, amorphous materials inherently host two-level systems (TLSs) that contribute to energy relaxation. To overcome these constraints, researchers have pursued alternative materials to replace conventional aluminum oxide, including and silicon [1] and crystalline epitaxial systems such as indium arsenide [2] and nitride compounds [3]. The relatively low critical temperature of aluminum ($T_c$ = 1.2 K) further restricts device operation, but recent efforts with higher-$T_c$ materials like niobium have begun to extend these limits [4]. However, exploration of new deposition-grown material platforms often entails extensive optimization of growth, oxidation, and substrate conditions, while epitaxial systems are typically constrained by strict lattice-matching requirements.

Van der Waals crystals differ fundamentally from epitaxial materials in their fabrication and assembly. These materials can be synthesized as bulk single crystals or thin films, and then mechanically assembled into designed heterostructures. Recent advances have significantly matured scalable growth, transfer techniques, and wafer-scale integration [5, 6, 7, 8, 9, 10, 11, 12]. The 2D materials platform offers unique advantages for quantum device applications. Because adjacent layers are held together by weak van der Waals forces rather than covalent bonds, atomically sharp interfaces can be formed without the lattice matching requirements of epitaxy. This property, combined with the intrinsic gate tunability of many 2D materials, enables exceptional control over quantum device properties. Recent experiments have begun to demonstrate this potential: transmon qubits incorporating van der Waals materials have successfully exhibited quantum coherence [13, 14, 15], drawing on the diverse library of available 2D superconductors, insulators, semiconductors, and semimetals. Given its versality, we envision 2D materials can expand the functionality of qubits and serve as prototypes when we explore variety of qubit architectures.

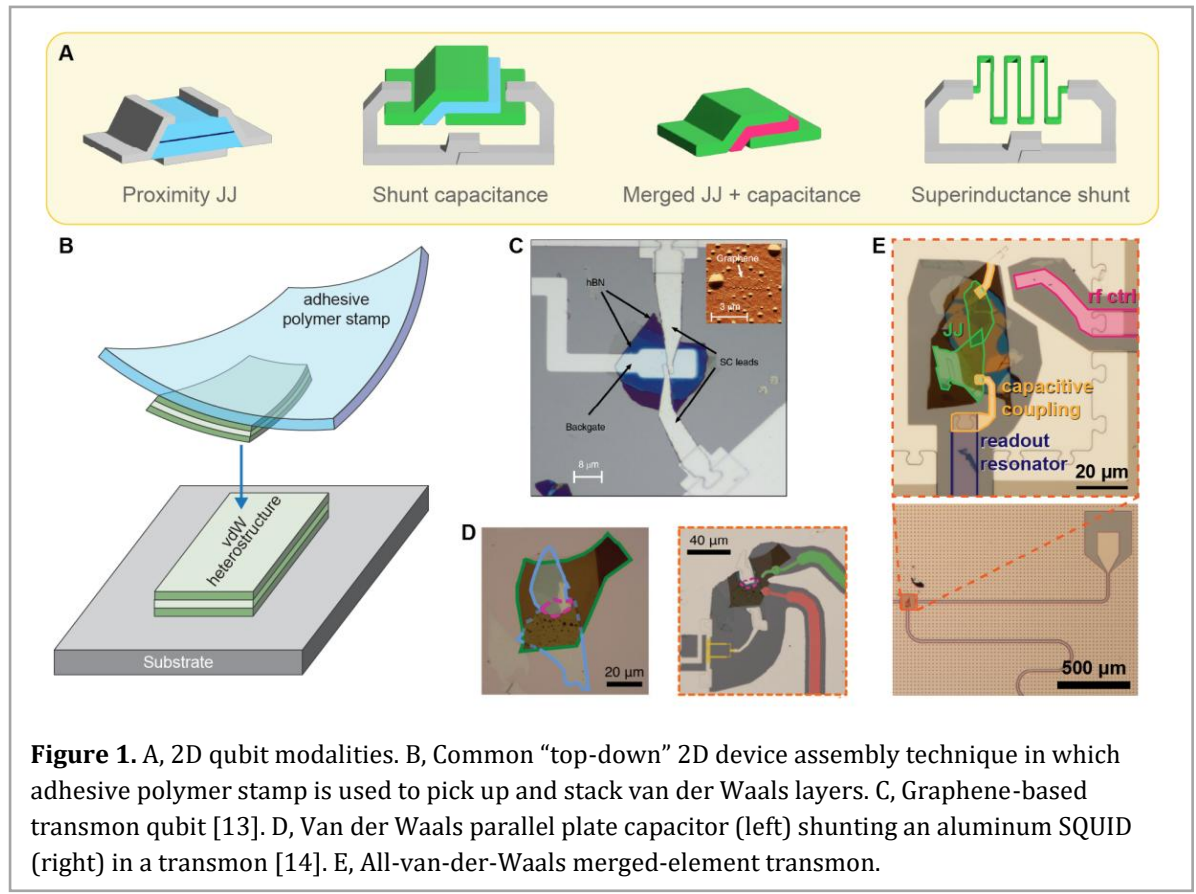

**Figure 1.** A, 2D qubit modalities. B, Common "top-down" 2D device assembly technique in which adhesive polymer stamp is used to pick up and stack van der Waals layers. C, Graphene-based transmon qubit [13]. D, Van der Waals parallel plate capacitor (left) shunting an aluminum SQUID (right) in a transmon [14]. E, All-van-der-Waals merged-element transmon.

To appreciate how 2D materials can enrich the functionalities of superconducting qubits, we consider various modalities, as depicted in Fig. 1A. The first is a lateral junction, where an atomically thin conductor serves as a weak link between two superconducting electrodes. Graphene, as the most widely studied example, can be encapsulated between layers of the van der Waals insulator hexagonal boron nitride (hBN) using a polymer stacking method (Fig. 1B). The stack is then etched to expose one-dimensional edges of the graphene, which are contacted via metal deposition. Proximitization of the graphene electrons allows supercurrent to flow across the channel, forming a Josephson junction (JJ). The 2D nature and small density of states of graphene permit continuous tuning of carrier density and conductivity via electrostatic gating using relatively small voltages. Similar behavior is expected in other semimetallic or semiconducting 2D weak links, making this hybridized qubit akin to the superconductor-semiconductor-superconductor (super-semi) JJ. Both approaches share the advantage of controllable JJ properties — namely, the critical current — through gate tuning [2].

Fig. 1C shows an aluminum-contacted graphene JJ from Ref. [13], incorporated into a transmon qubit architecture. A pre-deposited metal gate beneath the JJ tunes the conductivity of the graphene and thereby the Josephson energy and qubit frequency. This particular qubit showed frequency tunability over a 6 GHz range with relaxation time $T_1 \approx 36$ ns. Compared with the *in situ* control of qubit frequency from threading magnetic flux through a SQUID loop with a current-carrying wire, the voltage control in this modality has a more confined electromagnetic field profile and lower power consumption. This can enable smaller qubit footprints and minimize crosstalk as the number of qubits scales in future quantum processors.

The second modality preserves the high-quality aluminum-based JJ of a conventional transmon but replaces the shunt capacitance — typically realized with coplanar metal structures spanning hundreds of microns to millimeter scales — with a parallel plate capacitor (PPC) fabricated from van der Waals materials. The left panel of Fig. 1D shows a stacked vdW PPC, where multilayer flakes of the vdW superconductor $NbSe_2$ serve as the capacitor plates and a multilayer flake of hBN is the dielectric spacer. This structure provides a low-loss capacitance ranging from a few to hundreds of femtofarads, all within a compact footprint (typically $\sim 100 \ \mu m^2$). In the right panel, the vdW PPC is electrically connected in parallel with an aluminum-based SQUID (false-colored yellow) to form a compact, flux-tunable transmon qubit. Refs. [14] & [15] demonstrated relaxation and coherence times up to 25 μs using this architecture, approaching the competitive values seen in conventional transmons but with a 250 times smaller areal footprint.

The third modality employs a single vertical Josephson junction made from 2D materials to form a "merged-element" transmon (MET). This nomenclature reflects the fact that the two typically separate elements — the Josephson inductance and the large shunt capacitance — are merged into a single parallel-plate structure [16]. The top panel of Fig. 1E shows a vdW MET architecture, in which the vdW JJ (green) is capacitively coupled via aluminum (orange) to a readout resonator (dark blue) at one electrode and to ground at the other, while a nearby wire delivers qubit control pulses.

Early demonstrations of the MET used conventional aluminum-based JJs [17] or JJs with niobium electrodes and amorphous silicon tunnel barriers [16]. Van der Waals materials have been pursued as a platform for small-form-factor, low-loss JJs, with the potential for unprecedented uniformity and repeatability due to the atomically precise thicknesses of vdW tunnel barriers. These benefits can be further enhanced with semiconducting tunnel barriers, whose relatively small bandgaps allow for thicker weak links.

Lastly, 2D superconductors can be harnessed as "superinductors" in superconducting qubits, as well as in couplers or detectors. Because the kinetic inductance of a superconductor increases with reduced thickness and superfluid density [18], 2D superconductors can exhibit much larger inductances per unit length than thin films of granular aluminum or superconducting alloys [19, 20]. Consequently, 2D materials may provide compact, low-loss alternatives to sputtered nanowires or SQUID arrays in applications requiring large inductances, such as the superinductance shunts of fluxonium qubits (Fig. 1A).

In all modalities, utilizing 2D materials for superconducting qubit demands careful design, characterization, and control of the heterostructures. Early effort from Lee *et al.* characterized JJs consisting of flakes of the layered semiconductor $MoS_2$ stacked atop aluminum electrodes and capped with a deposited layer of aluminum to form the junction [21]. Placed in parallel with a large shunt capacitance, these JJs showed dispersive coupling to a microwave cavity. More recently, Balgley *et al.* realized an all-2D-material MET with $NbSe_2$ JJ electrodes and a $WSe_2$ semiconducting weak link [22]. The qubit was made flux-tunable via a subtractive etching process: the vdW JJ was first stacked as a single junction, then patterned lithographically and etched to define a SQUID loop of two JJs. Using a semiconducting weak link approximately 10 times thicker than traditional aluminum oxide barriers, the device achieved a qubit frequency around 5 GHz, within 10% agreement with predictions based on DC transport characterization of similar JJs. This correlation between DC transport and qubit performance establishes that new materials can be rapidly screened through transport measurements prior to the more demanding process of qubit fabrication and testing.

## Challenges and advances

While 2D materials have emerged as a promising platform for superconducting qubits, several challenges remain on the path toward broad implementation. The most pressing issues can be grouped into four categories: (1) material quality (intrinsic disorder within 2D crystals); (2) fabrication cleanliness (extrinsic disorder introduced during processing); (3) manufacturability (scalability of stacking, exfoliation, and growth techniques); (4) materials engineering (optimal material choices for contacts, band alignment, and microwave functionality). To illustrate the promises and challenges associated with 2D-materials-based qubits --- and to provide perspective on how this technology might evolve in a roadmap --- we make a qualitative comparison with conventional qubits in Fig. 2. This comparison should be regarded as an approximate guide rather than a definitive assessment.

**Material quality:** Although 2D materials — boasting single-crystallinity, atomically pristine interfaces, the absence of dangling bonds, and the ability to be encapsulated for protection against oxidation — appear advantageous for low-loss applications compared to polycrystalline materials grown by deposition techniques, all growth methods yield inevitably some degree of disorder. In polycrystalline films, grain boundaries and surfaces host dissipative two-level systems that induce qubit relaxation, whereas in 2D materials, defect formation during growth remains a challenge. Recent advances, however, have significantly reduced defect densities in both bulk and monolayer growth. For example, Liu *et al.* introduced a two-step flux synthesis method that dramatically lowered charged and isovalent defect densities in selenide and telluride transition metal dichalcogenides (TMDs) relative to conventional single-step flux or chemical vapor transport processes [11]. Likewise, the quality of sulfide TMDs has been markedly improved through a salt flux synthesis method [12].

**Fabrication cleanliness:** State-of-the-art vdW heterostructures are traditionally assembled using "top-down" stacking with adhesive polymer stamps (Fig. 1B). This approach enables deterministic control over flake ordering, orientation, and twist angle, as well as substrate placement. However, commonly used polymers like PDMS and PC can leave trace residues of silicon, carbon, and oxygen, even after solvent cleaning. These residues not only accumulate on stack surfaces, but can intercalate between flakes, creating "bubbles" that scattering carriers, impede tunneling, or alter capacitances in junctions.

While more complex cleaning procedures, such as contact AFM "brooming" or high-temperature anneals, can mitigate residues, they are not universally compatible with all vdW materials or are impractical for scalable device fabrication. The direct impact of such residues on microwave loss in 2D qubits has yet to be systematically quantified. To circumvent polymer contamination altogether, alternative transfer methods have recently been developed. Wang, *et al.* demonstrated a polymer-free technique based on ultrathin gold films deposited on flexible silicon nitride membranes, which naturally adhere to 2D layers for stacking [10]. Although this method requires more complex preparation and pre-transferred bottom flakes for deterministic placement, it has shown strong promise for producing large-area, high-quality, bubble-free heterostructures.

**Manufacturability:** Most vdW heterostructures are assembled via manual "top-down" stacking: exfoliating flakes from bulk crystals, selecting suitable candidates, and sequentially picking and placing them into heterostructures. While this method enables various device modalities described in this roadmap, it is labor-intensive, stochastic, and incompatible with large-scale production. To address this, researchers have developed robotic systems capable of scanning exfoliated yields for desirable flakes and autonomously stacking them into complex heterostructures [9]. While this can greatly improve the manufacturability of 2D devices, it has yet to be widely implemented.

Exfoliation itself has also advanced beyond traditional tape-based methods. New approaches can peel off entire monolayer surfaces from millimeter-scale parent crystals, even for materials that are otherwise difficult to thin down [8]. Concurrently, growth-based strategies are being refined to enable wafer-scale device arrays. Advances in chemical vapor deposition (CVD) have dramatically improved quality; for instance, Amontree *et al.* recently reported CVD-grown graphene monolayers with electronic properties matching exfoliated flakes [5]. Sophisticated CVD methods [7], as well as a new technique called hypotaxy [6], enable deterministic growth of TMDs with controlled thickness and geometry. Together, these advances suggest a clear path toward wafer-scale, reproducible, and manufacturable 2D qubit devices.

**Materials engineering:** The choice and engineering of materials at device interfaces play a critical role in qubit performance. In graphene-based superconducting devices, the choice of superconductor greatly influences the contact transparency, as work functions mismatches often produce p-n junctions at the contact interface which limit supercurrent flow across the junction. Various superconductors have been explored, including Ti, Ta, MoRe alloy, NbTiN, and aluminum [15, 23, 24, 25, 26, 27]. Recently, Jang *et al.* demonstrated that dual-gating architectures can produce highly transparent boundaries between superconductors and graphene [28].

In vertical vdW JJs, the work function of the superconducting electrodes is also pivotal. Balgley, *et al.* demonstrated that the overlap of the work function of the 2D superconductor $NbSe_2$ with the valence band of the 2D semiconductor $WSe_2$ results in proximitization of approximately 3 atomic layers of $WSe_2$ at the interface, enabling a crossover from proximity- to tunneling-type junction behavior as weak link thickness is varied [22]. By replacing $WSe_2$ with $MoS_2$, the band alignment was altered, leading to systematic changes in tunneling conductance, critical current, and the extent of proximitization in the weak link.

Finally, new techniques are needed to characterize 2D materials at microwave frequencies. A promising method employs high-$Q$ microwave resonators to extract the complex impedance of small 2D samples loaded at the resonator end [13, 14, 18, 19, 20, 29, 30]. This method enables precise measurements of reactive components such as kinetic inductance — relevant for qubits like the fluxonium (Fig. 1A) — and can also provide estimates of dielectric loss tangents. In this way, resonator-based measurements offer an efficient means to screen new materials for their applicability in 2D qubit technologies.

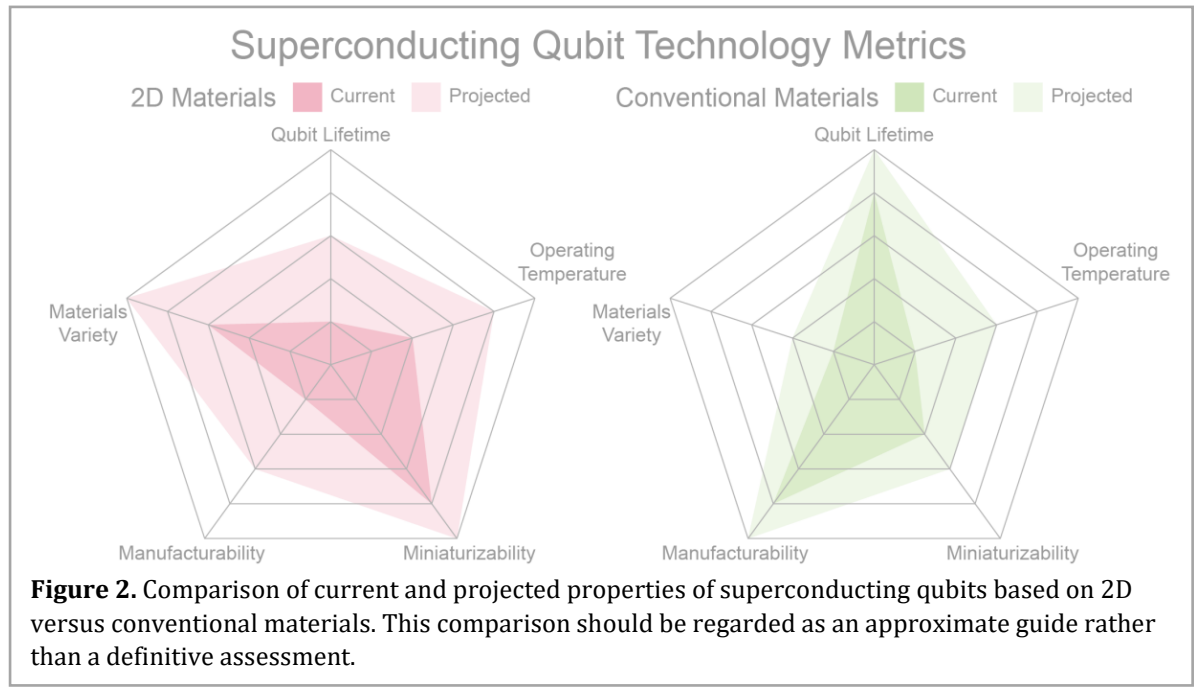

**Figure 2.** Comparison of current and projected properties of superconducting qubits based on 2D versus conventional materials. This comparison should be regarded as an approximate guide rather than a definitive assessment.

## Concluding remarks

In summary, we compare 2D-material-based qubits and conventional qubits across five metrics in Fig. 2: qubit lifetime, operating temperature, miniaturizability, manufacturability, and materials variety. State-of-the-art conventional qubits excel in achieving long lifetimes and in their readiness for wafer-scale manufacturing. However, their operating temperatures and energy scales are constrained by the limited palette of available materials. Moreover, reliance on amorphous materials has necessitated large device footprints, hindering miniaturization and, ultimately, scalability. In contrast, 2D-materials-based qubits — though currently limited by moderate lifetimes and immature manufacturability — offer a broad library of materials that expand the operable design space, support new functionalities, and hold potential for miniaturization. As the 2D-material-based and conventional qubits excel in different attributes, they can complement one another and may be integrated in the near term to harness the strengths of each material platform. In particular, the ease with which 2D materials can be stacked in diverse combinations of superconducting, conducting, and insulating layers makes them a versatile testbed for exploring new physics and building understanding that can be translated to conventional qubit architectures. At the same time, continued development of 2D-material qubits may open pathways to entirely new device concepts and next-generation superconducting qubits. In the longer term, sustained efforts to characterize and implement 2D materials into superconducting qubits will enhance the versatility of superconducting quantum technologies and facilitate their adoption as a competitive qubit platform.

## Acknowledgements

K.C.F. acknowledges support from the Army Research Office under Cooperative Agreement W911NF-24-1-0163. J.B. acknowledges support from the Army Research Office under Grant Number W911NF-24-1-0133. J.H. acknowledges support from the Army Research office under Contract W911NF-22-C-0021. The views and conclusions contained in this document are those of the authors and should not be interpreted as representing the official policies, either expressed or implied, of the Army Research Office or the U.S. Government. The U.S. Government is authorized to reproduce and distribute reprints for Government purposes notwithstanding any copyright notation herein.

**References**

[1] A. Goswami, A. P. McFadden, T. Zhao, H. Inbar, J. T. Dong, R. Zhao, M. C. R. H, R. W. Simmonds, C. J. Palmstrøm and D. P. Pappas, "Towards merged-element transmons using silicon fins: The FinMET," *Applied Physics Letters,* vol. 121, p. 064001, 2022.

[2] J. Shabani, M. Kjaergaard, H. J. Suominen, Y. Kim, F. Nichele, K. Pakrouski, T. Stankevic, R. M. Lutchyn, P. Krogstrup, R. Feidenhans'l, S. Kraemer, C. Nayak, M. Troyer, C. M. Marcus and C. J. Palmstrøm, "Two-dimensional epitaxial superconductor-semiconductor heterostructures: A platform for topological superconducting networks," *Phys Rev B,* vol. 93, p. 155402, 2016.

[3] C. J. K. Richardson, A. Alexander, C. G. Weddle, B. Arey and M. Olszta, "Low-loss superconducting titanium nitride grown using plasma-assisted molecular beam epitaxy," *J Appl Phys,* vol. 127, 2020.

[4] A. Anferov, S. P. Harvey, F. Wan, J. Simon and D. I. Schuster, "Superconducting Qubits above 20 GHz Operating over 200 mK," *PRX Quantum,* vol. 5, p. 030347, 2024.

[5] J. Amontree, X. Yan, C. S. DiMarco, P. L. Levesque, T. Adel, J. Pack, M. Holbrook, C. Cupo, Z. Wang, D. Sun, A. J. Biacchi, C. E. Wilson-Stokes, K. Watanabe, T. Taniguchi, C. R. Dean, A. R. Hight Walker, K. Barmak, R. Martel and J. Hone, "Reproducible graphene synthesis by oxygen-free chemical vapour deposition," *Nature,* vol. 630, p. 636–642, 2024.

[6] D. Moon, W. Lee, C. Lim, J. Kim, J. Kim, Y. Jung, H. Y. Choi, W. S. Choi, H. Kim, J. H. Baek, C. Kim, J. Joo, H. G. Oh, H. Jang, K. Watanabe, T. Taniguchi, S. Bae, J. Son, H. Ryu, J. Kwon, H. Cheong, J. W. Han, H. Jang and G. H. Lee, "Hypotaxy of wafer-scale single-crystal transition metal dichalcogenides," *Nature,* vol. 638, 2025.

[7] Z. Zhou, F. Hou, X. Huang, G. Wang, Z. Fu, W. Liu, G. Yuan, X. Xi, J. Xu, J. Lin and L. Gao, "Stack growth of wafer-scale van der Waals superconductor heterostructures," *Nature,* 2023.

[8] F. Liu, W. Wu, Y. Bai, S. H. Chae, Q. Li, J. Wang, J. Hone and X.-Y. Zhu, "Disassembling 2D van der Waals crystals into macroscopic monolayers and reassembling into artificial lattices," *Science (1979),* vol. 367, p. 903–906, 2020.

[9] S. Masubuchi, M. Morimoto, S. Morikawa, M. Onodera, Y. Asakawa, K. Watanabe, T. Taniguchi and T. Machida, "Autonomous robotic searching and assembly of two-dimensional crystals to build van der Waals superlattices," *Nat Commun,* vol. 9, p. 1413, 2018.

[10] W. Wang, N. Clark, M. Hamer, A. Carl, E. Tovari, S. Sullivan-Allsop, E. Tillotson, Y. Gao, H. de Latour, F. Selles, J. Howarth, E. G. Castanon, M. Zhou, H. Bai, X. Li, A. Weston, K. Watanabe, T. Taniguchi, C. Mattevi, T. H. Bointon, P. V. Wiper, A. J. Strudwick, L. A. Ponomarenko, A. V. Kretinin, S. J. Haigh, A. Summerfield and R. Gorbachev, "Clean assembly of van der Waals heterostructures using silicon nitride membranes," *Nat Electron,* 2023.

[11] S. Liu, Y. Liu, L. Holtzman, B. Li, M. Holbrook, J. Pack, T. Taniguchi, K. Watanabe, C. R. Dean, A. N. Pasupathy, K. Barmak, D. A. Rhodes and J. Hone, "Two-Step Flux Synthesis of Ultrapure Transition-Metal Dichalcogenides," *ACS Nano,* vol. 17, p. 16587–16596, 2023.

[12] F. A. Cevallos, S. Guo, H. Heo, G. Scuri, Y. Zhou, J. Sung, T. Taniguchi, K. Watanabe, P. Kim, H. Park and R. J. Cava, "Liquid Salt Transport Growth of Single Crystals of the Layered Dichalcogenides MoS 2 and WS 2," *Cryst Growth Des,* vol. 19, p. 5762–5767, 2019.

[13] J. I. J. Wang, D. Rodan-Legrain, L. Bretheau, D. L. Campbell, B. Kannan, D. Kim, M. Kjaergaard, P. Krantz, G. O. Samach, F. Yan, J. L. Yoder, K. Watanabe, T. Taniguchi, T. P. Orlando, S. Gustavsson, P. Jarillo-Herrero and W. D. Oliver,

"Coherent control of a hybrid superconducting circuit made with graphene-based van der Waals heterostructures," *Nat Nanotechnol,* vol. 14, p. 120–125, 2019.

[14] A. Antony, M. V. Gustafsson, G. J. Ribeill, M. Ware, A. Rajendran, L. C. G. Govia, T. A. Ohki, T. Taniguchi, K. Watanabe, J. Hone and K. C. Fong, "Miniaturizing Transmon Qubits Using van der Waals Materials," *Nano Lett,* vol. 21, p. 10122–10126, 2021.

[15] J. I. J. Wang, M. A. Yamoah, Q. Li, A. H. Karamlou, T. Dinh, B. Kannan, J. Braumüller, D. Kim, A. J. Melville, S. E. Muschinske, B. M. Niedzielski, K. Serniak, Y. Sung, R. Winik, J. L. Yoder, M. E. Schwartz, K. Watanabe, T. Taniguchi, T. P. Orlando, S. Gustavsson, P. Jarillo-Herrero and W. D. Oliver, "Hexagonal boron nitride as a low-loss dielectric for superconducting quantum circuits and qubits," *Nat Mater,* vol. 21, p. 398–403, 2022.

[16] R. Zhao, S. Park, T. Zhao, M. Bal, C. R. H. McRae, J. Long and D. P. Pappas, "Merged-Element Transmon," *Phys Rev Appl,* vol. 14, p. 064006, 2020.

[17] H. J. Mamin, E. Huang, S. Carnevale, C. T. Rettner, N. Arellano, M. H. Sherwood, C. Kurter, B. Trimm, M. Sandberg, R. M. Shelby, M. A. Mueed, B. A. Madon, A. Pushp, M. Steffen and D. Rugar, "Merged-Element Transmons: Design and Qubit Performance," *Phys Rev Appl,* vol. 16, p. 024023, 2021.

[18] M. Kreidel, X. Chu, J. Balgley, A. Antony, N. Verma, J. Ingham, L. Ranzani, R. Queiroz, R. M. Westervelt, J. Hone and K. C. Fong, "Measuring kinetic inductance and superfluid stiffness of two-dimensional superconductors using high-quality transmission-line resonators," *Phys Rev Res,* vol. 6, p. 043245, 2024.

[19] M. Tanaka, J. Î. Wang, T. H. Dinh, D. Rodan-Legrain, S. Zaman, M. Hays, A. Almanakly, B. Kannan, D. K. Kim, B. M. Niedzielski, K. Serniak, M. E. Schwartz, K. Watanabe, T. Taniguchi, T. P. Orlando, S. Gustavsson, J. A. Grover, P. Jarillo-Herrero and W. D. Oliver, "Superfluid stiffness of magic-angle twisted bilayer graphene," *Nature,* vol. 638, p. 99–105, 2025.

[20] A. Banerjee, Z. Hao, M. Kreidel, P. Ledwith, I. Phinney, J. M. Park, A. Zimmerman, M. E. Wesson, K. Watanabe, T. Taniguchi, R. M. Westervelt, A. Yacoby, P. Jarillo-Herrero, P. A. Volkov, A. Vishwanath, K. C. Fong and P. Kim, "Superfluid stiffness of twisted trilayer graphene superconductors," *Nature,* vol. 638, p. 93–98, 2025.

[21] K.-H. Lee, S. Chakram, S. E. Kim, F. Mujid, A. Ray, H. Gao, C. Park, Y. Zhong, D. A. Muller, D. I. Schuster and J. Park, "Two-Dimensional Material Tunnel Barrier for Josephson Junctions and Superconducting Qubits," *Nano Lett,* vol. 19, p. 8287–8293, 2019.

[22] J. Balgley, J. Park, X. Chu, E. G. Arnault, M. V. Gustafsson, L. Ranzani, M. Holbrook, K. Watanabe, T. Taniguchi, V. Perebeinos, J. Hone and K. C. Fong, "Crystalline superconductor-semiconductor Josephson junctions for compact superconducting qubits," *Phys Rev Appl,* vol. XX, p. 1–10, 2025.

[23] V. E. Calado, S. Goswami, G. Nanda, M. Diez, A. R. Akhmerov, K. Watanabe, T. Taniguchi, T. M. Klapwijk and L. M. K. Vandersypen, "Ballistic Josephson junctions in edge-contacted graphene," *Nat Nanotechnol,* vol. 10, p. 761–764, 2015.

[24] F. Amet, C. T. Ke, I. V. Borzenets, J. Wang, K. Watanabe, T. Taniguchi, R. S. Deacon, M. Yamamoto, Y. Bomze, S. Tarucha and G. Finkelstein, "Supercurrent in the quantum Hall regime," *Science (1979),* vol. 352, p. 966–969, 2016.

[25] F. E. Schmidt, M. D. Jenkins, K. Watanabe, T. Taniguchi and G. A. Steele, "A ballistic graphene superconducting microwave circuit," *Nat Commun,* vol. 9, p. 1–7, 2018.

[26] J. Sarkar, K. V. Salunkhe, S. Mandal, S. Ghatak, A. H. Marchawala, I. Das, K. Watanabe, T. Taniguchi, R. Vijay and M. M. Deshmukh, "Quantum-noise-limited microwave amplification using a graphene Josephson junction," *Nature Nanotechnology,* vol. 17, pp. 1147--1152, 2022.

[27]  G. Butseraen, A. Ranadive, N. Aparicio, K. Rafsanjani Amin, A. Juyal, M. Esposito, K. Watanabe, T. Taniguchi, N. Roch, F. Lefloch and J. Renard, "A gate-tunable graphene Josephson parametric amplifier," *Nature Nanotechnology,* vol. 17, pp. 1153--1158, 2022.

[28]  S. Jang, G.-H. Park, S. Park, H. Jeong, K. Watanabe, T. Taniguchi and G. Lee, "Engineering Superconducting Contacts Transparent to a Bipolar Graphene," *Nano Lett,* vol. 24, p. 15582–15587, 2024.

[29]  A. Antony, M. V. Gustafsson, A. Rajendran, A. Benyamini, G. Ribeill, T. A. Ohki, J. Hone and K. C. Fong, "Making high-quality quantum microwave devices with van der Waals superconductors," *Journal of Physics: Condensed Matter,* vol. 34, p. 103001, 2022.

[30]  X. Chu, J. Park, J. Balgley, S. Clemons, T. S. Chung, K. Watanabe, T. Taniguchi, L. Ranzani, M. V. Gustafsson, K. C. Fong and J. Hone, "Measuring Reactive-Load Impedance with Transmission-Line Resonators Beyond the Perturbative Limit," *arXiv,* p. 2511.14621, 2025.

# 11. Graphene-based single-photon detector

**Kin Chung Fong**[1,2,3]

[1] Department of Electrical and Computer Engineering, Northeastern University, Boston, MA 02115, USA
[2] Department of Physics, Northeastern University, Boston, MA 02115, USA
[3] Quantum Materials and Sensing Institute, Burlington, MA 01803, USA

E-mail: k.fong@northeastern.edu

**Status**

Graphene-based single-photon detectors (SPDs) [1–4] represent a recent addition to the SPD family. With the recent rise of quantum sensing, computation, and communication, it stands out for two unique advantages: its extremely low dark count, and its wide spectral coverage, from visible photons to low-energy microwave photons, spanning more than five orders of magnitude in wavelength. With its extremely low dark count, graphene-based SPDs could accelerate the search for axionic dark matter from decades-long averaging to merely a few days [5]. Detecting single far-infrared or microwave photons would enable next-generation astrophysics missions [6] such as the *Habitable Worlds Observatory* and allow for the remote qubit entanglement, enabling scalable quantum networks based on the superconducting qubit systems.

Graphene-based SPDs operate on a fundamentally different mechanism: sensing the thermal energy from an incident photon, rather than exciting carriers across an energy gap as in conventional paradigms. The possibility of detecting a single photon by graphene was first speculated in [4,7,8] (Fig. 1A). However, researchers quickly encountered a major challenge—resolving the minute rise in electron temperature from a single photon within an unexpectedly short, sub-µs timescale [9], due to disorder-enhanced electron–phonon coupling [10]. A breakthrough came with the proposal, building on earlier concepts [11] to incorporate graphene into a Josephson junction [1,12,13] (Fig. 1B). In this hybrid device, graphene serves simultaneously as the photon coupler and weak link in the Josephson junction. When the incident photon heats up the electrons in graphene, the Josephson junction can register the fleeting rise of electron temperature rise due to the suppression of its critical current. Since then, substantial progress has followed. For higher-energy photons, a graphene-based Josephson junction has shown the detection of single near-infrared photons via noise generated from the broken Cooper-pairs [2]. More recently, a proof-of-concept experiment demonstrated, unambiguously for the first time, the detection of heat from a single near-infrared photon by the massless Dirac fermions in graphene (Fig. 1D and E), achieving 87% (75%) quantum efficiency with 1 ($10^{-6}$) Hz dark count rate [3]. Detecting a lower-energy photon at microwave frequencies is more challenging and remains as one of the most exciting goals in the scientific community. Nevertheless, graphene–based bolometers have already achieved a noise equivalent power on the order of $10^{-19}$ W/Hz$^{1/2}$, reaching the fundamental limit imposed by thermodynamic fluctuation [12,14]. Overall, these state-of-the-art results are very encouraging, particularly since the devices used still have ample room for further improvement. These advances justify the growing confidence that graphene-based SPDs is emerging as a distinctive technology, complementary to other approaches based on the internal states of Josephson junctions, dissipative engineering, or breaking of Cooper-pair breaking mechanism [6,15].

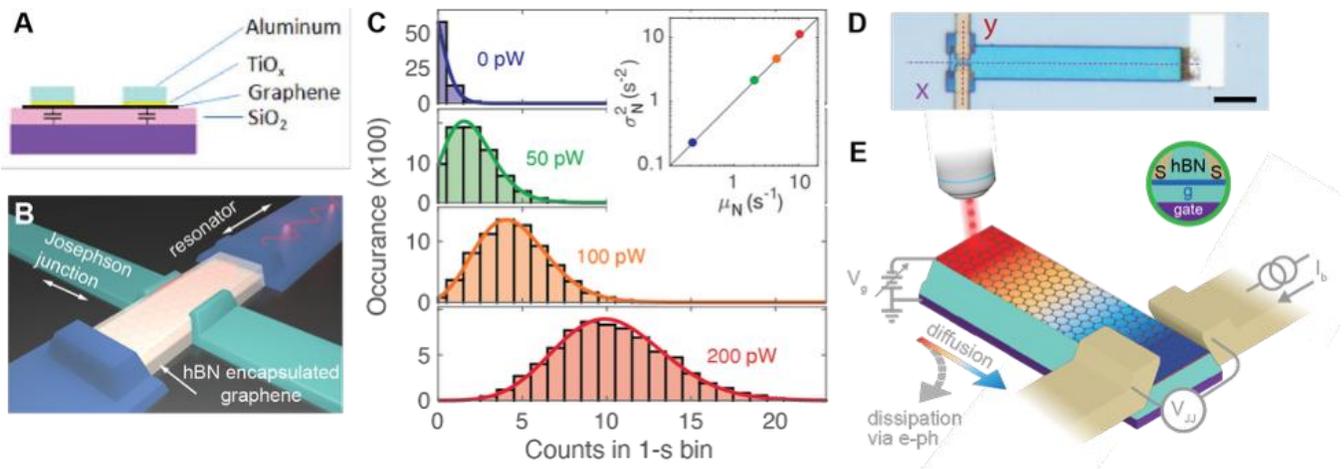

**Figure 1. A**. Initial concept of a graphene bolometer that can detect a single photon through a superconducting tunnel barrier [4]. **B**. Schematic of cross structure that can include both the input coupling through a resonator and Josephson junction [12]. **C**. Counting of the switching of Josephson junction with increasing laser power reveals the statistics of the coherent state of the photon source [2]. **D**. Optical micrograph of device showing a long graphene absorber with a Josephson junction fabricated on one end. **E**. Upon absorption, the heat from a single photon in graphene electron propagates through the sample and can be registered by the Josephson junction at one end of the sample [3]. All images reproduced with permissions.

The uniqueness of graphene-based SPDs thus originates from their orthogonal detection mechanism [1,9]. To appreciate how graphene-based SPDs uniquely situates within the broader landscape of quantum sensing, we can generally dissect the photon measurement process into three fundamental processes: absorption, magnification, and readout (see Fig. 2A). This framework also allows us to pinpoint where progress will be needed on the roadmap.

*Absorption* is the first step in which a photon interacts with the detector. Because the total detection efficiency is the product of absorption efficiency and intrinsic quantum efficiency, achieving high absorption is a prerequisite for high overall performance. Fortunately, the interaction of light with graphene is well understood through its AC conductivity, thanks to many significant contributions that have established a solid foundation for coupling strategies.

For higher-energy photons, e.g. infrared and visible light, graphene electrons absorb the photons by interband excitation, whereas for lower-energy photons, e.g. millimeter wave or microwave photons, intraband excitation. Since the coupling efficiency of graphene is about 2% when infrared photons are incident from the normal direction, a broadband resonator can substantially boost up the absorption efficiency of graphene-based SPDs [1]. After initial absorption, the photon energy will be distributed to the graphene electrons through e-e interactions. The fast e-e interaction in graphene is crucial to thermalize the hot electrons, and retain a sizable (~90%) portion of photon energy against the loss through optical phonons [10]. For microwave photons, the method of impedance matching is very effective because the graphene resistance is tunable by a gate voltage. Experiments have demonstrated an absorption efficiency of 99% in a transmission-line design [12]. The gapless band-structure enables graphene a wide spectrum SPD. Deploying to a specific system will require some designs of lenses, lenslets, waveguides, or antennas to guide the photon to the SPD to achieve high efficiency.

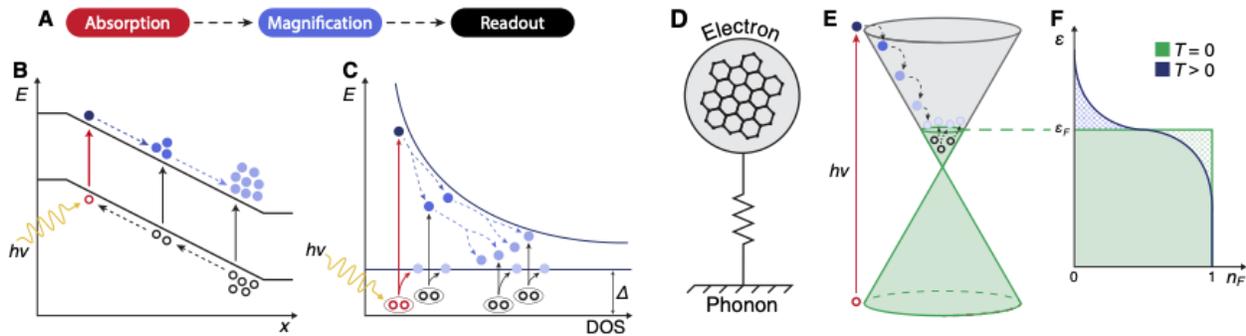

**Figure 2. A.** Three internal processes of single-photon detectors. **B.** In avalanche photodiodes, an input photon excites an electron-hole pair across the semiconductor bandgap. As the charge carriers accelerate by the biased electric field, impact ionization multiplies the carriers to generate a sizable readout current. **C.** A large class of SPD relies on the breaking of Cooper-pair in superconductor. Initial quasiparticles from the broken Cooper-pair are highly energetic and continue to break more. Large amount of quasiparticles right above the superconducting gap modifies the impedance, producing a readout register. **D.** Thermal diagram of the graphene-based SPD. Graphene electron functions as the bolometric material with a heat leakage through its phonon if the electron diffusion and radiation channels are quenched. **E.** The massless Dirac fermion in graphene has a minute specific heat. The internal energy of a single-photon couples into graphene by inter-(this figure) and intra-band excitation for optical and microwave frequencies, respectively [10]. **F.** Graphene electrons thermalize through strong electron-electron interaction with an elevated temperature.

*Magnification* determines how the absorbed photon energy is converted into a measurable signal. For instance, in avalanche photodiodes (APDs), an absorbed photon generates an electron–hole pair, which is then amplified through impact ionizations under a high voltage bias. In superconducting SPDs, photon absorption breaks a Cooper pair; the resulting quasiparticles undergo further pair-breaking until the photon energy is carried among the generated quasiparticles, which settle temporarily near the edge of the superconducting gap before recombination. This down-conversion process is typically ~60% efficient due to competition with phonon emission.

For graphene, the linear band structure of two-dimensional massless Dirac fermions leads to a heat capacity linearly proportional to its Fermi energy measured from the charge neutrality point [16]. In practice, the smallest achievable density of states, and hence heat capacity, is limited by inhomogeneity due to doping from impurities. At dilution refrigeration temperatures, e.g. 10 mK, 1 $\mu m^2$ graphene has only about 1 $k_B$, of heat capacity. This is to be compared with an estimate of >$10^5$ $k_B$ in transition-edge sensors, another SPD that employs thermal phenomena. The large temperature rise from even one single photon creates an unprecedented mechanism to detect a single photon. The thermal energy will eventually dissipate through the graphene phonons, after the diffusive and radiative thermal conductance channels are quenched. The thermal time constant is proportional to the thermal resistance through the electron-phonon coupling channel in the linear response regime [10].

In addition to confining photon energy within graphene electrons, the weak electron–phonon coupling also plays a crucial role in suppressing the background noise. Inevitable high-energy cosmic rays can generate athermal phonons through nuclear recoils in the substrate, producing errors in superconducting qubits and detectors [15]. For SPDs, such false positives appear as background dark counts, triggered when these phonons break Cooper pairs in the superconducting materials used for photon absorption. Graphene-based SPDs, however, are inherently immune to this problem: their exceptionally weak electron–phonon coupling prevents substrate phonons from exciting electrons, thus evading the otherwise inevitable dark counts induced by cosmic rays. This property is best illustrated in the recent experiment in which the measured dark count is exponentially suppressed [3]. As a trade-off, such detectors are also insensitive to phonon-mediated signals of interest, e.g. superlight dark matter or neutrinos [15]. For axionic dark matter searches and rare-event experiments where the dark count dominantly sets the sensitivity, graphene-based SPDs have a decisive advantage to fill the void of existing quantum sensors in 10-100 GHz, important especially in the search of dark matter [5,17].

**Current and Future Challenges**

While experiments [2,3,12,14] including the recent moiré graphene [18,19], have shown great promise, there are many challenges ahead before graphene-based SPD can detect low-energy microwave photons. These challenges include better control of electron-phonon coupling, stronger temperature response, lower-noise readout, and scalable fabrication.

Due to the low density of states, symmetry, and high acoustic phonon velocity, graphene electrons couple weakly to the remaining acoustic phonons at low temperatures. However, impurities can significantly expand the scattering-phase space, leading to a much larger electron-phonon coupling. The high circumference-to-area-ratio from a smaller piece of graphene, often required in SPD, aggravates the problem because the atomic imperfection along the edges can host the resonant scatterers [20]. More systematic studies of electron-phonon coupling dependence on the etching and graphene cutting processes by laser or atomic force microscopy could calibrate, improve, and control this thermal relaxation channel.

Minimizing heat leakage through phonons while measuring the electron temperature on the short thermal timescale remains a central challenge. Early proposals focused on Johnson-noise radiometry [7], but this approach loses sensitivity when the photon heats the graphene. A more promising solution has emerged by incorporating a Josephson junction: when graphene electrons form the weak link of the junction [1,11], the occupation of Andreev bound states becomes thermally dependent on the electronic temperature. In this proximity-junction configuration, the device effectively behaves as a hybrid quantum sensor, combining graphene electrons as the thermal absorber with the macroscopic superconducting phase of the Josephson junction. A deeper understanding of Andreev bound states—particularly in regimes with enhanced quasiparticle populations—will not only improve the temperature dependence of the Josephson critical current (and thus the impedance readout) but also align with the broader needs of qubit engineering. As an alternative, exploiting the intrinsic temperature dependence of graphene's resistance remains a viable pathway for readout [19].

A good readout design can suppress the noise, hence improving the signal-to-noise ratio that ultimately determines the SPD performances and utilities. The readout process relies on the temperature-dependent impedance, that could be measured by simply DC electrical transports or a resonator, used especially when the impedance response to a photon is reactive [11,13]. The lowest noise limit can be bounded by the amplifiers, extrinsic to the SPD mechanism, or by the intrinsic mechanism, i.e. the quasiparticle generation and recombination in kinetic inductance detector and the temperature fluctuation of graphene electron as a canonical ensemble [1,12]. Modeling and design of the readout can optimize the signal-to-noise ratio in the near future for the proof-of-principle experiment of the graphene-based SPD.

**Advances in science and technology to meet challenges**

Scalable fabrication of quantum devices on 2D material platforms could be challenging. Fortunately, high-quality wafer scale graphene is now available. Their mobility approaches 105 cm2/Vs at room temperatures, reaching a doping level that is clean enough for detecting near and mid-infrared photons. With the demonstration of the fabrication of arrays of graphene-based Josephson junctions [21], we expect the frequency-multiplexed readout and array of graphene-based SPD as quantum imager are achievable in short (1-2 year time scale) and mid-(3-5 years) term, respectively. Single-photon imaging arrays will have many applications in astronomy, e.g. exoplanet exploration, as well as quantum communication, especially with photon-number resolving capability which is also achievable in short term. For the long term 5-10 years time scale, we should be able to operate the graphene-based SPD in higher temperatures with CMOS-integration [22] with the exciting progress made continuously on the growth of 2D materials [23].

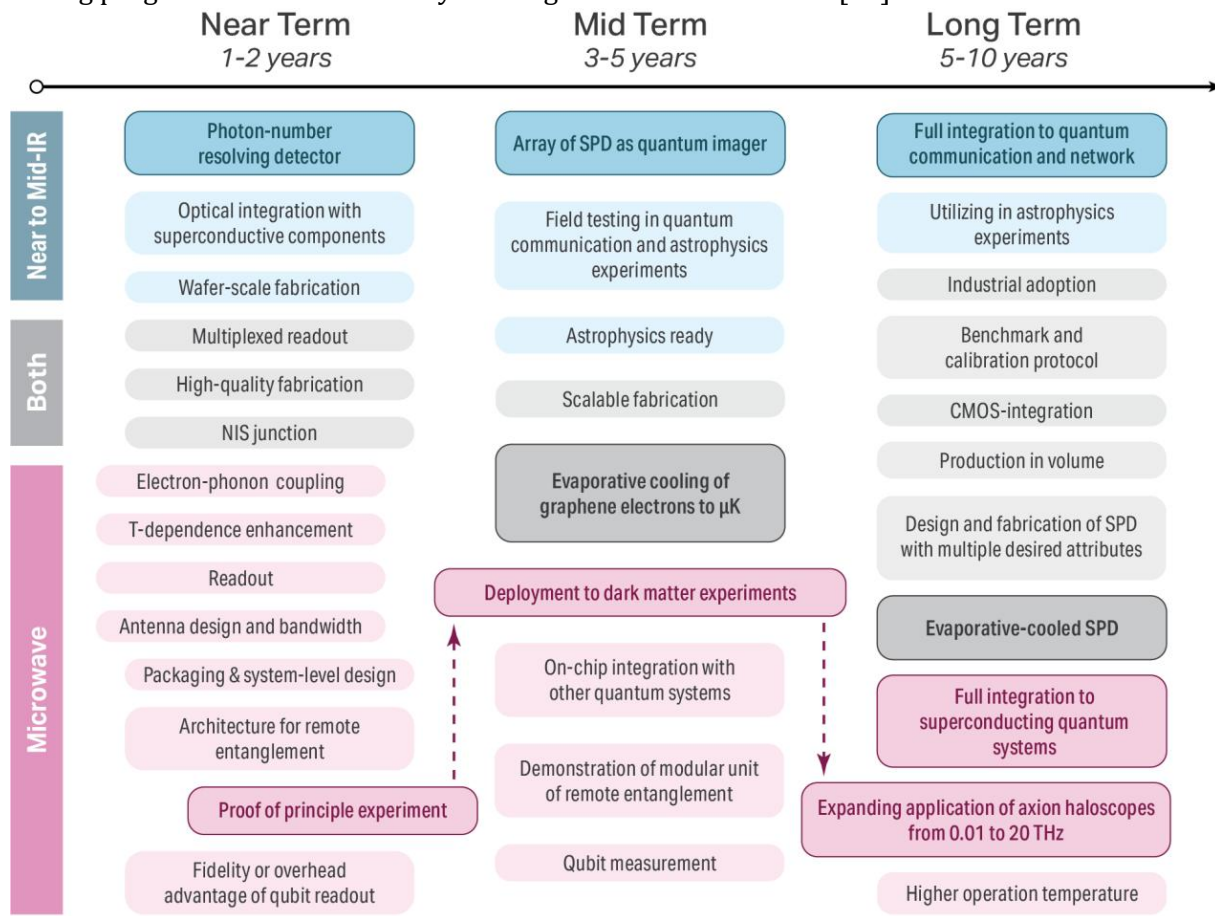

**Figure 3.** Graphene-based single-photon detector roadmap

Overcoming the challenges above will help us to achieve the remarkable goal of wide-bandwidth, microwave SPD. We can aim at a proof-of-principle experiment in short term based on photon statistics. Deployable in mid-term, the most exciting application will be its use in the search of dark matter. Hopefully that will lead to other photon-starved experiments, e.g. spectroscopy for astrophysics, in longer term. For its application to superconducting qubit systems, we can design the remote entanglement modular architecture in short term and implement that in mid-term when the graphene-based SPDs are optimized for qubits experiments. The availability of microwave SPD will hopefully inspire theorists to study its potential advantage in qubit measurement, especially in higher-frequency qubits at higher temperatures. Its optimization for and integration with larger superconducting quantum systems will be a major topic in its mid-term research because each application may demand different attributes of graphene-based SPDs. The goal is to expand the functionality of the conventional qubit system with integration techniques demonstrated with the 2D-material based qubits (see another chapter in this roadmap issue).

## Conclusions

Restricting the thermal mass within the electronic degree of freedom in graphene-based SPDs can also open up unprecedented opportunities. A less explored avenue is the cooling of electrons through the normal-insulator-superconductor (NIS) junction. Previous experiment has demonstrated the cooling of copper down to 1/3 of the starting temperature, limited only by the heat load through the phonons. The cooling of graphene electrons will be more efficient because of weaker electron-phonon coupling [24]. If we can develop a sharp, NIS-tunnel barrier for graphene in short term, a full electron-cooled SPD is possible in mid-term so that we can enjoy the benefit of SPD in higher operation temperatures!

To conclude, I would like to highlight a few less-traveled roads to SPD in the larger family of van der Waals (vdW) materials. For instance, BSCCO [25] and $NbSe_2$ [26] are superconducting 2D-materials that could be patterned to nanowires for SPD. The reduced volume and dimensional, carbon nanotube included, could have a comparatively small heat capacity at a higher superconducting-transition temperatures, esp. BSCCO, to allow SPD at augmented operation temperatures. The flat band in moiré materials could also enable SPD for lower energy photons [18,19]. Amplifier may also be integrated to the detector [27]. As the practical quantum computation and communication are likely going to be realized in many different systems, we will need SPDs operating under various conditions and regimes to meet the demand. Graphene-based SPDs is certainly a member providing unique capability that will grow strong within the SPD family.

## Acknowledgements

K.C.F. acknowledges support from the Department of Energy under Award Number DE-SC0025923. The views and opinions of authors expressed herein do not necessarily state or reflect those of the United States Government or any agency thereof. Fig. 2 image credits to E. Clifford. The views and conclusions contained in this document are those of the authors and should not be interpreted as representing the official policies, either expressed or implied, of the Asian Office of Aerospace Research and Development or the U.S. Government. The U.S. Government is authorized to reproduce and distribute reprints for Government purposes notwithstanding any copyright notation herein.

## References

[1] Walsh E D, Efetov D K, Lee G-H, Heuck M, Crossno J, Ohki T A, Kim P, Englund D and Fong K C 2017 Graphene-Based Josephson-Junction Single-Photon Detector *Phys. Rev. Appl.* **8** 024022

[2] Walsh E D, Jung W, Lee G-H, Efetov D K, Wu B-I, Huang K-F, Ohki T A, Taniguchi T, Watanabe K, Kim P, Englund D and Fong K C 2021 Josephson junction infrared single-photon detector *Science* **372** 409–12

[3] Huang B, Arnault E G, Jung W, Fried C, Russell B J, Watanabe K, Taniguchi T, Henriksen E A, Englund D, Lee G-H and Fong K C 2024 Graphene calorimetric single-photon detector *arXiv [cond-mat.mes-hall]*

[4] Vora H, Kumaravadivel P, Nielsen B and Du X 2011 Bolometric response in graphene based superconducting tunnel junctions *arXiv [cond-mat.mes-hall]*

[5] Liu J, Dona K, Hoshino G, Knirck S, Kurinsky N, Malaker M, Miller D W, Sonnenschein A, Awida M H, Barry P S, Berggren K K, Bowring D, Carosi G, Chang C, Chou A, Khatiwada R, Lewis S, Li J, Nam S W, Noroozian O, Zhou T X and BREAD Collaboration 2022 Broadband solenoidal haloscope for terahertz Axion Detection *Phys. Rev. Lett.* **128** 131801

[6] Day P K, Cothard N F, Albert C, Foote L, Kane E, Eom B H, Basu Thakur R, Janssen R M J, Beyer A, Echternach P M, van Berkel S, Hailey-Dunsheath S, Stevenson T R, Dabironezare S, Baselmans J J A, Glenn J, Bradford C M and Leduc H G 2024 A 25-micrometer Single-Photon-Sensitive Kinetic Inductance Detector *Phys. Rev. X.* **14**

[7] Fong K C and Schwab K C 2012 Ultrasensitive and Wide-Bandwidth Thermal Measurements of Graphene at Low Temperatures *Phys. Rev. X* **2** 031006

[8] Yan J, Kim M-H, Elle J A, Sushkov A B, Jenkins G S, Milchberg H M, Fuhrer M S and Drew H D 2012 Dual-gated bilayer graphene hot-electron bolometer *Nat. Nanotechnol.* **7** 472–8

[9] Du X, Prober D E, Vora H and Mckitterick C B 2014 Graphene-based Bolometers *Graphene 2D Mater.* **1**

[10] Massicotte M, Soavi G, Principi A and Tielrooij K-J 2021 Hot carriers in graphene - fundamentals and applications *Nanoscale* **13** 8376–411

[11] Giazotto F, Heikkilä T T, Pepe G P, Helistö P, Luukanen A and Pekola J P 2008 Ultrasensitive proximity Josephson sensor with kinetic inductance readout *Appl. Phys. Lett.* **92** 162507

[12] Lee G-H, Efetov D K, Jung W, Ranzani L, Walsh E D, Ohki T A, Taniguchi T, Watanabe K, Kim P, Englund D and Fong K C 2020 Graphene-based Josephson junction microwave bolometer *Nature* **586** 42–6

[13] Katti R, Arora H S, Saira O-P, Watanabe K, Taniguchi T, Schwab K C, Roukes M L and Nadj-Perge S 2023 Hot carrier thermalization and Josephson inductance thermometry in a graphene-based microwave circuit *Nano Lett.* **23** 4136–41

[14] Kokkoniemi R, Girard J-P, Hazra D, Laitinen A, Govenius J, Lake R E, Sallinen I, Vesterinen V, Partanen M, Tan J Y, Chan K W, Tan K Y, Hakonen P and Möttönen M 2020 Bolometer operating at the threshold for circuit quantum electrodynamics *Nature* **586** 47–51

[15] Chiles J, Charaev I, Lasenby R, Baryakhtar M, Huang J, Roshko A, Burton G, Colangelo M, Van Tilburg K, Arvanitaki A, Nam S W and Berggren K K 2022 New constraints on dark photon dark matter with superconducting nanowire detectors in an optical haloscope *Phys. Rev. Lett.* **128** 231802

[16] Aamir M A, Moore J N, Lu X, Seifert P, Englund D, Fong K C and Efetov D K 2021 Ultrasensitive Calorimetric Measurements of the Electronic Heat Capacity of Graphene *Nano Lett.* **21** 5330–7

[17] Brun P, MADMAX Collaboration, Caldwell A, Chevalier L, Dvali G, Freire P, Garutti E, Heyminck S, Jochum J, Knirck S, Kramer M, Krieger C, Lasserre T, Lee C, Li X, Lindner A, Majorovits B, Martens R, Matysek M, Millar A, Raffelt G, Redondo J, Reimann O, Ringwald A, Saikawa K, Schaffran J, Schmidt A, Schütte-Engel J, Steffen F, Strandhagen C and Wieching G 2019 A new experimental approach to probe QCD axion dark matter in the mass range above $$40\,\upmu\mathrm{eV$$ 40 μ eV *Eur. Phys. J. C Part. Fields* **79**

[18] Di Battista G, Fong K C, Díez-Carlón A, Watanabe K, Taniguchi T and Efetov D K 2024 Infrared single-photon detection with superconducting magic-angle twisted bilayer graphene *Sci. Adv.* **10** eadp3725

[19] Nowakowski K, Agarwal H, Slizovskiy S, Smeyers R, Wang X, Zheng Z, Barrier J, Barcons Ruiz D, Li G, Bertini R, Ceccanti M, Torre I, Jorissen B, Reserbat-Plantey A, Watanabe K, Taniguchi T, Covaci L, Milošević M V, Fal'ko V, Jarillo-Herrero P, Krishna Kumar R and Koppens F H L 2025 Single-photon detection enabled by negative differential conductivity in moiré superlattices *Science* **389** 644–9

[20] Halbertal D, Ben Shalom M, Uri A, Bagani K, Meltzer A Y, Marcus I, Myasoedov Y, Birkbeck J, Levitov L S, Geim A K and Zeldov E 2017 Imaging resonant dissipation from individual atomic defects in graphene *Science* **358** 1303–6

[21] Li T, Gallop J, Hao L and Romans E 2018 Ballistic Josephson junctions based on CVD graphene *Supercond. Sci. Technol.* **31** 045004

[22] Generalov A A, Viisanen K L, Senior J, Ferreira B R, Ma J, Möttönen M, Prunnila M and Bohuslavskyi H 2024 Wafer-scale CMOS-compatible graphene Josephson field-effect transistors *Appl. Phys. Lett.* **125**

[23]     Amontree J, Yan X, DiMarco C S, Levesque P L, Adel T, Pack J, Holbrook M, Cupo C, Wang Z, Sun D, Biacchi A J, Wilson-Stokes C E, Watanabe K, Taniguchi T, Dean C R, Hight Walker A R, Barmak K, Martel R and Hone J 2024 Reproducible graphene synthesis by oxygen-free chemical vapour deposition *Nature* **630** 636–42

[24]     Vischi F, Carrega M, Braggio A, Paolucci F, Bianco F, Roddaro S and Giazotto F 2020 Electron cooling with graphene-insulator-superconductor tunnel junctions for applications in fast bolometry *Phys. Rev. Appl.* **13**

[25]     Charaev I, Bandurin D A, Bollinger A T, Phinney I Y, Drozdov I, Colangelo M, Butters B A, Taniguchi T, Watanabe K, He X, Medeiros O, Božović I, Jarillo-Herrero P and Berggren K K 2023 Single-photon detection using high-temperature superconductors *Nat. Nanotechnol.* **18** 343–9

[26]     Zugliani L, Palermo A, Scaparra B, Patra A, Wietschorke F, Metuh P, Paralikis A, De Fazio D, Kastl C, Flaschmann R, Munkhbat B, Müller K, Finley J J and Barbone M 2025 Single-photon detection in few-layer NbSe$_2$ superconducting nanowires *arXiv [cond-mat.supr-con]*

[27]     Sarkar J, Maji K, Sunamudi A, Agarwal H, Samanta P, Bhattacharjee A, Rajkhowa R, Patankar M P, Watanabe K, Taniguchi T and Deshmukh M M 2025 Kerr non-linearity enhances the response of a graphene Josephson bolometer *Nat. Commun.* **16** 7043

# 12. Qubits in 2D quantum dots

**Lin Wang[1] and Guido Burkard[2]**

[1] Institute for Advanced Simulation (IAS-4), Forschungszentrum Jülich, Germany
[2] Department of Physics, University of Konstanz, D-78457 Konstanz, Germany

E-mail: l.wang@fz-juelich.de
E-mail: guido.burkard@uni-konstanz.de

**Status**

Semiconductor spin qubits use the intrinsic spin degree of freedom of electrons or holes bound to quantum dots to encode quantum information [1], offering the possibility of dense and scalable registers of high-fidelity qubits. Graphene [2] has emerged as a favorable host material system for spin qubits due to the intrinsic two-dimensional confinement of electrons in combination with the expected weakness of the magnetic noise in the dilute nuclear-spin environment and the weak spin-orbit coupling in carbon [3,4,5]. The theory of spin physics and spin-orbit effects in graphene is well developed [6]. It is understood how the dephasing time of electron spins due to the hyperfine interaction with the nuclear spin ½ of carbon-13 depends on the $^{13}$C concentration (naturally about 1%; $^{12}$C has nuclear spin 0) [7]. However, the absence of a bandgap in pristine monolayer graphene represents a challenge for the required confinement of electrons inside quantum dots (QDs) which is needed for the individual addressability of spin qubits. A possible solution is the use of further confinement into gapped one-dimensional structures such as graphene nanoribbons [3] or semiconducting carbon nanotubes. Following a different strategy, scanning-tip oxidation has provided a versatile method to form nanostructured quantum dots in which carriers are confined in all three dimensions [8]. Both graphene nanoribbons and nanostructured graphene have to deal with the extra disorder which is typically present at the edges of the structure. A very promising path forward which avoids this problem is provided by the electrically tunable bandgap available in bilayer graphene (BLG) [2]. The BLG system thus offers the possibility of a two-dimensional gapped graphene system allowing for electrostatically defined quantum dots [9-18]. Another alternative consists in replacing graphene by intrinsically semiconducting two-dimensional materials, such as the transition-metal dichalcogenides [19]. In recent years, the experimental progress towards qubits in 2D quantum dots has been most rapid in the area of quantum dots formed electrostatically in BLG [9-18]. Measurements have been reported of spin states [8], single-shot spin readout [14], the Pauli spin-blockade effect in electronic transport through a BLG double quantum dot, spin-valley coupling and Kramers pairs [12,16], as well as valley [9,15] and spin [13,14] lifetimes. Theoretical analysis shows that valley lifetimes can be explained by a combination of inter-valley scattering and phonon emission [10], while spin relaxation appears to be due to Rashba spin-orbit coupling combined with phonon emission [11].

**Current and future challenges**

Although spin and valley qubits based on BLG QDs are promising, there still remain some challenges both experimentally and theoretically:

(i) Coherence times: To encode quantum information, long relaxation times and coherence times are main characteristics of good qubits. Very recently, remarkably long spin and valley relaxation times have been reported in single-particle BLG QDs [9,13,14,16] and singlet-triplet BLG double QDs [15]. However, measurements on spin and valley coherence times in BLG QDs remain absent. To measure coherence times, precise state initialization, coherent manipulation and also accurate readout are required.

(ii) Measurement at low magnetic fields: Measurements in BLG QDs at low fields are difficult due to insufficient energy resolution or too high tunnel rates. At low fields, the spin and valley states are almost degenerate. This small energy splitting causes the fidelity of Elzerman readout to markedly drop due to finite temperature and charge noise. In addition, to achieve low tunneling rates at low fields, the fabrication of devices with sufficiently opaque tunneling barriers is required [13].

(iii) Manipulation of the valley degree of freedom: To realize valley qubits in BLG QDs, manipulating the valley degree of freedom is crucial. Different from spin or charge, valleys located at two opposite corners in momentum space are not easily addressable. The valley states can be mixed by the short-range disorder present in real devices. Additionally, the valleys are not directly coupled by uniform magnetic or electric fields. Recently, the valley magnetic moment was reported to couple to the out-of-plane magnetic field when the spatial inversion symmetry is broken by the out-of-plane electric field [9, 12-16]. Even so, how to coherently manipulate the valleys in BLG QDs remains an open question.

(iv) Relaxation mechanisms: Recently, both spin and valley relaxation times were measured in BLG QDs [9, 13-16]. To explain these experiments, theoretical work was carried out on both spin and valley relaxation [10,11]. The spin relaxation times $T_1$ at high magnetic fields can be explained by the Rashba spin-orbit coupling together with the electron-phonon coupling, shown on the left panel of Figure 1. However, at high fields, an observable difference between the experimental measurement and the theoretical calculation appears, which requires further research on spin relaxation. Valley lifetimes can be understood as a result of inter-valley coupling assisted by the electron-phonon scattering via deformation potential and bond-length change (right panel of the figure). However, valley relaxation at low fields needs further investigations both theoretically and experimentally.

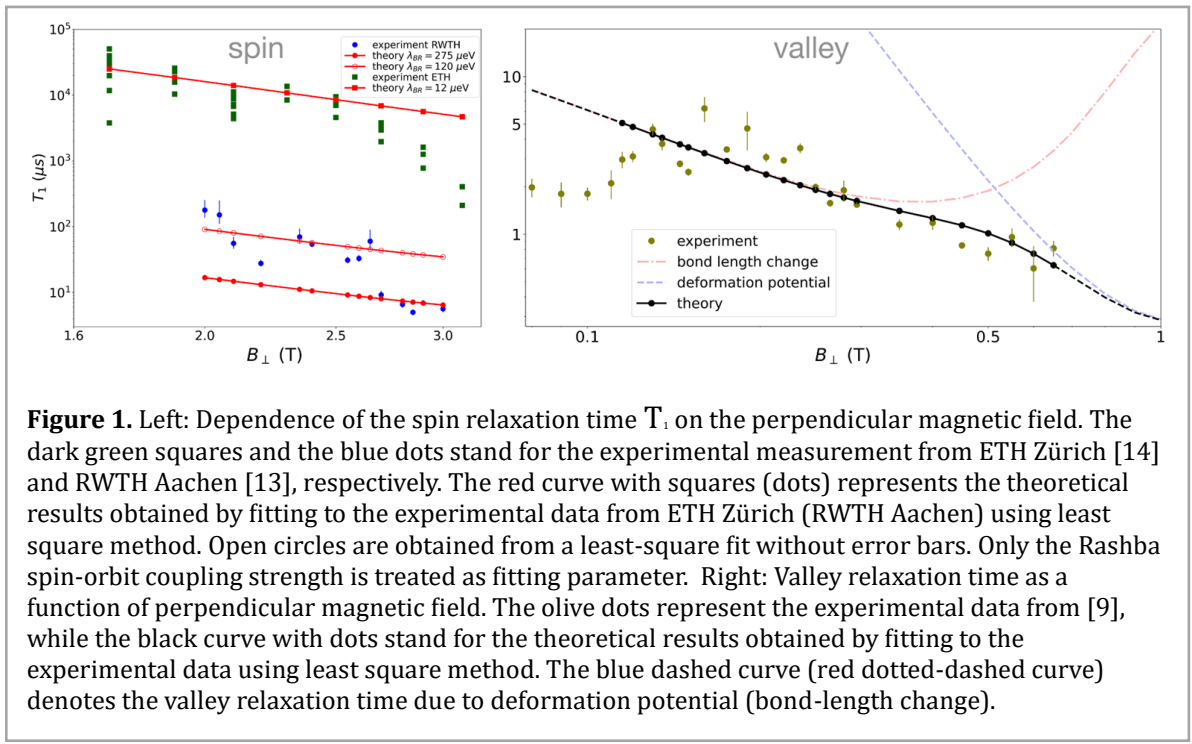

**Figure 1.** Left: Dependence of the spin relaxation time $T_1$ on the perpendicular magnetic field. The dark green squares and the blue dots stand for the experimental measurement from ETH Zürich [14] and RWTH Aachen [13], respectively. The red curve with squares (dots) represents the theoretical results obtained by fitting to the experimental data from ETH Zürich (RWTH Aachen) using least square method. Open circles are obtained from a least-square fit without error bars. Only the Rashba spin-orbit coupling strength is treated as fitting parameter. Right: Valley relaxation time as a function of perpendicular magnetic field. The olive dots represent the experimental data from [9], while the black curve with dots stand for the theoretical results obtained by fitting to the experimental data using least square method. The blue dashed curve (red dotted-dashed curve) denotes the valley relaxation time due to deformation potential (bond-length change).

**Advances in science and technology to meet challenges**

To address the current and future challenges, some key advances in science and technology enable the progress in spin and valley qubits based on BLG QDs.

(i) Charge, spin and valley readout: Single-shot readout is a crucial step towards building quantum computers, as well as implementing quantum algorithms and quantum error detection. Reading out charge, spin, and valley states in BLG QDs is the core challenge in quantum device development. Charge state readout in BLG QDs was achieved by using quantum point contact and radio-frequency reflectometry. Single-shot spin readout has been used to measure spin relaxation times in BLG QDs [14]. The Pauli spin (valley) blockade, a key mechanism for manipulating and reading out spin (valley) qubits, has been established in BLG double QDs. This paves the way for the future study of spin and valley qubits in BLG QDs.

(ii) Coherently drive spin: The electric dipole spin resonance allows to drive spin coherently by using an AC electric field. Recently, Rabi oscillations have been observed in graphene-transition metal dichalcogenide heterostructures. In addition, by pulsing gate voltages, coherent oscillations between singlet and triplet spin states can be driven in double QDs. Pauli spin blockade and tunable exchange has been achieved in BLG QDs.

(iii) Further work on double quantum dots: Pauli spin and valley blockade of tunable two-electron spin and valley states in BLG double QDs have been demonstrated. This paves the way to manipulate and read out spin and valley qubits. In addition, remarkably long valley relaxation times between the valley triplet and singlet states have been reported in BLG double QDs [15]. This demonstrates BLG QDs as a promising platform to host long-lived valley qubits. Theoretical work has started to explore the exchange interaction between spins in adjacent tunnel-coupled BLG quantum dots [20], and further work may provide guidance towards optimal control schemes that lend themselves to the operation of two-qubit entangled quantum gates.

(iv) Spin-orbit engineering: To enable the manipulation of spin degree of freedom, it is necessary to selectively enhance the spin-orbit coupling in BLG. This can be realized by applying perpendicular electric fields to induce Rashba spin-orbit coupling. In addition, stacking BLG in proximity to transition metal dichalcogenides can introduce large spin-orbit coupling in both Ising type and Rashba type [6]. This allows more possibilities to study spintronic and quantum devices.

(v) Theory and modeling: In addition to phonon emission, $1/f$ charge noise has also been taken into account to investigate the spin and valley relaxation [10,11]. $1/f$ charge noise plays an important role at low magnetic fields. This will help us to further understand the behavior of spin and valley relaxation at low fields.

## Concluding remarks

There has been remarkable progress in our understanding of the spin and valley degrees of freedom of single electrons or holes in graphene quantum dots. In particular, the results obtained recently for BLG quantum dots are promising first indicators that spin, valley, or combined spin-valley qubits in graphene may not only be possible, but may have highly favorable properties and may ultimately qualify as good qubits. Once improved procedures for readout, coherent driving, and further understanding of the exchange interaction between graphene quantum dots become available, we will be able to compare graphene-based qubits with their silicon and germanium counterparts, as well as other qubit platforms. Further experimental and theoretical work will be needed to firmly establish the coherence properties of graphene-based qubits, as well as mechanisms underlying both short-range and long-range interactions between qubits that are required for a scalable qubit architecture.

## Acknowledgements

LW was funded in part by the Deutsche Forschungsgemeinschaft (DFG, German Research Foundation) as part of the CRC 1639 NuMeriQS–project no. 511713970. We acknowledge support from Deutsche Forschungsgemeinschaft (DFG, German Research Foundation)—Project No. 425217212—SFB 1432.

## References

[1]     Loss D and DiVincenzo D P 1998 Quantum computation with quantum dots *Phys. Rev. A* **57** 120

[2]     Castro Neto A H, Guinea F, Peres N M R, Novoselov K S and Geim A K 2009 The electronic properties of graphene *Rev. Mod. Phys.* **81** 109

[3]     Trauzettel B, Bulaev D V, Loss D and Burkard G 2007 Spin qubits in graphene quantum dots *Nat. Phys.* **3** 192–196

[4]     Droth M and Burkard G 2016 Spintronics with graphene quantum dots *Phys. Status Solidi RRL* **10** 75-90

[5]     Recher P and Trauzettel B 2010 Quantum dots and spin qubits in graphene *Nanotechnology* **21** 302001

[6]     Zollner K, Kurpas M, Gmitra M. and Fabian J 2025 First-principles determination of spin–orbit coupling parameters in two-dimensional materials *Nat. Rev. Phys.* **7** 255–269

[7]     Fuchs M, Rychkov V and Trauzettel B 2012 Spin decoherence in graphene quantum dots due to hyperfine interaction *Phys. Rev. B* **86** 085301

[8]     Güttinger J, Frey T, Stampfer C, Ihn T and Ensslin K 2010 Spin states in graphene quantum dots *Phys. Rev. Lett.* **105** 116801

[9]		Banszerus L, Hecker K, Wang L, Möller S, Watanabe K, Taniguchi T, Burkard G, Volk C and Stampfer C 2025 Phonon-limited valley life times in single-particle bilayer graphene quantum dots *Phys. Rev. B* **112** 035409

[10]		Wang L and Burkard G 2024 Valley relaxation in a single-electron bilayer graphene quantum dot *Phys. Rev. B* **110** 035409

[11]		Wang L and Burkard G 2025 Spin relaxation in a single-electron bilayer graphene quantum dot *Phys. Rev. Research* **7** 043061

[12]		Banszerus L, Möller S, Steiner C, Icking E, Trellenkamp S,  Lentz F, Watanabe K, Taniguchi T, Volk C and Stampfer C 2021 Spin-valley coupling in single-electron bilayer graphene quantum dots *Nat. Commun.* **12** 5250

[13]		Banszerus L, Hecker K, Möller S, Icking E, Watanabe K, Taniguchi T, Volk C and Stampfer C 2022 Spin relaxation in a single-electron graphene quantum dot *Nat. Commun.* **13** 3637

[14]		Gächter L M, Garreis R, Gerber J D, Ruckriegel M J, Tong C, Kratochwil B, de Vries F K, Kurzmann A, Watanabe K, Taniguchi T, Ihn T, Ensslin K and Huang W W 2022 Single-shot spin readout in graphene quantum dots *PRX Quantum* **3** 020343

[15]		Garreis R, Tong C, Terle J, Ruckriegel M J, Gerber J D, Gächter L M, Watanabe K, Taniguchi T, Ihn T, Ensslin K and Huang W W 2024  Long-lived valley states in bilayer graphene quantum dots *Nat. Phys.* **20** 428

[16]		Denisov A O, Reckova V, Cances S, Ruckriegel M J, Masseroni M, Adam C, Tong C, Gerber J D, Huang W W, Watanabe K, Taniguchi T, Ihn T, Ensslin K and Duprez H 2025 Spin–valley protected Kramers pair in bilayer graphene *Nat. Nanotechnol.* **20** 494–499

[17]		Eich M, Herman F, Pisoni R, Overweg H, Kurzmann A, Lee Y, Rickhaus P, Watanabe K, Taniguchi T, Sigrist M, Ihn T and Ensslin K 2018 Spin and valley states in gate-defined bilayer graphene quantum dots *Phys. Rev. X* **8** 031023

[18]		Banszerus L, Möller S, Hecker K, Icking E, Watanabe K, Taniguchi T, Hassler F, Volk C and Stampfer C 2023 Particle-hole symmetry protects spin-valley blockade in graphene quantum dots *Nature* **618** 51

[19]		Kormányos A, Zólyomi V, Drummond N D and Burkard G 2014 Spin-orbit coupling, quantum dots, and qubits in monolayer transition metal dichalcogenides *Phys. Rev. X* **4** 011034

[20]		Knothe A and Burkard G 2024 Extended Hubbard model describing small multidot arrays in bilayer graphene *Phys. Rev. B* **109**  245401

# 13. Quantum simulation with 2D Moiré systems

**Yihang Zeng[1]**

[1] Department of Physics and Astronomy, Purdue University, West Lafayette, United States
E-mail: zeng334@purdue.edu

**Status**

Emergent phenomena in strongly correlated systems have long stood at the forefront of condensed matter physics. Understanding these phenomena not only reveals fundamental aspects of nature but also holds transformative potential for technology—ranging from energy systems to quantum information processing, next-generation computing, and memory devices. However, the strong interaction terms in many-body Hamiltonians often render perturbative approaches ineffective, presenting profound challenges to theoretical analysis. On the computational front, simulating quantum many-body systems on classical computers remains severely constrained by the exponential scaling of the Hilbert space with system size.

A compelling alternative, proposed by Richard Feynman, is the use of quantum simulators—engineered quantum systems governed by model Hamiltonians of interest[1]. By tuning system parameters, these simulators naturally exhibit the quantum states corresponding to those regimes. Efforts in optical lattices leveraged the individual addressability of atoms; however, their ~mm length scale limited energy scales (e.g., hopping integrals and interaction energies), making it difficult to stabilize low-temperature correlated phases. In contrast, conventional materials with atomic-scale periodicity (<1 Å) offer larger intrinsic energy scales but lack tunability, as their band structures are fixed by atomic lattices. Two-dimensional (2D) artificial lattices—particularly moiré superlattices formed by stacking two slightly mismatched 2D crystals—strike a remarkable balance. These systems feature periodicities of a few to tens of nanometers, yielding intermediate energy scales (~meV) that are large compared to achievable cryogenic temperatures and sample disorder, yet small enough to allow electrostatic gating and band structure engineering.

Moiré systems host flat bands, where quenched kinetic energy amplifies the role of Coulomb interactions. Over the past five years, this platform has revealed a plethora of correlated and topological phenomena: Mott insulators[2], unconventional superconductivity[3], generalized Wigner crystals[4], excitonic insulators[5] and density waves[6], Kondo insulators[7], quantum anomalous Hall[8] and quantum spin Hall effects[9], and fractional Chern insulators (FCI)[10,11]. These discoveries demonstrate not only the richness of emergent phases but also the unprecedented level of control moiré systems offer—via twist angle, displacement fields, magnetic fields, and proximity effect—, showcasing the potential for a moiré quantum simulator[12].

**Current and future challenges**

Despite the remarkable progress, several challenges limit the scalability, reproducibility, and robustness of quantum simulation using moiré materials[13]. Chief among them is sample-to-sample variability, arising from uncontrolled strain, twist angle inhomogeneity, and disorder. Correlated phases in moiré systems are exquisitely sensitive to local stacking configurations and electrostatic environments, complicating reproducibility across devices. Strain engineering presents both an obstacle and an opportunity: while inadvertent strain can suppress uniformity, deliberately applied strain may be used to stabilize novel quantum phases or manipulate electronic structures. Device yield remains a significant bottleneck. The fabrication of dual-gated, atomically clean, and precisely aligned moiré structures with integrated contacts is still confined to a limited number of research groups worldwide. Furthermore, large-area, high-quality moiré superlattices—critical for scattering-based experiments and THz spectroscopy—remain difficult to produce with uniformity.

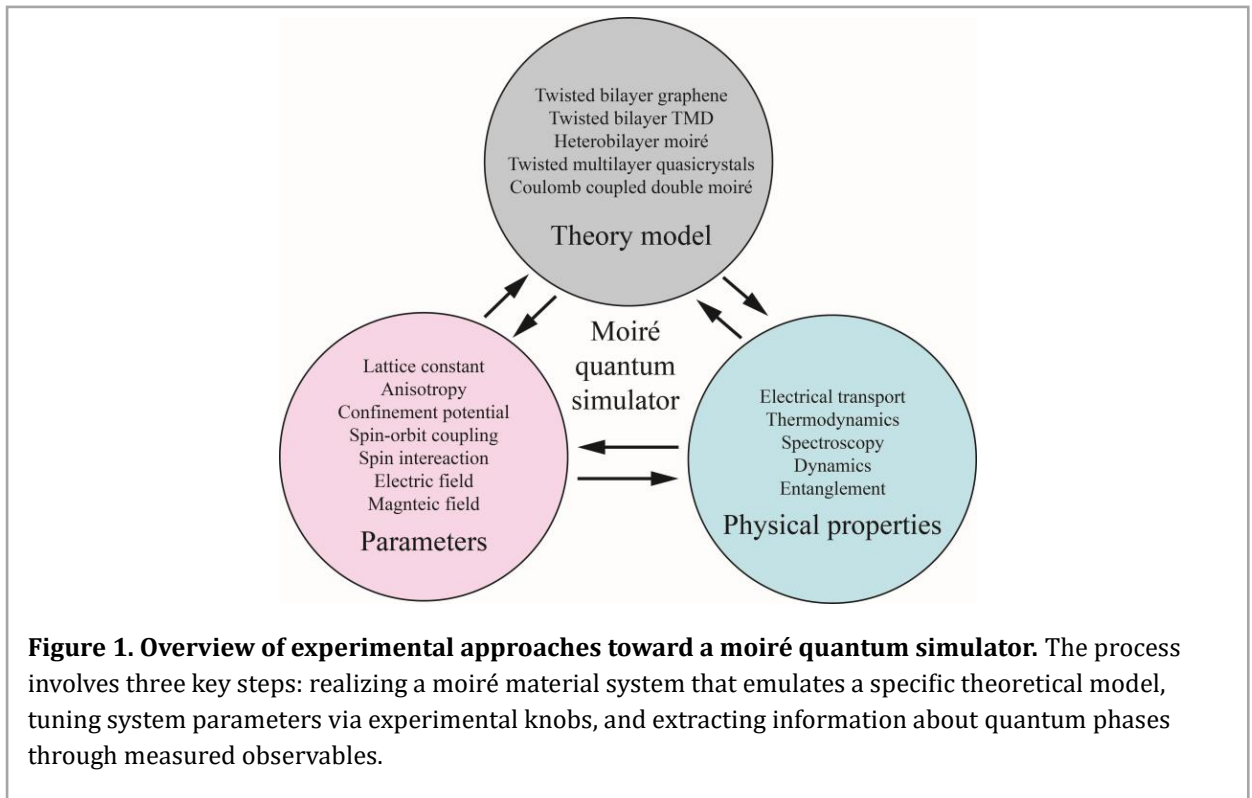

**Figure 1. Overview of experimental approaches toward a moiré quantum simulator.** The process involves three key steps: realizing a moiré material system that emulates a specific theoretical model, tuning system parameters via experimental knobs, and extracting information about quantum phases through measured observables.

The experimental toolbox for probing moiré materials is still maturing. Conventional transport measurements, while insightful, often fall short of revealing the full character of quantum phases. Thermodynamic equation of state measurements (e.g., chemical potential, magnetization, polarization, entropy) offer another angle but are often limited in spatial resolution. Optical techniques face two major hurdles: (1) they provide only indirect evidence of correlated electronic phases; (2) generating sufficient photon flux in the far-infrared to THz regime (the energy scale of moiré mini bands and correlated insulators) for device-scale measurements remains technically difficult. Scanning probe microscopies, offer powerful local diagnostics but are typically incompatible with top-gated structures, limiting their utility across the full range of phase space. Conventional scattering experiments (e.g., neutron or X-ray scattering, ARPES) are similarly constrained by the small lateral size and volume of exfoliated 2D samples, as well as the limited energy resolution of available probes. This lack of comprehensive and scalable detection techniques hampers the crucial feedback loop between theory and experiment, slowing progress in understanding and controlling moiré quantum matter.

On the theoretical side, describing correlated and topological phenomena in moiré superlattices poses deep challenges. The Hilbert space of flat bands with multiple internal degrees of freedom (spin, valley, sublattice) and long-range Coulomb interactions is extremely large, rendering exact solutions infeasible except in the smallest systems. Traditional condensed matter techniques, such as mean-field theory or perturbation theory, often fail to capture the full complexity of the strongly interacting regime. While numerical approaches like exact diagonalization, density matrix renormalization group (DMRG), and tensor networks are making progress, they remain limited by system size and geometry. Furthermore, accurate modelling of the electronic band structure, especially considering the effect of lattice relaxation— is computationally intensive and requires careful regularization. Bridging the gap between model Hamiltonians and experimentally relevant parameters remains a key open problem. A more integrated effort that combines experimental data, symmetry analysis, and large-scale simulation will be critical to advance predictive theory and guide future materials design.

**Advances in science and technology to meet challenges**

Addressing these challenges requires both innovative device manipulation techniques and the development of advanced experimental probes. On the mechanical front, recent studies have demonstrated both in-situ and pre-determined tuning of twist angle and strain in moiré superlattices using atomic force microscopy (AFM)-based manipulation[14,15]. The development of the Quantum Twist Microscope (QTM) has enabled spatially resolved characterization of twist angle, band topology, and correlated phases over large areas—an essential step toward mitigating inhomogeneity and device variability[16]. Additionally, MEMS-based dynamically twisting platforms have shown great promise for continuously tuning the quantum properties of moiré systems[17].

The fabrication of large-area moiré superlattices via metal-assisted exfoliation has also made it feasible to perform conventional scattering experiments and terahertz spectroscopy[18]. These emerging manipulation tools offer precise and reversible control over twist angle, interlayer spacing, and in-plane strain. Simultaneously, device fabrication protocols are being routinely optimized to increase the yield and consistency of high-quality moiré devices.

On the detection side, emerging techniques are addressing the critical need for sensitive and non-invasive measurements of correlated states. A recently developed chemical potential microscope offers a contactless method to probe thermodynamic properties of moiré materials with sub-micron spatial resolution[19]. This technique holds potential for quantitatively studying light-induced phase transitions and non-equilibrium dynamics. A novel approach combining scanning tunneling microscopy (STM) with a graphene sensor layer has enabled thermodynamic measurements at the atomic scale, allowing real-space observation of a 2D generalized Wigner crystal for the first time[20]. SQUID-on-tip magnetometry has been employed to directly image nanoscale magnetization, with improved sensitivity capable of detecting fragile magnetic orders unique to moiré superlattices. Scanning microwave impedance microscopy (sMIM), integrated with specialized device architecture, has recently been used to detect edge states associated with the FCI in twisted $MoTe_2$ devices[21]. Furthermore, nano-ARPES and photocurrent-based terahertz spectroscopy are opening new avenues for probing local band structures and many-body energy gaps.

With continued advances in experimental techniques and growing volumes of data, theoretical modeling of moiré band structures—and their relationship to emergent quantum phenomena—is becoming increasingly refined. The integration of these cutting-edge tools with high-fidelity fabrication methods, novel moiré materials, and improved theoretical frameworks is poised to significantly advance the potential of 2D moiré systems as programmable quantum simulators.

## Concluding remarks

Two-dimensional moiré materials are transforming our ability to simulate and explore emergent quantum phenomena. As the field evolves from discovery to control, innovations in both device engineering and measurement science will be essential. By combining mechanical manipulation with next-generation detection techniques, researchers are beginning to unlock the full potential of these platforms. The goal is not only to emulate known models of correlated matter but to uncover entirely new regimes of topological, fractional, and non-equilibrium quantum physics—accessible only through the unique flexibility of moiré systems. Continued progress in reproducibility, scalability, and probe development will determine how far moiré quantum simulation can go, both as a scientific tool and as a stepping stone toward programmable quantum devices.

## References

[1]        Feynman R P 1982 Simulating physics with computers *Int J Theor Phys* **21** 467–88

[2]        Tang Y, Li L, Li T, Xu Y, Liu S, Barmak K, Watanabe K, Taniguchi T, MacDonald A H, Shan J and Mak K F 2020 Simulation of Hubbard model physics in WSe2/WS2 moiré superlattices *Nature* **579** 353–8

[3]        Cao Y, Fatemi V, Fang S, Watanabe K, Taniguchi T, Kaxiras E and Jarillo-Herrero P 2018 Unconventional superconductivity in magic-angle graphene superlattices *Nature* **556** 43–50

[4]        Xu Y, Liu S, Rhodes D A, Watanabe K, Taniguchi T, Hone J, Elser V, Mak K F and Shan J 2020 Correlated insulating states at fractional fillings of moiré superlattices *Nature* **587** 214–8

[5]        Gu J, Ma L, Liu S, Watanabe K, Taniguchi T, Hone J C, Shan J and Mak K F 2022 Dipolar excitonic insulator in a moiré lattice *Nat. Phys.* **18** 395–400

[6]        Zeng Y, Xia Z, Dery R, Watanabe K, Taniguchi T, Shan J and Mak K F 2023 Exciton density waves in Coulomb-coupled dual moiré lattices *Nat. Mater.* **22** 175–9

[7]        Zhao W, Shen B, Tao Z, Han Z, Kang K, Watanabe K, Taniguchi T, Mak K F and Shan J 2023 Gate-tunable heavy fermions in a moiré Kondo lattice *Nature* **616** 61–5

[8]        Serlin M, Tschirhart C L, Polshyn H, Zhang Y, Zhu J, Watanabe K, Taniguchi T, Balents L and Young A F 2020 Intrinsic quantized anomalous Hall effect in a moiré heterostructure *Science* **367** 900–3

[9]        Zhao W, Kang K, Zhang Y, Knüppel P, Tao Z, Li L, Tschirhart C L, Redekop E, Watanabe K, Taniguchi T, Young A F, Shan J and Mak K F 2024 Realization of the Haldane Chern insulator in a moiré lattice *Nat. Phys.* **20** 275–80

[10]       Zeng Y, Xia Z, Kang K, Zhu J, Knüppel P, Vaswani C, Watanabe K, Taniguchi T, Mak K F and Shan J 2023 Thermodynamic evidence of fractional Chern insulator in moiré MoTe2 *Nature* **622** 69–73

[11]       Cai J, Anderson E, Wang C, Zhang X, Liu X, Holtzmann W, Zhang Y, Fan F, Taniguchi T, Watanabe K, Ran Y, Cao T, Fu L, Xiao D, Yao W and Xu X 2023 Signatures of fractional quantum anomalous Hall states in twisted MoTe2 *Nature* **622** 63–8

[12]       Kennes D M, Claassen M, Xian L, Georges A, Millis A J, Hone J, Dean C R, Basov D N, Pasupathy A N and Rubio A 2021 Moiré heterostructures as a condensed-matter quantum simulator *Nat. Phys.* **17** 155–63

[13]       Lau C N, Bockrath M W, Mak K F and Zhang F 2022 Reproducibility in the fabrication and physics of moiré materials *Nature* **602** 41–50

[14]       Ribeiro-Palau R, Zhang C, Watanabe K, Taniguchi T, Hone J and Dean C R 2018 Twistable electronics with dynamically rotatable heterostructures *Science* **361** 690–3

[15]       Kapfer M, Jessen B S, Eisele M E, Fu M, Danielsen D R, Darlington T P, Moore S L, Finney N R, Marchese A, Hsieh V, Majchrzak P, Jiang Z, Biswas D, Dudin P, Avila J, Watanabe K, Taniguchi T, Ulstrup S, Bøggild P, Schuck P J, Basov D N, Hone J and Dean C R 2023 Programming twist angle and strain profiles in 2D materials *Science* **381** 677–81

[16]       Inbar A, Birkbeck J, Xiao J, Taniguchi T, Watanabe K, Yan B, Oreg Y, Stern A, Berg E and Ilani S 2023 The quantum twisting microscope *Nature* **614** 682–7

[17]       Tang H, Wang Y, Ni X, Watanabe K, Taniguchi T, Jarillo-Herrero P, Fan S, Mazur E, Yacoby A and Cao Y 2024 On-chip multi-degree-of-freedom control of two-dimensional materials *Nature* **632** 1038–44

[18]        Johnson A C, Georgaras J D, Shen X, Yao H, Saunders A P, Zeng H J, Kim H, Sood A, Heinz T F, Lindenberg A M, Luo D, da Jornada F H and Liu F 2024 Hidden phonon highways promote photoinduced interlayer energy transfer in twisted transition metal dichalcogenide heterostructures *Science Advances* **10** eadj8819

[19]        Xia Z, Zeng Y, Shen B, Dery R, Watanabe K, Taniguchi T, Shan J and Mak K F 2024 Optical readout of the chemical potential of two-dimensional electrons *Nat. Photon.* **18** 344–9

[20]        Li H, Li S, Regan E C, Wang D, Zhao W, Kahn S, Yumigeta K, Blei M, Taniguchi T, Watanabe K, Tongay S, Zettl A, Crommie M F and Wang F 2021 Imaging two-dimensional generalized Wigner crystals *Nature* **597** 650–4

[21]        Ji Z, Park H, Barber M E, Hu C, Watanabe K, Taniguchi T, Chu J-H, Xu X and Shen Z-X 2024 Local probe of bulk and edge states in a fractional Chern insulator *Nature* **635** 578–83

# 14. Quantum communication using 2D materials

**Tobias Heindel[1], Serkan Ateş[2] Tobias Vogl[3] and Igor Aharonovich[4, 5]**

[1] Department for Quantum Technology, University of Münster, Heisenbergstraße 11, 48149 Münster, Germany
[2] Faculty of Engineering and Natural Sciences, Sabanci University, 34956, Istanbul, Turkey
[3] TUM School of Computation, Information and Technology, Technical University of Munich, 80333 Munich, Germany
[4] School of Mathematical and Physical Sciences, University of Technology Sydney, Ultimo, New South Wales 2007, Australia
[5] ARC Centre of Excellence for Transformative Meta-Optical Systems, University of Technology Sydney, Ultimo, New South Wales 2007, Australia

E-mail: igor.aharonovich@uts.edu.au

**Status**

Pioneering work on single-photon emission from localized excitons in monolayers of transition metal dichalcogenides (TMDCs) and from defects in hexagonal boron nitride (hBN) opened up a new field of research of quantum information with 2D materials[1]. The suitability of these materials for applications in quantum communications has only recently been demonstrated, with first quantum key distribution (QKD) experiments using defects in hBN and strain engineered localized excitons in WSe$_2$ [2-4]. Figure 1 (a,b) show a typical spectrum from an hBN quantum emitter and a corresponding $g^{(2)}(\tau)$ function, while figure 1(c) shows performance of the WSe$_2$ source in a QKD. A notable difference between these sources is that defects in hBN can operate at room temperature, while single-photon emission in WSe$_2$ has so far only been observed at cryogenic temperatures. TMDCs, on the other hand, can cover all three telecom windows cantered around 850, 1300, and 1550 nm, including wavelengths compatible with atomic transitions used for quantum memories[5]. Nevertheless, the performance of both sources in QKD is comparable. The WSe$_2$-based single-photon source, implemented in a QKD testbed emulating the BB84 protocol[6], achieved sifted key rates of 33.4 kbit/s, a raw $g^{(2)}(0)$ of 0.127, and an average quantum bit error ratio (QBER) down to 0.52% at a clock-rate of the excitation laser of 5 MHz. For hBN, the first QKD experiment employed the B92 protocol (using two non-orthogonal states, offering simpler implementation) with 1 MHz active polarization encoding, resulting in a sifted key rate of 238 bit/s at a QBER of 8.95%. Using the same protocol running at a clock-rate of 40 MHz, the performance could be pushed to a sifted (secure) key rate of 17.5 kbit/s (7 kbit/s) with a QBER of 6.49%[7]. Moreover, hBN has also been utilised to implement the BB84 protocol, reporting a raw (secure) key rate of 432 bit/s (6 bit/s) with a QBER of 6%. All demonstrations were performed in lab-scale QKD experiments using a free-space optical quantum channel and single photons between 630 nm (hBN) and 807 nm (WSe$_2$) wavelength and mean photon numbers per pulse of approximately $\mu = 0.012$. These results, summarized in Table 1, highlight the rapid progress and strong potential of 2D based quantum emitters for integrated quantum communication systems. Noteworthy, none of the 2D material devices used so far for single-photon QKD exploited microcavity approaches to enhance the photon collection efficiency and reduce the radiative lifetime.

**Current and future challenges**

The first experiments on single-photon QKD using emerging 2D materials revealed a performance readily competitive with early QKD-experiments using alternative quantum emitter species[8-10]. To fully exploit the advantage of the deterministic single-photon generation process for QKD as compared to probabilistic sources (i.e., attenuated lasers and SPDCs), however, substantial improvements are required in terms of the mean photon number per pulse $\mu$ in the quantum channel and the maximally achievable tolerable losses. Moreover, a technological challenge concerns the fast,

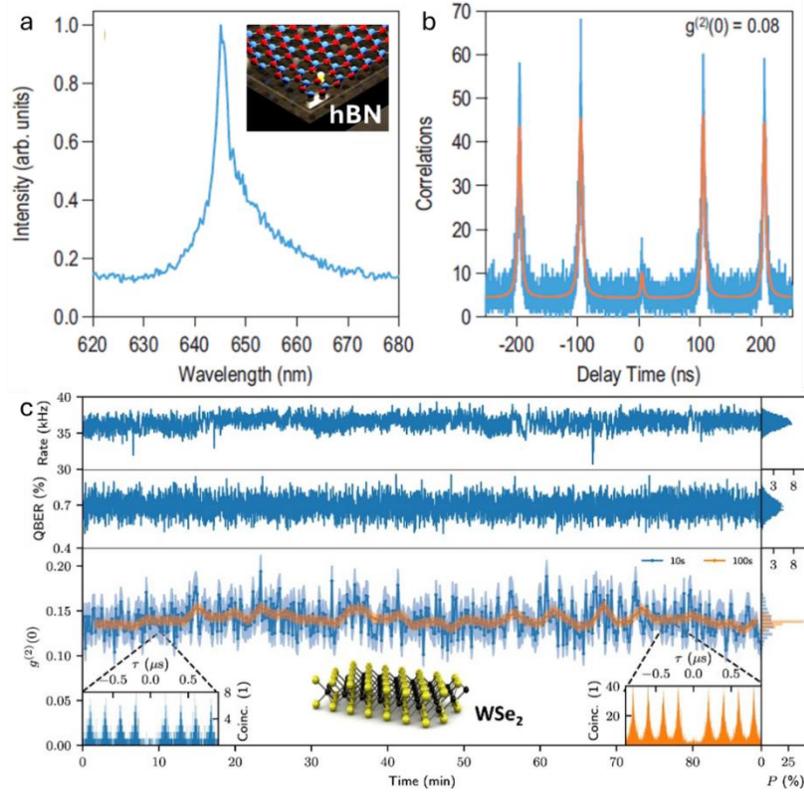

**Figure 1.** *Source performance for hBN (a, b) and TMDC (c) that have been utilised for QKD. In both cases, source stability, purity and brightness are evident, positioning these materials as key contenders for viable QKD technologies. Adapted from Ref [2, 3]*

dynamic polarization switching for preparing the four BB84-states at clock-rates beyond 40 MHz at low QBER, as the extinction ratio of electro-optical modulators heavily depends on the emitter's spectral linewidth. To advance QKD with 2D materials beyond direct point-to-point links, the major challenges to address concern the achievement of high photon indistinguishability[11] and the demonstration of entanglement. While correlated photon pair emission has been observed in WSe$_2$[12], a demonstration of polarization entanglement, is still missing. Also, first work started to explore the photon-indistinguishability, being a crucial quantum resource for advanced protocols in quantum information including measurement-device-independent (MDI) and fully device-independent (DI) QKD allowing for both a higher level of practical security and new network topologies. A major challenge in this regard concerns the improvement of the coherence properties and a reduction of the spectral linewidth of the quantum emitters still being far of the Fourier limit.

Quantum emitters in 2D materials offer key advantages for QKD, including high brightness, and compatibility with on-chip integration. However, several challenges must be overcome to transition of the single-photon sources from proof-of-concept demonstrations to scalable and deployable QKD systems. A major obstacle is the inability to deterministically create emitters with consistent properties[13]. Specifically for hBN, current fabrication techniques, such as thermal annealing or mechanical exfoliation, rely on stochastic processes, resulting in emitters with random spatial distributions and variability in emission characteristics. This challenge has been recently partially rectified within the blue spectral range, with deterministic engineering of the B centre emitting at 436 nm[14-16]. However, this emission wavelength is not ideal for quantum communications.

Photon extraction efficiency is another limiting factor for both defects in hBN and excitons in WSe$_2$. QKD demonstrations to date used devices without nanophotonic structures, resulting in limited photon collection efficiency and limited scalability. Polarization encoding in current systems depends on bulk electro-optic modulators, which hinders miniaturization and integration. Overall, progress is needed in deterministic emitter control, spectral and coherence properties and stability, optical integration, on-chip encoding, and telecom compatibility. This goal is feasible, given the recent progress of telecom emitters in gallium nitride[17], for example.

**Table 1.** QKD experiments employing single-photon source based on emerging 2D materials.

| 2D Material | QKD Protocol | Clock-Rate | Polarization Switching | Sifted /Secure Key Rate (kBit/s) | QBER (%) | g$^{(2)}$(0) | Ref. |
|---|---|---|---|---|---|---|---|
| TMDC (WSe$_2$) | BB84 | 5 MHz | static | 33.4 / NA | 0.52 | 0.127 | [2] |
| hBN | B92 | 1 MHz | dynamic | 0.238/- | 8.95 | 0.120 | [4] |
| hBN | BB84 | 0.5 MHz | dynamic | 0.096/0.006 | 6.00 | 0.080 | [3] |
| hBN | B92 | 40 MHz | dynamic | 17.5/7 | 6.49 | 0.240 | [6] |

The spectral and quantum-optical properties of both hBN and TMDC-based quantum emitters have to be further improved using encapsulation techniques, electrostatic gating, substrate engineering, and charge-control. More generally, controlling of the immediate dielectric environment of these systems is crucial to achieve a deterministic coherent source. Further pushing the emitter linewidth towards the lifetime limit, will in turn enable advance in the coherent resonant driving, which is expected to significantly boost the photon indistinguishability. In combination with implementing spectral tuneability of individual site-controlled quantum emitters, this opens the route for quantum communication in multi-user networking scenarios.

Moving beyond fibre based QKD links, a satellite constellation could be used as the transmission losses in the atmosphere become negligible above 10 km. So far, realizations of quantum communication with satellites only use laser-based light sources[18]. Single photon sources are usually not considered due to their complexity and operational requirements. Quantum emitters in hBN, however, can operate at room temperature, are chemically stable over very long timeframes, insensitive to space radiation[19], and can be packaged onto a CubeSat platform (a commercial pico-class satellite standard)[20].

**Advances in science and technology to meet challenges**

Photon extraction stands often as the key bottleneck in photonic technologies. As such, embedding emitters in optical cavities or photonic waveguides is a promising direction[21, 22] not only for enhancing the photon collection, and hence μ, but also to reduce the emitter lifetime via the Purcell effect allowing for higher clock-rates. This will enable significant improvements in the achievable secure key rates and the tolerable losses. A route for further increasing the end-to-end efficiency in quantum communication relates to the direct coupling of the single-photon emission to optical fibers or waveguide circuits[23]. Developing compact, high-speed modulation schemes remain a critical engineering challenge. Additionally, most hBN emitters operate in the visible spectral range (600–700 nm), which is poorly matched to the low-loss window of optical fibers used in long-distance QKD. To facilitate practical deployment and overcome compatibility issues with existing telecom infrastructure, quantum frequency conversion to telecom C-band wavelengths is a promising route, which is successfully employed for alternative quantum emitter species.

It has been predicted that for typical low Earth orbit altitudes of quantum communication satellites, hBN emitters can already outperform laser-based protocols in terms of the achievable key rate. This, however, comes at the expense of the light source, even for hBN, still being much more complex in comparison to a simple laser diode. In addition, the quantum state manipulation (e.g., setting a specific polarization for the BB84 or B92 protocol) onboard a satellite has not been demonstrated yet and so far, no space-compatible component for this exists. One could instead spatially overlap multiple sources (e.g., how the *Micius* satellite has done this for laser-based sources), but this must be done in a loss-less way to maintain the key rate advantage of single photon sources. Furthermore, this also increases the complexity of the QKD transmitter, as one needs 4 (for BB84) or 2 (for B92) identical single photon sources in all degrees of freedom.

While this is a high price to pay, once this technical challenge has been resolved, one could exploit the large spread of emission wavelengths of hBN quantum emitters and choose one with an emission wavelength at a Fraunhofer line in the solar spectrum. It has been proposed if this wavelength is resonantly enhanced (i.e., the emission is funnelled into a linewidth narrower than the Fraunhofer line), then one can spectrally separate the single photons from the solar background in the space-to-ground scenario and transmit quantum information during daylight conditions. Emitters based on 2D materials are unique for such applications, as the emission wavelength can be tuned via strain over a larger range compared to 3D crystals.

## Concluding remarks

The birth of quantum emitters in 2D systems revived the interest in realisation of quantum communications with these sources. Addressing the challenges discussed above will be critical for realizing practical and scalable QKD systems using quantum light source based on 2D material. Once achieved, this will enable direct point-to-point QKD in free-space optical communication scenarios via telescopes for ad-hoc secure communication, including air-to-ground links via airplanes or satellites.

## Acknowledgements

TH acknowledges financial support by the German Federal Ministry of Research, Technology and Space (BMFTR) via the joint project "tubLAN Q.0" (Grant No. 16KISQ087K) and EU Horizon's 2020 QuantERA II project "COMPHORT" (Grant No. 16KIS2106K). SA acknowledges funding from the EU Horizon 2020 research and innovation programme under the QuantERA II programme with Grant Agreement No. 101017733 (Comphort), and from the Scientific and Technological Research Council of Türkiye (TÜBİTAK) under Grant Agreement Nos. 124N115 and 124N110. IA acknowledges the support from the Australian Research Council (CE200100010, FT220100053, DP250100973) and the the Air Force Office of Scientific Research under award number FA2386-25-1-4044. TV acknowledges funding from the Munich Quantum Valley, which is supported by the Bavarian state government with funds from the Hightech Agenda Bayern Plus and the Deutsche Forschungsgemeinschaft under Germany's Excellence Strategy - EXC-2111-390814868.

## References

[1]      Liu, X.; Hersam, M. C. Nature Reviews Materials *2D materials for quantum information science* 4, 669-684 (2019)

[2]      Gao, T.; von Helversen, M.; Antón-Solanas, C.; Schneider, C.; Heindel, T. npj 2D Materials and Applications *Atomically-thin single-photon sources for quantum communication* 7, 4 (2023)

[3]      Al-Juboori, A.; Zeng, H. Z. J.; Nguyen, M. A. P.; Ai, X.; Laucht, A.; Solntsev, A.; Toth, M.; Malaney, R.; Aharonovich, I. Advanced Quantum Technologies *Quantum Key Distribution Using a Quantum Emitter in Hexagonal Boron Nitride* 6, 2300038 (2023)

[4]      Samaner, Ç.; Paçal, S.; Mutlu, G.; Uyanık, K.; Ateş, S. Advanced Quantum Technologies *Free-Space Quantum Key Distribution with Single Photons from Defects in Hexagonal Boron Nitride* 5, 2200059 (2022)

[5]      Zhao, H.; Pettes, M. T.; Zheng, Y.; Htoon, H. Nat. Commun. *Site-controlled telecom-wavelength single-photon emitters in atomically-thin MoTe2* 12, 6753 (2021)

[6]      Gisin, N.; Ribordy, G. G.; Tittel, W.; Zbinden, H. Reviews of Modern Physics *Quantum cryptography* 74, 145-195 (2002)

[7]      *Tapşın Ö S, Ağlarcı F, Pousa R G, Oi D K L, Gündoğan M, and Ateş S 2025 Secure Quantum Key Distribution Using a Room-Temperature Quantum Emitter arXiv:2501.13902*,

[8]      Vajner, D. A.; Rickert, L.; Gao, T.; Kaymazlar, K.; Heindel, T. Advanced Quantum Technologies *Quantum Communication Using Semiconductor Quantum Dots* 5, 2100116 (2022)

[9]      Y Zhang et al. Phys. Rev. Lett. *Experimental single-photon quantum key distribution surpassing the fundamental weak coherent-state rate limit,* (2025)

[10]     Zahidy, M.; Mikkelsen, M. T.; Müller, R.; Da Lio, B.; Krehbiel, M.; Wang, Y.; Bart, N.; Wieck, A. D.; Ludwig, A.; Galili, M.; Forchhammer, S.; Lodahl, P.; Oxenløwe, L. K.; Bacco, D.; Midolo, L. npj Quantum Information *Quantum key distribution using deterministic single-photon sources over a field-installed fibre link* 10, 2 (2024)

[11]     Fournier, C.; Roux, S.; Watanabe, K.; Taniguchi, T.; Buil, S.; Barjon, J.; Hermier, J.-P.; Delteil, A. Physical Review Applied *Two-Photon Interference from a Quantum Emitter in Hexagonal Boron Nitride* 19, L041003 (2023)

[12]     He, Y.-M.; Iff, O.; Lundt, N.; Baumann, V.; Davanco, M.; Srinivasan, K.; Höfling, S.; Schneider, C. Nat. Commun. *Cascaded emission of single photons from the biexciton in monolayered WSe2* 7, 13409 (2016)

[13]     Klein, J.; Lorke, M.; Florian, M.; Sigger, F.; Sigl, L.; Rey, S.; Wierzbowski, J.; Cerne, J.; Müller, K.; Mitterreiter, E.; Zimmermann, P.; Taniguchi, T.; Watanabe, K.; Wurstbauer, U.; Kaniber, M.; Knap, M.; Schmidt, R.; Finley, J. J.; Holleitner, A. W. Nat. Commun. *Site-selectively generated photon emitters in monolayer MoS2 via local helium ion irradiation* 10, 2755 (2019)

[14]     Fournier, C.; Plaud, A.; Roux, S.; Pierret, A.; Rosticher, M.; Watanabe, K.; Taniguchi, T.; Buil, S.; Quelin, X.; Barjon, J.; Hermier, J.-P.; Delteil, A. Nat. Commun. *Position-controlled quantum emitters with reproducible emission wavelength in hexagonal boron nitride* 12, 3779 (2021)

[15]     Gale, A.; Li, C.; Chen, Y.; Watanabe, K.; Taniguchi, T.; Aharonovich, I.; Toth, M. Acs Photonics *Site-Specific Fabrication of Blue Quantum Emitters in Hexagonal Boron Nitride* 9, 2170-2177 (2022)

[16]     Scognamiglio, D.; Gale, A.; Al-Juboori, A.; Toth, M.; Aharonovich, I. Materials for Quantum Technology *On-demand quantum light sources for underwater communications* 4, 025402 (2024)

[17]     Zhang, H.; Zhang, X.; Eng, J.; Meunier, M.; Yang, Y.; Ling, A.; Zúñiga-Pérez, J.; Gao, W. Physical Review Applied *Metropolitan quantum key distribution using a $\mathrm{Ga}\mathrm{N}$-based room-temperature telecommunication single-photon source* 23, 054022 (2025)

[18]     Yin, J.; Cao, Y.; Li, Y.-H.; Liao, S.-K.; Zhang, L.; Ren, J.-G.; Cai, W.-Q.; Liu, W.-Y.; Li, B.; Dai, H.; Li, G.-B.; Lu, Q.-M.; Gong, Y.-H.; Xu, Y.; Li, S.-L.; Li, F.-Z.; Yin, Y.-Y.; Jiang, Z.-Q.; Li, M.; Jia, J.-J.; Ren, G.; He, D.; Zhou, Y.-L.; Zhang, X.-X.; Wang, N.; Chang, X.; Zhu, Z.-C.; Liu, N.-L.; Chen, Y.-A.; Lu, C.-Y.; Shu, R.; Peng, C.-Z.; Wang, J.-Y.; Pan, J.-W. Science *Satellite-based entanglement distribution over 1200 kilometers* 356, 1140-1144 (2017)

[19]     Vogl, T.; Sripathy, K.; Sharma, A.; Reddy, P.; Sullivan, J.; Machacek, J. R.; Zhang, L.; Karouta, F.; Buchler, B. C.; Doherty, M. W.; Lu, Y.; Lam, P. K. Nat. Commun. *Radiation tolerance of two-dimensional material-based devices for space applications* 10, 1202 (2019)

[20]     Ahmadi, N.; Schwertfeger, S.; Werner, P.; Wiese, L.; Lester, J.; Da Ros, E.; Krause, J.; Ritter, S.; Abasifard, M.; Cholsuk, C.; Krämer, R. G.; Atzeni, S.; Gündoğan, M.; Sachidananda, S.; Pardo, D.; Nolte, S.; Lohrmann, A.; Ling, A.; Bartholomäus, J.; Corrielli, G.; Krutzik, M.; Vogl, T. Advanced Quantum Technologies *QUICK$^3$ - Design of a Satellite-Based Quantum Light Source for Quantum Communication and Extended Physical Theory Tests in Space* 7, 2300343 (2024)

[21]     Nonahal, M.; Horder, J.; Gale, A.; Ding, L.; Li, C.; Hennessey, M.; Ha, S. T.; Toth, M.; Aharonovich, I. Nano Lett. *Deterministic Fabrication of a Coupled Cavity–Emitter System in Hexagonal Boron Nitride* 23, 6645-6650 (2023)

[22]     Iff, O.; Buchinger, Q.; Moczala-Dusanowska, M.; Kamp, M.; Betzold, S.; Davanco, M.; Srinivasan, K.; Tongay, S.; Anton-Solanas, C.; Hofling, S.; Schneider, C. Nano Lett. *Purcell-Enhanced Single Photon Source Based on a Deterministically Placed WSe2 Monolayer Quantum Dot in a Circular Bragg Grating Cavity* 21, 4715-4720 (2021)

[23]     Errando-Herranz, C.; Schöll, E.; Picard, R.; Laini, M.; Gyger, S.; Elshaari, A. W.; Branny, A.; Wennberg, U.; Barbat, S.; Renaud, T.; Sartison, M.; Brotons-Gisbert, M.; Bonato, C.; Gerardot, B. D.; Zwiller, V.; Jöns, K. D. ACS Photonics *Resonance Fluorescence from Waveguide-Coupled, Strain-Localized, Two-Dimensional Quantum Emitters* 8, 1069-1076 (2021)